\documentclass[aps,twocolumn,nofootinbib,preprintnumbers,superscriptaddress,eqsecnum]{revtex4}

\usepackage[bookmarks=true,colorlinks,linkcolor=red,urlcolor=blue,citecolor=blue]{hyperref}

\usepackage[dvips]{color}
\usepackage[normalem]{ulem}
\usepackage{amsmath}
\usepackage{amssymb}
\usepackage{amscd}
\usepackage{enumerate}
\usepackage{amsfonts}
\usepackage{epsfig}
\usepackage{yfonts}

\usepackage{graphicx}
\usepackage{bm}

\definecolor{davecolor}{rgb}{0.95,  0.5,  0.2}

\def\eg{{\it e.g.}}
\def\ie{{\it i.e.}}

\def\({\left(}
\def\){\right)}
\def\[{\left[}
\def\]{\right]}
\def\<{\langle}
\def\>{\rangle}

\def\CC{{\cal C}}

\def\CO{{\cal O}}

\def\tr{\mathop{\rm tr}}

\newcommand\half{{\ensuremath{\frac{1}{2}}}}
\newcommand\p{\ensuremath{\partial}}

\newcommand\field[1]{{\ensuremath{\mathbb{{#1}}}}}

\newcommand\vev[1]{{\ensuremath{\left\langle{#1}\right\rangle}}}

\newcommand{\RR}{\field{R}}

\newcommand{\be}{\begin{equation}}
\newcommand{\ee}{\end{equation}}
\newcommand{\bea}{\begin{eqnarray}}
\newcommand{\eea}{\end{eqnarray}}
\newcommand{\bwt}{\begin{widetext}}
\newcommand{\ewt}{\end{widetext}}

\newcommand{\bi}{\begin{itemize}}
\newcommand{\ei}{\end{itemize}}
\newcommand{\ben}{\begin{enumerate}}
\newcommand{\een}{\end{enumerate}}
\newcommand{\bca}{\begin{cases}}
\newcommand{\eca}{\end{cases}}
\newcommand{\bln}{\begin{align}}
\newcommand{\eln}{\end{align}}
\newcommand{\bst}{\begin{split}}
\newcommand{\est}{\end{split}}

\newcommand\al{{\alpha}}
\newcommand\ep{\epsilon}
\newcommand\sig{\sigma}
\newcommand\Sig{\Sigma}
\newcommand\lam{\lambda}
\newcommand\Lam{\Lambda}
\newcommand\om{\omega}
\newcommand\Om{\Omega}

\newcommand\ga{{\ensuremath{{\gamma}}}}
\newcommand\Ga{{\ensuremath{{\Gamma}}}}
\newcommand\de{{\ensuremath{{\delta}}}}

\newcommand\ze{{\zeta}}

\newcommand\da{{\dagger}}

\def\th{{\theta}}

\newcommand\ov{\over}
\newcommand\ha{{\half}}

\def\le{\left}
\def\ri{\right}

\newcommand\sC{{\ensuremath{{\mathcal C}}}}
\newcommand\sD{{\ensuremath{{\mathcal D}}}}

\newcommand\sI{{\ensuremath{{\mathcal I}}}}
\newcommand\sG{{\ensuremath{{\mathcal G}}}}

\newcommand\sL{{\ensuremath{{\mathcal L}}}}
\newcommand\sM{{\ensuremath{{\mathcal M}}}}
\newcommand\sN{{\ensuremath{{\mathcal N}}}}
\newcommand\sO{{\ensuremath{{\mathcal O}}}}
\newcommand\sR{{\ensuremath{{\mathcal R}}}}

\newcommand\bpsi{{\bar \psi}}

\renewcommand{\Im}{\textrm{Im}\,}
\renewcommand{\Re}{\textrm{Re}\,}

\newcommand{\vk}{{\vec k}}

\newcommand\ut{{\underline{t}}}
\newcommand\ur{{\underline{r}}}
\newcommand\ui{{\underline{i}}}
\newcommand\uj{{\underline{j}}}

\newcommand\uy{{\underline{y}}}

\newcommand\psinorm{\boldsymbol{\psi}}

\newcommand\Phinorm{\boldsymbol{\Phi}}
\def\vertexZ{\Lambda}
\def\Q{\mathcal{Q}}

\newcommand\Psinon{\mathfrak{Y}}

\def\dk{{d^{d-1} k \ov (2 \pi)^{d-1}}}
\newcommand\ca{\mathfrak{a}}
\newcommand\cb{\mathfrak{b}}

\def\SS{{\mathcal{S}}}

\begin{document}

\title{Charge transport by holographic Fermi surfaces}
\preprint{MIT-CTP/4306, CERN-PH-TH/2013-150, UCSD/PTH 13-10,NSF-KITP-13-121}

\author{Thomas Faulkner}
\affiliation{Institute for Advanced Study, Princeton, NJ, 08540}
\author{Nabil Iqbal}
\affiliation{KITP, Santa Barbara, CA 93106}
\author{Hong Liu}
\affiliation{Center for Theoretical Physics,
Massachusetts
Institute of Technology,
Cambridge, MA 02139 }
\author{John McGreevy\footnote
{On 
leave from: Department of Physics, MIT,
Cambridge, Massachusetts 02139, USA.}}
\affiliation{Department of Physics, 
University of California at San Diego,
La Jolla, CA 92093} 
\author{David Vegh}
\affiliation{Theory Group, Physics Department, CERN, CH-1211 Geneva 23, Switzerland}

\begin{abstract}

We compute the contribution to the conductivity
from holographic Fermi surfaces obtained from probe
fermions in an AdS charged black hole. This requires calculating a certain part of the one-loop correction to a vector propagator on the charged black hole geometry.   
We find that the current dissipation is as efficient as possible and the 
transport lifetime coincides with the single-particle lifetime. In particular, 
in the case where the spectral density is that of a marginal Fermi liquid,
the resistivity is linear in temperature.

\end{abstract}


\maketitle
\tableofcontents

\section{Introduction}

For over fifty years our understanding of the low-temperature
properties of metals has been based on Laudau's theory of Fermi liquids.
In Fermi liquid theory, the ground state of an interacting fermionic system is characterized by a Fermi surface in momentum space, and the low energy excitations are weakly interacting fermionic quasiparticles near the Fermi surface. This picture of well-defined quasiparticles close to the Fermi surface provides a powerful tool for obtaining low temperature properties of the system and has been very successful in explaining most metallic states observed in nature, from liquid $^{3}He$ to heavy fermion behavior in rare earth compounds.

Since the mid-eighties, however,  there has been an
accumulation of metallic materials whose thermodynamic and transport properties differ significantly from those predicted by Fermi liquid theory~\cite{varmareview,stewart}. A prime example of these so-called non-Fermi liquids is the strange metal phase of the high $T_c$ cuprates, a funnel-shaped region in the phase diagram emanating from optimal doping at $T=0$, the understanding of which is believed to be essential for deciphering the mechanism for high $T_c$ superconductivity. The anomalous behavior of the strange metal---perhaps most prominently the simple and robust linear temperature dependence of the resistivity---has resisted a satisfactory
theoretical explanation for more than 20 years (see~\cite{phil} for a recent attempt).
While photoemission experiments
in the strange metal phase do reveal
a sharp Fermi surface in momentum space,
various anomalous behavior, including the ``marginal Fermi liquid''~\cite{varma} form of the spectral function and the linear-$T$ resistivity imply that quasiparticle description
breaks down for the low-energy excitations near the Fermi surface~\cite{varma,anderson,kivelson-emery}. Other non-Fermi liquids include heavy fermion systems near a quantum phase transition~\cite{Gegenwart,coleman06}, where similar anomalous behavior to the strange metal has also been observed.

The strange metal behavior of high $T_c$ cuprates and heavy fermion systems
challenges us to formulate a low energy theory of an interacting fermionic
system with a sharp Fermi surface but without quasiparticles (see also~\cite{Senthil:0803,Senthil:0804}).

Recently, techniques from the AdS/CFT correspondence \cite{AdS/CFT} have been used to find
a class of non-Fermi liquids~\cite{Lee:2008xf,Liu:2009dm,Cubrovic:2009ye,Faulkner:2009wj,Faulkner:2010da,Faulkner:2011tm} (for a review see~\cite{Iqbal:2011ae}).
The low energy behavior of these non-Fermi liquids was shown to be governed by a nontrivial infrared (IR) fixed point which exhibits nonanalytic scaling behavior only in the time direction. In particular, the nature of low energy excitations around the Fermi surface
is found to be governed by the scaling dimension $\nu$ of the fermionic operator in the IR
fixed point. For $\nu >\ha$ one finds a Fermi surface with long-lived quasiparticles
while the scaling of the self-energy is in general different from that of the Fermi liquid.
For $\nu \leq \ha$ one instead finds a Fermi surface without quasiparticles.
At $\nu=\ha$ one recovers the ``marginal Fermi liquid'' (MFL)
which has been used to describe the strange metal phase of cuprates. 

In this paper we 
extend the analysis of~\cite{Lee:2008xf,Liu:2009dm,Cubrovic:2009ye,Faulkner:2009wj} to address 
the question of charge transport. We compute 
the contribution to low temperature optical and DC conductivities from such a non-Fermi liquid. We find that the optical and DC conductivities have a scaling form which 
is again characterized by the scaling dimension $\nu$ of the fermionic operators in the IR. 
The behavior of optical conductivity gives an independent confirmation of the absence of 
quasiparticles near the Fermi surface. In particular we find for $\nu = \ha$, which corresponds to MFL, the linear-T resistivity is recovered. 
A summary of the qualitative scaling behavior has been presented earlier in~\cite{Faulkner:2010da}. Here we provide a systematic 
exposition of the rather intricate calculation behind them and also give the numerical prefactors. 

There is one surprise in the numerical results for the prefactors: for certain parameters of the bulk model (co-dimension one in parameter space),
the leading contribution to the DC and optical conductivities vanishes, i.e. the actual conductivities are higher order in temperature than
that presented~\cite{Faulkner:2010da}. This happens because the effective vertex determining the coupling between the fermionic operator and the external DC gauge field vanishes at leading order. The calculation of the leading non-vanishing order for that parameter subspace is complicated 
and will not be attempted here.

While the underlying UV theories in which our models are embedded most likely
have no relation with the UV description of the electronic system underlying the strange metal
behavior of cuprates or a heavy fermion system, it is tantalizing that they share striking similarities in terms of infrared phenomena associated with a Fermi surface without quasiparticles. This points to a certain ``universality'' underlying the low energy behavior of these systems.  The
emergence of an infrared fixed point and the associated scaling phenomena, which dictate the
electron scattering rates and transport, could provide important hints in formulating a low energy theory describing interacting fermionic systems with a sharp surface but no quasiparticles.

\subsection{Set-up of the calculation}

In the rest of this introduction, we describe the set-up of our calculation. Consider a $d$-dimensional conformal field theory (CFT) with a global $U(1)$ symmetry that has an AdS gravity dual
(for reviews of 
applied holography see \eg~\cite{Hartnoll:2009sz, McGreevy:2009xe,CasalderreySolana:2011us,Adams:2012th}). Examples of such theories include the $\sN=4$ super-Yang-Mills (SYM) theory in $d=4$, ABJM theory in $d=3$~\cite{Bagger:2007vi,Gustavsson:2007vu,Aharony:2008ug}, and many others with less supersymmetry. These theories essentially consist of elementary bosons and fermions interacting with non-Abelian gauge fields. The rank $N$ of the gauge group is mapped to the gravitational
constant $G_N$ of the bulk gravity such that ${1 \ov G_N} \propto  N^2$. A typical theory may also contain another coupling constant which is related to 
 the ratio of the curvature radius and the string scale. 
The classical gravity approximation in the bulk corresponds to the large $N$ and strong coupling limit in the boundary theory. The spirit of the discussion of this paper will be similar to that of~\cite{Liu:2009dm,Faulkner:2009wj}; we will not restrict to any specific theory. Since Einstein gravity coupled to matter fields captures universal features of a large class of field theories with a gravity dual, we will simply work with this universal sector, essentially scanning many possible CFTs.
This analysis does involve an important assumption about the spectrum
of charged operators in these CFTs, as we elaborate below.

One can put such a system at a finite density by turning on a chemical potential $\mu$ for 
the $U(1)$ global symmetry. On the gravity side, such a finite density system is described by a charged black hole in $d+1$-dimensional anti-de Sitter spacetime
(AdS$_{d+1}$)~\cite{Romans:1991nq,Chamblin:1999tk}. The conserved current $J_\mu$ of the boundary global $U(1)$ is mapped to a bulk $U(1)$ gauge field $A_M$. The black hole is charged under this gauge field, resulting in a nonzero classical background for the electrostatic potential $A_t(r)$. 
As we review further in the next subsection, in the limit of zero temperature,
the near-horizon geometry of this black hole is of the form $AdS_2 \times \RR^{d-1}$,
exhibiting an emergent scaling symmetry which acts on time but not on space.

The non-Fermi liquids discovered in~\cite{Liu:2009dm,Faulkner:2009wj} were
found by calculating the fermionic response functions in the finite density 
state. This is done by solving Dirac equation for a probe fermionic field in the black hole 
geometry~\eqref{bhmetric1b}. 
The Fermi surface has 
a size of order $O(N^0)$ and various arguments in~\cite{Liu:2009dm,Faulkner:2009wj}
indicate that the fermionic charge density associated with a Fermi surface should also be of order $O(N^0)$. In contrast, the charge density carried by the black hole
is given by the classical geometry, giving rise to a boundary theory density of order $\rho_0 \sim O(G_N^{-1}) \sim O(N^2)$. Thus in the large $N$ limit, we will be studying a small part of a large system, with the background $O(N^2)$ charge density essentially providing a bath for the much smaller $O(N^0)$ fermionic system we are interested in. 
This ensures that we will obtain a well-defined conductivity despite the translation symmetry. See Appendix~\ref{sec:deltaargument} for further elaboration on this point.

\begin{figure}[h]
\begin{center}
\includegraphics[scale=0.3]{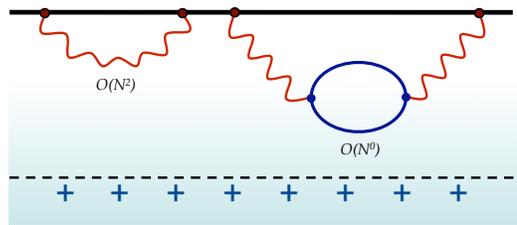}
\end{center}

\caption{Conductivity from gravity. The horizontal solid line denotes the boundary spacetime, 
and the vertical axis denotes the radial direction of the black hole,
which is the direction extra to the boundary spacetime. The dashed line denotes the black hole horizon. 
The black hole
spacetime asymptotes to that of AdS$_4$ near the boundary and factorizes into AdS$_2 \times \RR^2$ near the horizon. The current-current correlator in~\eqref{kubo1} can be obtained from the propagator of the gauge field $A_y$ with endpoints on the boundary. Wavy lines correspond to gauge field propagators and the dark line denotes the bulk propagator for
the probe fermionic field. The left diagram is the tree-level propagator for $A_y$, while the right diagram includes the contribution from a loop of fermionic quanta. The contribution from the Fermi surface associated with corresponding boundary fermionic operator can be extracted from
the diagram on the right.
\label{twopA}
}
\end{figure}

The optical conductivity of the system can be obtained from the Kubo formula 
\be \label{kubo1}
\sig (\Omega) = {1 \ov i \Omega} \vev{J_y (\Omega) J_y (-\Omega)}_{\rm retarded}
\equiv {1 \ov i \Omega} G_R^{yy}
\ee
where $J_y$ is the current density for the global $U(1)$ in $y$ direction at zero spatial
momentum. The DC conductivity is given by\footnote{Note that we are considering a time-reversal invariant system which satisfies $\sig (\Om) = \sig^* (-\Om)$. Thus the definition below is guaranteed to be real.}
\be \label{kubo2}
\sig_{\rm DC} = \lim_{\Om \to 0}  \sig (\Om) \ .
\ee
The right hand side of~\eqref{kubo1} can be computed on the gravity side from the propagator of the gauge field $A_y$ with endpoints on the boundary, as in Fig.~\ref{twopA}. In a $1/N^2$ expansion, the leading contribution---of $O(N^2)$---comes from the background black hole geometry. This reflects the presence of the charged bath, and not the fermionic subsystem in which we are interested. Since these fermions have a density of $O(N^0)$, and will give a contribution to $\sig$ of order $O(N^0)$. Thus to isolate their contribution we must perform a one-loop calculation on the gravity side
as indicated in Fig.~\ref{twopA}. Higher loop diagrams can be ignored since they are suppressed by higher powers in $1/N^2$. 

The one-loop contribution to~\eqref{kubo1} from gravity contains many contributions and we are 
only interested in the part coming from the Fermi surface, which can be unambiguously 
extracted.  This is possible because conductivity from independent channels is additive,
and, as we will see, the contribution of the Fermi surface
is {\it non-analytic} in temperature $T$ as $T \to 0 $.
We emphasize that the behavior of interest to us
is {\it not} the conductivity that one could measure most easily
if one had an experimental instantiation of this system and could hook up a
battery to it.
The bit of interest is swamped by the contribution from the $\CO(N^2)$ charge density,
which however depends analytically on temperature.
In the large-$N$ limit which is well-described by classical gravity,
these contributions appear at different orders in $N$ and can be
clearly distinguished.
In cases where there are multiple Fermi surfaces,
we will see that the `primary' Fermi surface
(this term was used in \cite{Faulkner:2009wj} to denote
the one with the largest $k_F$)
makes the dominant contribution to the conductivity.

\begin{figure}[h]
\begin{center}
\includegraphics[scale=0.4]{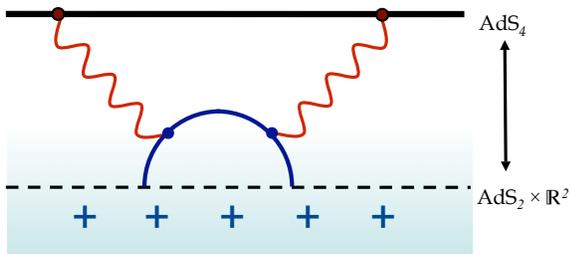}
\end{center}
\caption{The imaginary part of the current-current correlator~\eqref{kubo1} receives its dominant contribution from diagrams in which the fermion loop goes into the horizon. This also gives an intuitive picture that the dissipation of current is controlled by the decay
of the particles running in the loop, which in the bulk occurs by falling into the black hole.
 \label{adiss}}
\end{figure}

Calculations of one-loop Lorentzian processes in a black hole geometry are notoriously subtle. Should one integrate the interaction vertices over the full black hole geometry or only region outside the black hole? How to treat the horizon? How to treat diagrams 
(such as that in figure~\ref{adiss}) in which a loop is cut into half by the horizon? 

One standard strategy is to compute the corresponding correlation function in Euclidean signature, where these issues do not arise and then obtain the Lorentzian expression using analytic continuation  (for  $\Om_l > 0$) 
 \be \label{anaLy}
 G_R^{yy} (\Om) = G_E^{yy} (i\Om_l = \Om + i \ep) \ .
 \ee
This, however, requires precise knowledge of the Euclidean correlation function which 
is not available, given the complexity of the problem. We will adopt a hybrid approach.
We first write down an integral expression for the two-point
current correlation function $G_E^{yy} (i \Om_l)$ in Euclidean signature. We then perform analytic continuation~\eqref{anaLy}  to Lorentzian signature {\it inside the integral}. 
This gives an intrinsic Lorentzian expression for the conductivity.  
The procedure can be considered as the generalization of the procedure discussed in~\cite{Iqbal:2009fd} for tree-level amplitudes to one-loop. 
After analytic continuation, the kind of diagrams indicated in Figure~\ref{adiss}
are included unambiguously. In fact they are the dominant contribution to the dissipative part of
current correlation function, which gives resistivity. 

Typical one-loop processess in gravity also contain UV divergences, which in Fig.~\ref{twopA} 
happen when the two bulk vertices come together. 
We are, however, interested in
the leading contributions to the conductivity from excitations around the Fermi surface, which are 
insensitive to short distance physics in the bulk.\footnote{For such a contribution the two vertices are always far apart along boundary directions.}
Thus we are computing a UV-safe quantity, and short-distance issues will not be relevant.
This aspect is similar to other one-loop applied holography calculations~\cite{Denef:2009kn, CaronHuot:2009iq}.



The plan of the paper is follows. 
In \S2, we first briefly review the physics of the finite density state and 
discuss the leading $O(N^2)$ conductivity which represents
a foreground to our quantity of interest.
\S3 outlines the structure of the one-loop calculation.
\S4 derives a general formula for the DC and optical conductivity respectively  in terms of 
the boundary fermionic spectral function and an effective vertex which can in turn be obtained from an integral 
over bulk on-shell quantities. 
\S5 discusses in detail the leading low temperature behavior of the effective vertices 
for the DC and optical conductivities. In \S6 we first derive the scaling behavior 
of DC and optical conductivities and then  
presents numerical results for the prefactors. \S7 concludes with some further discussion of the main results. 
A number of appendices contain additional background material and
fine details. 



\section{Black hole geometry and $O(N^2)$ conductivity}

In this section we first give a quick review of the AdS charged black hole geometry and then 
consider the leading $O(N^2)$ contribution
to the conductivity\footnote{This was also considered in
\cite{Edalati:2009bi,Paulos:2009yk,Cai:2009zn}.}.
Our motivation is twofold.
Firstly, this represents a `background' from which
we need to extract the Fermi surface contribution;
we will show that this contribution to the conductivity is
analytic in $T$, unlike the Fermi surface contribution.
Secondly, the bulk-to-boundary propagators which determine this answer
are building-blocks of the one-loop calculation required
for the Fermi surface contribution. 


\subsection{Geometry of a charged black hole}

We consider a finite density state for CFT$_d$ by 
turning on a chemical potential $\mu$ for 
a $U(1)$ global symmetry. 
On the gravity side, 
in the absence of other bulk matter fields to take up the charge density, 
such a finite density system is described by a charged black hole in $d+1$-dimensional anti-de Sitter spacetime
(AdS$_{d+1}$)~\cite{Romans:1991nq, Chamblin:1999tk}\footnote
{
For other finite density states
(with various bulk matter contents, corresponding to various operator contents 
of the dual QFT), 
see e.g. 
\cite{Goldstein:2009cv, 
Goldstein:2010aw, Iizuka:2011hg, 
Hartnoll:2010gu, Hartnoll:2010xj, Hartnoll:2010ik, Hartnoll:2011dm, Charmousis:2010zz, Gouteraux:2011ce, cubrovic, cubrovic2, sk1, sk2, Harrison:2011fs, Sachdev:2011ze} and~\cite{Hartnoll:2011fn} for an overview.  
}. The conserved current $J_\mu$ of the boundary global $U(1)$ is mapped to a bulk $U(1)$ gauge field $A_M$. The black hole is charged under this gauge field, resulting in a nonzero classical background for the electrostatic potential $A_t(r)$. For definiteness we take the charge of the black hole to be positive. 
The action for a vector field $A_M$ coupled to AdS$_{d+1}$ gravity can
be written
 \be \label{eq:grac}
 S = {1 \ov 2 \kappa^2} \int d^{d+1} x \,
 \sqrt{-g} \le[\sR +  { d (d-1) \ov R^2} - {R^2 \ov g_F^2} F_{\mu \nu} F^{\mu \nu} \ri]
 \ee
with the black hole geometry given by~\cite{Romans:1991nq, Chamblin:1999tk}
  \bea \label{bhmetric1b}
 ds^2 &\equiv & - g_{tt} dt^2 + g_{rr} dr^2 + g_{ii} d \vec x^2 \cr\cr
& = & {r^2 \ov R^2} (-h dt^2 + d\vec x^2)  + {R^2 \ov r^2} {dr^2 \ov h}
 \eea
 with
 \be \label{bhga2b}
 h = 1 + { Q^2 \ov r^{2d-2}} - {M \ov r^d}, \qquad A_t = \mu \le(1- {r_0^{d-2} \ov  r^{d-2}}\ri) \ .
 \ee
 Note that here we define $g_{tt}$ (and $g^{tt}$) with a positive sign.
$r_0$ is the horizon radius determined by the largest positive root of the redshift factor
\be \label{hord}
h(r_0) =0, \qquad \to \qquad M = r_0^d + {Q^2 \ov r^{d-2}_0}
\ee
and
 \be \label{chem2}
\mu \equiv  {g_F Q \ov c_d R^2 r_0^{d-2}}, \qquad c_d \equiv \sqrt{2 (d-2) \ov d-1} \ .
 \ee
 It is useful to parametrize the charge of the black hole by a length scale $r_*$, defined by
  \be
 Q \equiv \sqrt{d \ov d-2} \, r_*^{d-1} \ .
 \ee
 In terms of $r_*$, the density, chemical potential and temperature
 of the boundary theory are 
  \bea \label{chard}
 \rho ={1 \ov \kappa^2}   \le({r_* \ov R} \ri)^{d-1} {1 \ov e_d}, \qquad \\
\mu  =  {d (d-1) \ov d-2} {r_* \ov R^2} \le({r_* \ov  r_0} \ri)^{d-2} e_d,
\\
 T = {d r_0 \ov 4
\pi R^2} \le(1 - {r_*^{2d-2} \ov  r_0^{2d-2}} \ri) \quad
 \eea
where we have introduced
 \be \label{defed}
 e_d \equiv {g_F \ov \sqrt{2d (d-1)}}  \ .
 \ee

In the extremal limit $T = 0$, $r_0 = r_*$ and the geometry near the horizon (i.e. for ${r-r_* \ov r_*} \ll 1$) is $AdS_2 \times \RR^{d-1}$:
\be
\label{eq:NHgeom}
{ds^2 } = {R_2^2 \ov \ze^2 } \le(- d t^2 +  d \ze^2 \ri) + {r_*^2 \ov R^2} d\vec x^2
 \ee
 where  $R_2$ is the curvature of $AdS_2$, and  
 \be \label{conEl}
 R_2 = {1 \ov \sqrt{d (d-1)}}  R  , \quad \zeta =  \frac{R_2^2}{r-r_*}, \quad  A_t 
 =  {e_d \ov  \ze} \ .
 \ee
In the extremal limit the chemical potential and energy density are given by
\be\label{mudefJ}
\mu  =  {d (d-1) \ov d-2} {r_* \ov R^2} e_d  ,
\quad  \ep={R^{d-1} \ov  \kappa^2} {(d-1)^2 \ov d-2} \le({R^2 \ov r_*}\ri)^d \ 
\ee
with charge density still given by~\eqref{chard}. 

At a finite temperature $T \ll \mu$, ${r_0 - r_* \ov r_*} \ll 1$ and 
the near-horizon metric becomes that of a black hole in  AdS$_2$ times $\RR^{d-1}$
\be \label{ads2T}
 ds^2 =  {R^2_2 \ov \ze^2} \le( - \le(1- {\ze^2 \ov \ze_0^2} \ri)  dt^2 +
 {d \ze^2 \ov 1- {\ze^2 \ov \ze_0^2}} \ri)
 +{r_*^2 \ov R^2} d \vec x^2
 \ee
 with
\be
A_t =  {e_d  \ov \ze}  \le(1-{\ze \ov \ze_0} \ri), \qquad  \ze_0 \equiv  {R_2^2 \ov r_0 - r_*}
\ee
and the temperature 
 \be
 T ={1 \ov 2 \pi \ze_0} \ .
 \ee
 

In this paper we will be interested in 
extracting the leading temperature dependence in the limit $T \to 0$ (but with $T \neq 0$)
of various physical quantities. For this purpose it will be convenient to introduce 
dimensionless variables 
\be \label{nescV}
\xi \equiv {T \ze} = {T R_2^2 \ov r-r_*}, \qquad \xi_0 \equiv T \ze_0 = {1 \ov 2 \pi} , \qquad 
\tau \equiv T t 
\ee
after which~\eqref{ads2T}  becomes
\be \label{ads2T1}
 ds^2 =  { R^2_2 \ov \xi^2} \le( -  \le(1- {\xi^2 \ov \xi_0^2} \ri)  d\tau^2 +
 {d \xi^2 \ov 1- {\xi^2 \ov \xi_0^2}} \ri)
 +{r_*^2 \ov R^2} d \vec x^2
 \ee
 with
\be \label{atu}
A_\tau =  {e_d  \ov \xi}  \le(1-{\xi \ov \xi_0} \ri) \ . 
\ee
Equation~\eqref{ads2T1} can also be directly obtained from~\eqref{bhmetric1b} via a formal decoupling limit with 
\be \label{decoiL}
\xi, \tau = {\rm finite}, \qquad T \to 0 \ . 
\ee
Note that in this limit 
the system is still at a nonzero temperature as $\xi_0 = {1 \ov 2 \pi}$ remains finite.  
 An advantage of~\eqref{ads2T1}--\eqref{atu} is that in terms of these dimensionless variables, $T$ 
 completely drops out of the metric. 


\subsection{Vector boundary to bulk propagator} \label{app:btoB}

We now calculate the conductivity of the finite density state described by~\eqref{bhmetric1b} using   
the Kubo formula~\eqref{kubo1}, to leading order in $T$ in the limit 
\be \label{rioe}
T \to 0, \qquad s \equiv {\Om \ov T} = {\rm fixed} \ .
\ee

To calculate the two-point function of the boundary current $J_y$, we need to consider 
small fluctuations of the gauge field $\delta A_y \equiv a_y$ which is dual to $J_y$ with a nonzero
frequency $\Om$ and $k=0$. In the background of a charged black hole, 
such fluctuations of $a_y$ mix with
the vector fluctuations $h_{ty}$ of the metric, as
we discuss in detail in~Appendix~\ref{app:mix}. 
This mixing has a simple boundary interpretation; acting on  a system with net charges with an electric field causes momentum flows in addition to charge flows. 

After eliminating $h_{ty}$ 
from the equations for $a_y$, we find that (see Appendix~\ref{app:mix} for details)
\begin{equation}
\label{gaugefieldmassiveeqn}
\partial_r\left( \sqrt{-g} g^{rr} g^{yy} \partial_r a_y\right)
- \sqrt{-g} g^{yy}  \left( m_{\rm eff}^2  -  g^{tt} \Omega^2 \right) a_y = 0
\end{equation}
where $a_y$  acquires an $r$-dependent mass given by 
\begin{equation}
\label{eq:meff}
m_{\rm eff}^2 R_2^2 = 2 \left(\frac{r_*}{r} \right)^{2d-2} \ .
\end{equation}

The small frequency and small temperature limit of the solution to 
equation~\eqref{gaugefieldmassiveeqn} can be obtained by the matching technique of~\cite{Faulkner:2009wj};
the calculation parallels closely that of \cite{Faulkner:2009wj},
and was also performed in \cite{Edalati:2009bi}.
The idea is to divide the geometry into two regions 
in each of which the equation can be solved approximately; 
at small frequency, these regions overlap and the approximate solutions
can be matched. For this purpose, we  introduce a crossover radius $r_c$, which satisfies 
\be \label{cross1}
{r_c - r_* \ov r_*} \ll 1, \qquad  \xi_c \equiv 
{ T R_2^2 \ov r_c - r_* }  \ll 1 \ 
\ee
and will refer to the region $r > r_c$ as the {\it outer} (or UV) region and the region $r_0< r < r_c$ as the {\it inner} (or IR) region. In particular, in the $T \to 0$ limit~\eqref{decoiL}, $r_c$ should satisfy 
\be \label{cross}
r_c - r_* \to 0, \qquad \xi_c \to 0  \ .
\ee

To leading order in $T$ in the limit of~\eqref{rioe}, the inner region is simply described by the near-horizon metric~\eqref{ads2T1} with $s$ as the frequency conjugate to $\tau$. In~\eqref{gaugefieldmassiveeqn}, $g_{yy}$ becomes a constant, and  the $r$-dependent effective
mass term in \eqref{eq:meff} goes to a constant value 
\be
 m_{\rm eff}^2 R_2^2 = 2, \qquad r \to r_* \ .
\ee
As a result the IR region differential equation becomes the same
as that of a neutral scalar field in $AdS_2$ with this mass.
Thus at $k=0$, the CFT mode $J_y$ to which the gauge field couples
flows to a scalar operator in the IR. It then follows 
that  the IR scaling dimension of $J_y$ is given by~(see (56) of~\cite{Faulkner:2009wj}) $\Delta_{IR} 
= \half + \nu = 2 $ as
\begin{equation} \label{nuval}
\nu =  \sqrt{ {1 \ov 4} + m_{\rm eff}^2 R_2^2 (r_*) } = {3 \ov 2}  \ .
\end{equation}
Using current conservation this translates into that $\Delta_{IR} (J^t) =1$, i.e.
$J^t$ is a marginal operator in the IR, which is  expected as we are considering a 
compressible system.\footnote{Note that here the IR marginality of $J^t$ only applies to zero momentum. 
That $J^t$ must be marginal in the IR for any compressible system leads to general statements of 
the two-point function of $J^y$ in the low frequency limit~\cite{Hartnoll:2012rj}. 
For example for a system whose IR limit is characterized by a dynamical exponent $z$, 
then $J^y$ must have IR dimension $2 + {d-2 \ov z}$. For $d=3$ this further implies for any $z$
the two point function of $J^y$ must scale like $\Om^3$ at zero momentum.}

Near the boundary of the inner region (i.e. $\xi \to 0$), 
the solutions of~\eqref{gaugefieldmassiveeqn} behave as $a_y \sim \xi^{\ha \pm \nu} \sim \xi^{\ha \pm {3 \ov 2}}$.
We will choose a basis of solutions which are specified as (which also fixes their normalization)
\be \label{innerB}
\eta_I^\pm ( \xi;s) \to \le({r-r_* \ov T R_2^2}\ri)^{-\ha \pm {3 \ov 2}}
= {\xi }^{\ha \mp {3 \ov 2}}, \qquad \xi \to 0 \ .
\ee
Note that since the metric~\eqref{ads2T1} has no explicit $T$-dependence, as a function of $s$~\eqref{rioe}, 
$\eta_I^\pm (s, \xi)$ also have no explicit $T$-dependence. This will be important for our discussion in Sec.~\ref{sec:vertex}. 

The retarded solution (i.e. $a_y$ is in-falling at the horizon) for the inner region can be written as~\cite{Faulkner:2009wj}
\be
\label{eq:retardedIR}
a_y^{\rm (ret)}  ( \xi;s) =  \eta_I^+  + \sG_y (s) \eta_I^- ~.
\ee
where $\sG_y$ is the retarded function for $a_y$ in the AdS$_2$ region, which can be
obtained by setting $\nu={3 \ov 2}$ and $q=0$ in equation (D27) of Appendix D of~\cite{Faulkner:2009wj}\footnote{We copy it here for convenience
\be
\label{eq:ircftG}
 \sG_R (s) =   
 (4 \pi)^{2 \nu} \frac{ \Gamma (-2\nu ) \Gamma \left(\frac{1}{2} +\nu-\frac{i s }{2 \pi }+
i q e_d\right)\Gamma \left(\frac{1}{2}+ \nu-i q e_d\right)}{\Gamma
(2\nu )\Gamma \left(\frac{1}{2}-\nu -\frac{i s }{2 \pi  }+i q e_d
\right)\Gamma \left(\frac{1}{2}- \nu-i q e_d\right)   } \ .
\ee
Note that the above equation differs from of Appendix D of~\cite{Faulkner:2009wj} by a factor of $T^{2\nu}$ due to normalization difference in~\eqref{innerB}. 
}
\begin{equation} \label{irretr}
\mathcal{G}_{y}(s) =  \frac{ i s}{3} \left( s^2 + (2 \pi)^2 \right) ~.
\end{equation}

In the outer region we can expand the solutions to~\eqref{gaugefieldmassiveeqn} 
in terms of analytic series in $\Omega$ and $T$. In particular, the zero-th order equation is 
obtained by setting in~\eqref{gaugefieldmassiveeqn}  $\Om =0$ and $T=0$ (i.e. the background metric becomes that of the extremal black hole). Examining the behavior the resulting equation near $r=r_*$, one finds that  $a_y \sim (r-r_*)^{-\ha \pm {3 \ov 2}}$, which matches with those of in the inner region 
in the crossover region~\eqref{cross}. It is convenient to use the basis of 
the zeroth-order solutions $\eta^{(0)}_{\pm} (r)$ which are specified by the boundary condition 
\be
\label{eq:bulknorm}
\eta_{\pm}^{(0)}(r) 
 \to \({r-r_* \over R_2^2}\)^{- \half \pm {3\over2}}, ~~r \to r_*.
\ee
Note that in this normalization in the overlapping region we have the matching 
\be \label{eq:mat}
\eta^{(0)}_+ \leftrightarrow  T \eta_I^+, \qquad \eta^{(0)}_- \leftrightarrow  T^{-2} \eta_I^- \ .
\ee


Near the AdS$_{d+1}$ boundary, $\eta_\pm^{(0)}$ can be expanded as 
\be \label{woel}
\eta_{\pm}^{(0)}  \buildrel{r\to \infty}\over{\approx} \ca_\pm^{(0)} + \cb_\pm^{(0)} r^{2-d}~~
\ee
with $\ca_\pm^{(0)}, \cb_\pm^{(0)}$ some functions of $k$ (and does not depend $\Om, T$). 
We can now construct the bulk-to-boundary
(retarded) propagator to leading order in the limit~\eqref{rioe}, which we will denote as $K_A (r; \Om)$ with boundary condition $K_A (r; \Om)  \rightarrow 1$ at the $AdS_{d+1}$ boundary ($r \to \infty$). From~\eqref{eq:retardedIR}, the matching~\eqref{eq:mat}, and~\eqref{woel}, we thus find the full bulk-to-boundary propagator is then
\begin{equation}
\label{eq:aysolution}
K_A (r;\Om) = \bca \frac{ \eta_{+}^{(0)} (r) + \mathcal{G}_{y}  (s) T^3 \eta_{-}^{(0)} (r) }{ \ca_+^{(0)} + \mathcal{G}_{y} (s) T^3 \ca_-^{(0)}} &  {\rm outer \; region} \cr\cr
 T \frac{ \eta_I^+ + \mathcal{G}_{y} (s)  \eta_{I}^- }{ \ca_+^{(0)} + \mathcal{G}_{y} (s) T^3 \ca_-^{(0)}} &  {\rm inner \; region}  
 \eca  \ .
\end{equation}
Note that in the above expression all the $T$-dependence is made manifest. 

The leading order solutions $\eta^{(0)}_\pm$ in the outer region can be determined analytically~\cite{Edalati:2009bi}; with
\be
\label{eq:omegazerogaugefield}
  \eta_{+}^{(0)} (r) =  \frac{r_*}{(d-2)R_2^2} \(1 - \(\frac{r_*}{r}\)^{d-2} \) 
  \ee
and thus 
\be \label{neoe}
\ca_+^{(0)}  = \frac{r_*}{(d-2)R_2^2}  , \quad {\cb_+^{(0)} \ov \ca_+^{(0)}} = - r_*^{d-2} \ .
\ee

Useful relations among $\ca_\pm^{(0)}, \cb_\pm^{(0)}$ can be obtained  from the constancy in $r$ of the Wronskian
\be W[a_1, a_2] \equiv a_1 \sqrt g g^{yy}g^{rr} \partial_r a_2 -
a_2 \sqrt g g^{yy}g^{rr} \partial_r a_1 \ee
where $a_{1,2}$ are solutions to \eqref{gaugefieldmassiveeqn}.
In particular, equating 
$W[\eta_{+}^{(0)}, \eta_{-}^{(0)}]$ at boundary and horizon gives 
\begin{equation}
\label{gaugefieldwronskian}
 \cb_{-}^{(0)} \ca_+^{(0)} - \ca_-^{(0)} \cb_{+}^{(0)}=
{ 2\nu \over d-2}r_*^{d-3} R^2 = {3 \ov d-2} r_*^{d-3} R^2~.
\end{equation}
Note that this equation assumes the normalization of 
the gauge field specified in \eqref{eq:bulknorm}.

\subsection{Tree-level conductivity}
\label{subsec:treeconductivity}

We now proceed to study the low-frequency and low-temperature
conductivity at tree level in the charged black hole,
using the boundary to bulk propagator just discussed.

The $O(N^2)$ conductivity is given by the boundary value of the canonical momentum conjugate to $a_y$ (in terms of $r$ foliation) evaluated at the solution~(\ref{eq:aysolution}) 
\be
\sigma(\Omega) = - \lim_{r\to\infty}
\frac{2R^2}{ g_F^2 \kappa^2} 
{ \sqrt{-g} g^{yy}g^{rr} \partial_r K_A \over  i \Omega} 
\ee
which gives 
\begin{equation}
\sigma(\Omega) = 
(d-2)\frac{2R^{3-d}}{ g_F^2 \kappa^2} {1 \ov i\Omega} \frac{\cb_+^{(0)} + \mathcal{G}_{y} T^3 \cb_-^{(0)}}{ \ca_+^{(0)} +
\mathcal{G}_{y}T^3  \ca_-^{(0)} }~~.
\end{equation}
We thus find
\def\cck{{\cal K}}
\bwt
\bea
\label{treeconductivity}
\sigma(\Omega) &= & 
(d-2)\frac{2R^{3-d}}{ g_F^2 \kappa^2}{1 \ov i \Om} \( 
\frac{\cb_+^{(0)}}{\ca_+^{(0)}}
+ \mathcal{G}_{y} T^3 \frac{ \cb_-^{(0)} \ca_+^{(0)} - \ca_-^{(0)} \cb_+^{(0)} }{(\ca_+^{(0)})^2}
+ \cdots \) \\
& = & {\cal K}  (d-2) \le({r_* \ov R^2}\ri)^{d-2} {i \ov  \Om }+  {\cck} \( \frac{d-2}{d(d-1)}\)^{d-3} \( \frac{\mu}{e_d} \)^{d-5}  \left( \Omega^2 + (2 \pi T)^2 \right)
+ \cdots 
\label{finaltreeresult}
\eea
\ewt
where we have used~\eqref{neoe},~\eqref{gaugefieldwronskian} as well as~\eqref{mudefJ} and 
introduced $\cck \equiv \frac{2 R^{d-1}}{g_F^2 \kappa^2}$. $\cck$, which also appears in the 
vacuum two-point function, specifies the normalization of the boundary current and scales as $O(N^2)$. 

Using~\eqref{mudefJ} and~\eqref{chard}, the first term in~\eqref{finaltreeresult}  can also be written as 
\be \label{ballis}
\sig (\Om) = \frac{\rho^2}{\epsilon+P} \frac{i}{\Omega} + \cdots 
\ee
where $P = {\ep \ov d-1}$ is the pressure.  When supplied with the standard $i \epsilon$ prescription, this term gives rise to a contribution proportional to $\delta (\Om)$ in the real part of $\sig (\Om)$. 
This delta function follows entirely from kinematics and  represents a ballistic contribution to the conductivity for a clean charged system with translational and boost invariance, as we review in Appendix~\ref{sec:deltaargument}.  It is also interesting to note that from the bulk perspective the delta function in the conductivity is a direct result of the fact that the fluctuations of the bulk field $a_{y}$ are {\it massive}, as is clear from the perturbation equation~\eqref{gaugefieldmassiveeqn}. This is not a breakdown of gauge-invariance; rather the gauge field acquired a mass through its mixing with the graviton. In the absence of such a mass term the radial equation of motion is trivial in the hydrodynamic limit (as was shown in~\cite{Iqbal:2008by}) and there is no such delta function.

In~\eqref{finaltreeresult}, the important dynamical part is the second term, which gives 
{\it dissipative} part of the conductivity. This part, being proportional to $\Om^2 + (2 \pi T)^2$, is analytic in both $T$ and $\Om$ and the DC conductivity goes to zero in the $T \to 0$ limit. This has a simple physical interpretation; the dissipation of the current arises from the neutral component of the system, whose density goes to zero in the $T \to 0$ limit, leaving us with only the ballistic part~\eqref{ballis} for a clean charged system. We also note that for a clean system such as this, there is no
nontrivial heat conductivity or thermoelectric coefficient at $k=0$,
independent of the charge conduction.
This is because momentum conservation is exact,
and just as in the discussion of Appendix~\ref{sec:deltaargument},
there can therefore be no dissipative part of these
transport coefficients.


\section{Outline of computation of $O(1)$ conductivity}

As explained in the introduction, in order to obtain the contribution of a
Fermi surface to the conductivity, we need to extend the
tree-level gravity calculation of the previous section to the one-loop level with the
corresponding bulk spinor field running in the loop.
This one-loop calculation is rather complicated and is spelled out in detail in the next section.
In this section we outline the main ingredients of the computation 
suppressing the technical details.

\begin{figure}[h]
\begin{center}
\includegraphics[scale=0.3]{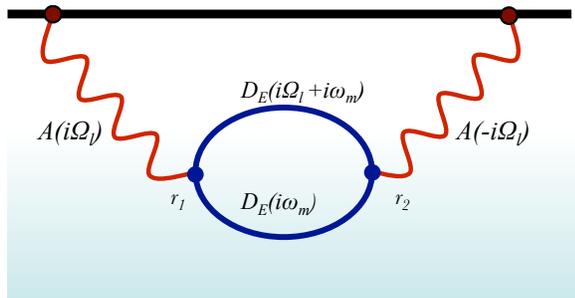}
\caption{
\label{fig:polarization_ads}
The bulk Feynman diagram by which the spinor contributes to the conductivity. 
}
\end{center}
\end{figure}

\subsection{Cartoon description}
\label{subsec:toymodel}

In this subsection we will describe
a toy version of the one-loop conductivity. We will assume that the {\it boundary theory}
retarded Green's function of a fermionic operator
has a Fermi-surface-like pole at $\omega=0$
and $k = k_F$ of the kind
described in \cite{Liu:2009dm,Faulkner:2009wj}.
We will  neglect many ``complications,'' including spinor indices, matrix structures, gauge-graviton mixing, and a host of other important details, which turn out to be inessential in understanding the structure of the calculation. 

The important bulk Feynman diagram 
is depicted in Fig.~\ref{fig:polarization_ads}.
Note that it is structurally very similar to the particle-hole bubble
which contributes to the Fermi liquid conductivity (see e.g. \cite{Schakel, Mahan}):
an external current source creates a fermion-antifermion pair,
which then recombine. The calculation differs from the standard Fermi liquid calculation 
in two important ways. 
The first, obvious difference is that the gravity amplitude
involves integrals over the extra radial dimension of the bulk geometry.
It turns out, however, that these integrals can be
packaged into factors in the amplitude (called $\vertexZ$ below) that play the role
of an effective vertex. 
The second main difference is that actual vertex correction diagrams in the bulk
are suppressed by further powers of $N^{-2}$
and are therefore negligible, at least in the large $N$ limit in which we work. 


We now proceed to outline the computation. While it is more convenient to perform the tree-level calculation of the last section in the Lorentzian signature, for the one-loop calculation it is far simpler to work in Euclidean signature, where one avoids thorny conceptual issues regarding the choice of vacuum and whether interaction vertices should be integrated through the horizon or not.  Our strategy is to
first write down an integral expression for the Euclidean two-point function function\footnote{We put the $i$ in the argument of all Euclidean correlation functions to eventually make
the analytic continuation to Lorentzian signature more natural.} $G_E^{yy} (i \Om_l)$ and then analytically continue to Lorentzian signature inside the integral (for  $\Om_l > 0$)
  \be
 G_R^{yy} (\Om) = G_E^{yy} (i\Om_l = \Om + i \ep) 
 \ee
which will then give us the conductivity via the Kubo formula~\eqref{kubo1} and~\eqref{kubo2}. 
 
 We now turn to the evaluation of the diagram in Figure~\ref{fig:polarization_ads}, which works out to have the structure
 \bwt
\be
G^{yy}_E (i\Om_l)   \sim   T\sum_{i\om_n} \int d \vec k  dr_2 dr_2 
D_E(r_1,r_2;i\Om_l+i\om_n,\vk) K_A(r_1;i\Om_l) D_E(r_2,r_1;i\om_n,\vk) K_A(r_2;- i\Om_l)  \ .
\label{exp1}
\ee
\ewt
The ingredients here require further explanation. $D_E(r_1,r_2;i \om_n,k)$ is the spinor propagator in Euclidean space. $K_A(r;i\Om_l)$ is the boundary-to-bulk propagator for the gauge 
field (i.e. Euclidean analytic continuation of~\eqref{eq:aysolution}); it takes a gauge field source localized at the boundary and propagates it inward, computing its strength at a bulk radius $r$. The vertices have a great deal of matrix structure that we have suppressed, and the actual derivation of this expression from the fundamental formulas of AdS/CFT requires a little bit of manipulation that is discussed in next section, but its structure should appear plausible. The radial integrals $dr$ should be understood as including the relevant metric factors to make the expression covariant, and $d \vec k $ denotes integration over spatial momentum along boundary directions. 

We would now like to perform the Euclidean frequency sum. This is conveniently done using the spectral representation of the Euclidean green's function for the spinor,
\be
D_E(r_1,r_2;i\om_m,\vk) = \int \frac{d{\om}}{(2\pi)} \frac{\rho(r_1,r_2;\om,\vk) }{i\om_m - \om},
\label{spectreprept}
\ee
where $\rho(r_1,r_2;\om,\vk)$ is the bulk-to-bulk spectral density. As we discuss in detail in 
Appendix~\ref{app:bulkP}, $\rho(r_1,r_2;\om,\vk)$ can be further written in terms of boundary theory spectral density $\rho_B(\om, k)$ as (again schematically, suppressing all indices)
\be \label{spefn}
\rho(r,r';\om, \vk) = \psinorm(r; \om, \vk)\rho_B(\om, k)\overline{\psinorm}(r';\om,\vk)
\ee
where $\psinorm(r)$ is the normalizable spinor wavefunction\footnote{As discussed in Appendix~\ref{app:bulkP}, up to normalization, there is a unique normalizable solution with given $\om, \vk$.} to the Dirac equation in the {\it Lorentzian} black hole geometry.  Equation~\eqref{spefn} can be somewhat surprising to some readers and we now pause to discuss it. The bulk-to-bulk spectral density {\it factorizes} in the radial direction; thus in some sense the density of states is largely determined by the analytic structure of the boundary theory spectral density $\rho_B(\om, \vk)$. We will see that this means that despite the presence of the extra radial direction in the bulk, the essential form of one-loop calculations in this framework will be determined by the boundary theory excitation spectrum, with all radial integrals simply determining the structure of interaction vertices that appear very similar to those in field theory. 

 We now turn to the evaluation of the expression \eqref{exp1}. 

\subsection{Performing radial integrals}

\begin{figure}[h]
\begin{center}
\includegraphics[scale=0.4]{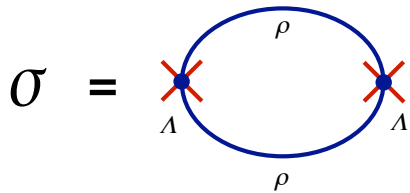}
\caption{
\label{fig:eff}
Final formula for conductivity; radial integrals only determine the effective vertex $\Lam$, with exact propagator for boundary theory fermion running in loop. 
}
\end{center}
\end{figure}

Inserting~\eqref{spectreprept} into~\eqref{exp1} we
can now use standard manipulations from finite-temperature field theory to rewrite the Matsubara sum in \eqref{exp1} in terms of an integral over Lorentzian spectral densities. The key identity is (see  
Appendix~\ref{app:useF} for a discussion), 
\be\label{sunm1}
 T\sum_{\om_m} {1 \ov i(\om_m + \Om_l)- \om_1}\frac{1}{i\om_m  - \om_2}
 = \pm {f (\om_1 ) - f(\om_2) \ov \om_1 - i \Om_l - \om_2} \ 
 \ee
with
 \be
 f(\om) = {1 \ov e^{\beta \om} \pm 1}
 \ee
where the upper (lower) sign is for fermion (boson). Using the above identity~\eqref{exp1} can now be written as 
\bea
&&G^{yy}_E (i\Om_l)  \sim \int d \om_1 d \om_2 d \vec k dr_1 dr_2 \frac{f(\om_1) - f(\om_2)}{\om_1 - \om_2 - i \Om_l } \times \nonumber \\
&& \rho(r_1,r_2;\om, \vk)K_A(r_1;i\Om_l) \rho(r_2,r_1; i \Om_l, \vk) K_A(r_2;-i\Om_l) \ .  \nonumber
\\
\label{sunm2}
\eea
We then analytically continue the above expression to the Lorentzian signature by setting $i\Om_l = \Om + i\epsilon$.  We now realize the true power of the spectral decomposition~\eqref{spectreprept} and~\eqref{spefn}; the bulk-to-bulk propagator factorizes into a product of spinor wavefunctions, allowing us to do each radial integral {\it independently.} In this way all of the radial integrals can be repackaged into an effective vertex
\be \label{effv}
\vertexZ(\om_1,\om_2,\Om, \vk) = \int dr \, \overline{\psinorm}(r;\om_1, \vk)K_A(r;\Om)\psinorm(r;\om_2,\vk),
\ee
where the propagator $K_A(r;\Om)$ has now become a Lorentzian object that propagates infalling waves, and we are left with the formula 
\bwt
\be
G_R^{yy} (\Om) \sim \int d \om_1 d \om_2 d \vk \, {f(\om_1) - f(\om_2 ) \ov \om_1 - \om_2 -\Om - i \ep}
\rho_B(\om_1,\vk)\vertexZ(\om_1,\om_2,\Om, \vk )\rho_B(\om_2, \vk)\vertexZ(\om_2,\om_1,\Om,\vk) \ .
 \label{doncD}
\ee
\ewt
This formula involves only integrals over the {\it boundary} theory spectral densities; the radial integral over spinor and gauge field wavefunctions simply provides an exact derivation of the effective vertex $\vertexZ$ that determines how strongly these fluctuations couple to the external field theory current, as shown in Fig.~\ref{fig:eff}. 

Computing the effective vertex requires a complete solution to the bulk wave equations; however, we will show  that in the low temperature and low frequency limit, the conductivity~\eqref{doncD} is dominated by
the contribution near the Fermi surface, where we can simply replace $\Lam$ by a constant. 
Thus if one is interested in extracting low temperature DC and optical conductivities in the low frequency regime,  the evaluation of~\eqref{doncD} reduces to a familiar one as that in a Fermi liquid (without vertex corrections).   


\section{Conductivity from a spinor field} \label{sec:spinorresistivity}

In this section we describe in detail the calculation leading to~\eqref{doncD}, paying attention to all subtleties. For readers who wants to skip the detailed derivation, the final results for the optical and DC conductivities are given by~\eqref{necon1} and~\eqref{eq:finalsig}, with the relevant vertices given by (for $d=3$)~\eqref{exm11}--\eqref{zzz1}.

Before going into details, it is worth mentioning some important complications which we ignored in the last section:

\ben 

\item As already discussed in Sec.~\ref{app:btoB}, the gauge field perturbations on the black hole geometry mix with the graviton perturbations;
a boundary source for the bulk gauge field will also lead to perturbations in metric, and as a result the propagator $K_A$ in~\eqref{exp1} should be supplemented by a graviton component. Thus the effective vertex $\vertexZ$ is rather more involved than the schematic form given in~\eqref{effv}.

\item Another side effect of the mixing with graviton is that, in addition to Fig.~\ref{fig:polarization_ads}, the conductivity  also receives a contribution from the `seagull' diagram of Fig.~\ref{fig:seagull}, coming 
from quartic vertices involving  terms quadratic in metric perturbations (given in Appendix~\ref{app:cou}). We will show in  Appendix~\ref{app:other} that such contributions give only subleading corrections in the low temperature limit and will be omitted. 

\item The careful treatment of spinor fields and associated indices will require some care.

\een
\begin{figure}[h!]
\centerline{\hbox{\psfig{figure=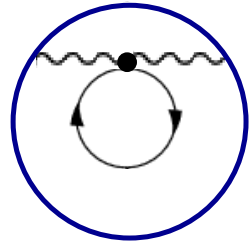, scale=.7}}}
\caption{The `seagull' diagram coming from quartic vertices involving terms quadratic in bosonic perturbations. The boundary is represented by a circle. 
}
\label{fig:seagull} 
\end{figure}

\subsection{A general formula}

We consider a free spinor field in~\eqref{bhmetric1b} with an action
 \be \label{Dirac}
 S = - \int d^{d+1} x \sqrt{-g} \, i (\bar \psi \Ga^M \sD_M \psi  - m \bpsi \psi)
 \ee
where $\bpsi = \psi^\da \Ga^\ut$ and
 \be
 \sD_M  = \p_M + {1 \ov 4} \om_{ab M} \Ga^{ab} - i q A_M
 \ee
The abstract spacetime indices are $M , N \cdots$ and
the abstract tangent space indices are $a,b, \cdots$. The index with an underline
denotes that in tangent space. Thus $\Ga^a$ to denote gamma matrices
in the tangent frame and $\Ga^M$ those in curved coordinates. According to this convention, for example, 
 \be
 \Ga^M = \Ga^a e_a{^M} , \qquad \Ga^r = \sqrt{g^{rr}} \Ga^{\ur} \ .
 \ee
 We will make our discussion slightly more general, applicable to a background metric 
given by the first line of~\eqref{bhmetric1b}  with $g_{MN}$ depending on $r$ only. Also for notational simplicity we will denote 
\def\dk{\int_k}
\be 
\dk = \int {d^{d-1} k \ov (2 \pi)^{d-1}}
\ee

We are interested in computing the one-loop correction to the retarded two-point function
of the boundary vector current due to a bulk spinor field.
In the presence of a background gauge field profile,
the fluctuations of the bulk gauge field mix with those
of the metric. Consider small perturbations in $a_j \equiv \delta A_j $ and  $h_t^j = \de g_t^j$.
In Appendix~\ref{app:cou} we find that the corrections to the Dirac action are given at cubic order by
\bwt
 \be
 \delta S_3[a_j, h^j_t, \psi] =- i \sum_j \int d^{d+1} x \sqrt{-g} \bar \psi \le(-  h_t^j \Ga^{t} \p_j + {1 \ov 8}  g_{jj} \, \p_r h_t^j \,
  \Ga^{r t j} - i  q a_j \Ga^{j} \ri) \psi \ 
 \ee
 \ewt
 where $\Ga^{r t j} \equiv \Ga^t \Ga^r \Ga^j$. Note that in the above equation and below the summations over boundary spatial indecies will be indicated explicitly (with no summation associated with repeated indices). 
There are also quartic corrections (involving terms quadratic in bosonic fluctuations) which are given in Appendix~\ref{app:cou}. These terms give only subleading corrections, as we discuss in Appendix~\ref{app:other}.

Now we go to Euclidean signature via
\be
t \to  -i t_E \qquad \om \to  i\om_E \qquad  A_t \to  iA_{\tau} \qquad iS \to  -S_E \ .
\ee
It's helpful to keep in mind that $\psi$ and $\psi^{\dagger}$ don't change under the continuation, and we do not change $\Ga^{\ut}$. The Euclidean spinor action can then be written as
 \be
 S_E = i \int d^{d+1}x \sqrt{g} \, \bpsi \left( \Gamma^M D_M^{(0)}  - m \ri) \psi
 + \delta S_3 + \cdots
 \ee
where $\delta S_3$ can be written in momentum space as
 \bea&&
 \de S_3 = - T^2 \sum_{\om_m} \sum_{\Om_l} 
 \dk \int dr \sqrt{-g} \,
\cr
&&
~\times  \bar \psi (r; i\om_m + i\Om_l, \vk) \, B (r; i\Om_l, \vk) \, \psi (r;i\om_m, \vk)\ .
 \eea
Here the kernel $B(r;i\Om_l,\vec{k})$ contains all dependence on the gauge and metric fluctuations, which we have also Fourier expanded:
\bwt
 \bea \label{kern}
 &&B(r; i\Om_l, \vk) = 
 - i \sum_j \le(- i  k_j   h_t^j (r;i\Om_l) \Ga^{t} + {g_{jj} \ov 8}  \p_r h_t^j (r;i\Om_l) \, \Ga^{r t j} - i  q a_j (r;i\Om_l) \Ga^{j}  \ri) \ .
 \eea
\ewt
Note that since we are only interested in calculating the conductivity at zero spatial momentum, we have taken $a_j$ and $h_t^j$ to have zero spatial momentum.

We now evaluate the one-loop determinant by integrating out the fermion field. We seek the quadratic dependence on the gauge and graviton fields; the relevant term in the effective action is given by the Feynman diagram in Figure \ref{fig:polarization_ads} and is
 \bea
&& \Ga[a_j,h^j_t] =  -{T^2 \ov 2} \sum_{\om_m, \Om_l} 
\dk \,
 \int dr_1 \sqrt{g(r_1)} dr_2 \sqrt{g(r_2)} \, \cr
&& \quad \times \tr \le(D_E(r_1,r_2;i\om_m + i\Om_l,\vec{k}) B (r_2; \Om_l, \vk)
\right.\cr &&\left.
\qquad \quad D_E(r_2,r_1; i\om_m,\vec{k})
B (r_1; -\Om_l, \vk) \ri)
\label{efac}
 \eea
where the tr denotes the trace in the bulk spinor indices and $D_E(r_1,r_2;i\om_m,\vec{k})$ denotes
the bulk spinor propagator in the Euclidean signature. As always we suppress bulk spinor indicies. The bulk spinor propagator is discussed in some detail in Appendix \ref{app:bulkP}. The single most important property that we use is its spectral decomposition
 \be
D_E(r_1,r_2;i\om_m,\vk) = \int \frac{d{\omega}}{2\pi} \frac{\rho(r_1,r_2;\omega,\vk) }{i\om_m - \omega},
\label{spectrep}
\ee
where $\rho(r_1,r_2;\omega,\vk)$ is the bulk spectral function. As is shown in show in \eqref{ImGspinorbulk} of Appendix~\ref{app:bulkP}, the bulk spectral function can be written in terms of
that of the boundary theory as
  \begin{equation}
  \label{specBB}
\rho(r,r';\omega,\vk) = \psinorm_{\alpha}(r) \, \rho_B^{\alpha \gamma} (\om, \vk) \, \overline{\psinorm_{\gamma}} (r')
\end{equation}
where the boundary spectral function $\rho_B$ is Hermitian and $\psinorm$ is the normalizable {\it Lorentzian} wave function for the free Dirac equation in the black hole geometry\footnote{
For a precise definition of the normalizable Lorentzian wave function, see again Appendix~\ref{app:bulkP}. In this section, to avoid clutter, we will use the boldface font to denote the normalizable solution.  The non-normalizable solution will not appear.}.
$\rho_B(\om,k)$ is the boundary theory spectral density of the holographic non-Fermi liquid, and was discussed in detail in \cite{Faulkner:2009wj} (see Appendix~\ref{app:spd} for a review). Note that in~\eqref{specBB} bulk spinor indices are suppressed and $\alpha,\gamma$ in $\psinorm$ label independent normalizable solutions and as discussed in Appendix~\ref{app:bulkP} can be interpreted as the boundary spinor indices.

 Now introduce various momentum space Euclidean {\it boundary} to bulk propagators
 for the gauge field and graviton. 
 \bea
 \label{bulktoboundarypropagation}
&& a_j (r; i\Om_l) = K_A (r;i\Om_l) A_j (i\Om_l), \cr
&&
 \quad h_t^j (r; i\Om_l) = K_h (r;i\Om_l) A_j (i\Om_l)
 \eea
where $A_j (i\Om_l)$ is the source for the boundary conserved current
in Euclidean signature.  These are objects which propagate a gauge field source at the boundary
 to a gauge field or a metric fluctuation in the interior, and so
 should perhaps be called $K_A^A$ and $K_h^A$; we drop the second `$A$' label
 since we will never insert metric sources in this paper.

 $K_A (r;i\Om_l)$ and $K_h (r;i\Om_l)$ go to zero (for any nonzero $\Om_l$) at the horizon
and they do not depend on the index $j$ due to rotational symmetry. $K_h$ and $K_A$ are not independent; in~Appendix~\ref{app:mix} we show that 
 \be
 \label{eq:KAKh}
     \p_r K_h  = - \CC \sqrt{g_{rr} g_{tt}} g_{ii}^{-{d+1 \ov 2} }  K_A , \quad 
 \ee
 with
 \be \label{defcc}
  \CC =2 \kappa^2 \rho
  \ee
where $\rho$ is the background charge density~\eqref{chard}. 

Now using the definition of the propagators \eqref{bulktoboundarypropagation} we can write the kernel $B(r;i\Om_l, \vk)$ in ~\eqref{kern} in terms of a new object $Q^j(r;, i\Om_l, \vk)$ that has the source explicitly extracted:
 \be \label{Bre}
 B  (r; i\Om_l, \vk) = \sum_j Q^j (r; i\Om_l, \vk)  A_j (i\Om_l)
 \ee
with
\bwt
  \be \label{kern1}
 Q^j (r; i\Om_l, \vk) =- i \le(- i  k_j  K_h (r;i\Om_l) \Ga^{t} + {g_{jj} \ov 8}   \p_r K_h (r;i\Om_l) \Ga^{r t j} - i  q K_A (r;i\Om_l) \Ga^{j}  \ri) \ .
 \ee
Plugging~\eqref{Bre} into~\eqref{efac}, we can now express the entire expression in terms of the boundary gauge field source $A_j(i\Om_l)$. Taking two functional derivatives of this expression with respect to $A_j$, we find that the the boundary Euclidean current correlator can now be written as
 \be
 G_E^{ij} (i\Om_l) = -T \sum_{\om_m} \dk  
 \int dr_1 \sqrt{g(r_1)} dr_2 \sqrt{g(r_2)} 
\tr \le(D_E(r_1,r_2;i\om_m + i\Om_l,\vec{k}) Q^i (r_2; i\Om_l, \vk) D_E(r_2,r_1; i \om_m,\vec{k})
Q^j (r_1; -i\Om_l, \vk) \ri) 
\label{GEx}
 \ee
Note that all objects in here are entirely well-defined and self-contained; we now need only evaluate this expression. 

To begin this process we first plug~\eqref{spectrep} into~\eqref{GEx} and then perform the sum over $\om_m$ using the techniques that was outlined earlier in~\eqref{sunm1}--\eqref{sunm2}. We find that
equation~\eqref{GEx} can be further written as
  \bea
 G_E^{ij} (i\Om_l) &=&   - \dk
 \int {d \om_1 \ov 2 \pi} {d \om_2 \ov 2 \pi}  \int dr_1 \sqrt{g(r_1)} dr_2 \sqrt{g(r_2)} \, {f(\om_1 ) - f(\om_2) \ov \om_1 -i \Om_l - \om_2 }
 \cr
 && \quad \times 
 \tr \le(\rho (r_1,r_2;\om_1 ,\vec{k})  Q^i (r_2; i\Om_l, \vk) \rho (r_2,r_1; \om_2,\vec{k})
Q^j (r_1; -i\Om_l, \vk) \ri) \label{grex},
 \eea
Now  plugging~\eqref{specBB} into~\eqref{grex} we find that
   \bea
 G_E^{ij} (i\Om_l) &=&   -
 \dk
 \int {d \om_1 \ov 2 \pi} {d \om_2 \ov 2 \pi}  \, {f(\om_1) - f(\om_2) \ov \om_1 -i \Om_l - \om_2} 
\rho_B^{\alpha \beta} (\om_1 ,\vec{k}) \, \vertexZ_{\beta \gamma}^i (\om_1, \om_2, i\Om_l, \vk) \, \rho_B^{\gamma \delta} (\om_2 ,\vec{k}) \,
 \vertexZ_{\delta \alpha}^j (\om_2, \om_1, -i\Om_l, \vk) \label{retx2}
\eea
\ewt
with
 \bea
 && \vertexZ_{\beta \gamma}^i (\om_1, \om_2, i\Om_l,\vk) = \nonumber \\ 
 &&  \int dr \sqrt{g} \, \overline{\psinorm_\beta} (r; \om_1,\vk) \, Q^i (r;  i\Om_l; \vk) \, \psinorm_\gamma (r; \om_2, \vk) \ .
 \eea
Let us now pause for a moment to examine this expression. In the last series of manipulations we replaced the interior frequency sum with an integral over real Lorentzian frequencies; however by doing this we exploited the fact that the spectral density {\it factorizes} in $r$. This allowed us to absorb all radial integrals into $\vertexZ$, which should be thought of as an effective vertex for the virtual spinor fluctuations. Now only the {\it boundary theory} spectral density $\rho_B$ appears explicitly in the expression. This form for the expression is perhaps not surprising from a field-theoretical point of view; however it is interesting that we have an {\it exact} expression for the vertex, found by evaluating radial integrals over normalizable wave functions. 

We now obtain the retarded Green function for the currents by starting with $G_E ^{ij}(i\Om_l)$ for $\Om_l > 0$ and analytically continuing $G_E^{ij} (i\Om_l)$ to Lorentzian signature via
 \be
 G_R^{ij} (\Om) = G_E^{ij} (i\Om_l = \Om + i \ep) \
 \ee
We will suppress the $i \ep$ in equations below for notational simplicity but it is crucial to keep it in mind. For simplicity of notations we will denote the Lorentzian analytic continuation of various quantities only by their argument, \eg\
\be
 K_A (r; i\Om_l)|_{i \Om_l = \Om}  \to K_A (r; \Om)
 \ee
We also make the analogous replacements also for $K_h, Q^i$ and $\vertexZ_{\alpha \beta}$.
Note that $K_A (r; \Om)$ and $K_h (r; \Om)$ have now become {\it retarded} functions which are in-falling at the horizon and satisfy
 \be \label{orp2}
 K_A^* (r, \Om) = K_A (r, -\Om), \qquad K_h^* (r, \Om) = K_h (r, -\Om)  \ .
 \ee
We thus find that the retarded Green function for the currents can be written as
    \bea
    \label{retx3}
&& G_R^{ij} (\Om) =
 - \dk
  \int {d \om_1 \ov 2 \pi} {d \om_2 \ov 2 \pi}  \, {f(\om_1) - f(\om_2) \ov \om_1 - \Om  - \om_2-i \ep} \, \times
  \cr
&&  \rho_B^{\alpha \beta} (\om_1 ,\vec{k}) \, \vertexZ_{\beta\gamma}^i (\om_1, \om_2, \Om,\vk) \, \rho_B^{\gamma \delta} (\om_2 ,\vec{k}) \,
 \vertexZ_{\delta \alpha}^j (\om_2, \om_1, \Om,\vk) \nonumber \\
\eea
where
 \be \label{effZ}
 \vertexZ_{\beta \gamma}^i (\om_1, \om_2, \Om,\vk) =  \int dr \sqrt{g} \, \overline{\psinorm_\beta} 
 \, Q^i 
 \, \psinorm_\gamma 
\ee
with
  \bea
 \label{kernL}
&& Q^j (r; \Om,\vk) =  - i \le(- i  k_j  K_h (r;\Om) \Ga^{t}   \ri. \cr
 && \le. + {g_{jj} \ov 8}   \p_r K_h (r;\Om) \Ga^{r t j} 
 - i  q K_A (r;\Om) \Ga^{j}  \ri) \ .
 \eea
Note that in~\eqref{retx2} both $\Lam^i (\om_1, \om_2, \pm i \Om_l, \vk)$ analytically continue to  $\Lam^i (\om_1, \om_2, \Om, \vk)$.\footnote{This is due to that $K_A (- i \Om_l) = K_A (i \Om_l)$.}

The complex, frequency-dependent conductivity is
\be
 \sigma^{ij}(\Omega) \equiv  {G_R^{ij} (\Om)\over i \Omega} \ 
 \ee
which through~\eqref{retx3} is expressed in terms of intrinsic boundary quantities;
$\vertexZ$ can be interpreted as an effective vertex. Note that from~\eqref{orp2} 
one can readily check that
 \be
 (Q^i (r; \Om, \vk))^\da = \Ga^{\ut} Q^i (r; -\Om, \vk) \Ga^{\ut}
 \ee
which implies that
 \be \label{proZ}
 \vertexZ_{\beta \gamma}^{i*} (\om_1, \om_2, \Om,\vk)  = \vertexZ_{\gamma \beta}^i (\om_2 ,\om_1, -\Om,\vk) \ .
 \ee

We now make a further manipulation on the expression for the effective vertex~\eqref{effZ} to rewrite  
the first term there in terms of $\p_r K_h$, which can then be related simply to $K_A$ via~\eqref{eq:KAKh}. 
 To proceed note that the wave function $\psinorm$ satisfies 
the Dirac equation~\eqref{diraceqn}, which implies that\footnote{For $\om_1 =\om_2$ the equation below reduces to the conservation of fermionic number.}~(see Appendix~\ref{sec:backs} for details)
\bea
&& \overline{\psinorm_\beta}(r;\om_1)\Gamma^t\psinorm_\gamma(r;\om_2) \cr
&&=  \frac{i}{\om_1 - \om_2}{1 \ov \sqrt{g}} \partial_r\left(\sqrt{g} \, \overline{\psinorm_\beta}(r;\om_1)\Gamma^{r}\psinorm_\gamma(r;\om_2)\right) .
\label{inepw}
\eea
 We now use this identity in the first term of $Q^j$ in~\eqref{kernL} and integrate by parts. We can drop both boundary terms: the term at infinity vanishes since the $\psinorm$ are normalizable, and the term at the horizon vanishes because the graviton wavefunction $h^i_t$ (and thus $K_h$) vanishes there.\footnote{This is analogous to the well-known statement that $A_t$ vanishes at black hole horizons, and is similarly most transparent in Euclidean signature, where a nonzero $h_t^i$ at the shrinking time cycle indicates a delta-function contribution to the Einstein tensor.} We then find that~\eqref{kernL} can be rewritten as  
\bea
&&Q^j(r;\Om,\vec{k}) 
=-i\left(-\frac{k_j}{\om_1 - \om_2}\partial_r K_h(r;\Om)\Gamma^r
\right. \\ \nonumber &&\left.
+\frac{g_{ii}}{8}\partial_r K_h(r;\Om)\Gamma^{rtj} - iqK_A(r;\Om)\Gamma^j\right),
\eea
where it is important to note that this expression makes sense only when sandwiched between the two on-shell spinors in $\vertexZ$.
This manipulation replaced the $K_h$ with its radial derivative $\p_r K_h$, 
and one can now use the relation between gauge and graviton propagators~\eqref{eq:KAKh} to eliminate $\p_r K_h$ in favor of $K_A$, leaving us with 
\be
Q^j(r; \Om, \vk) = K_A (r;\Om)  \le(Y_1 \Ga^{\uj} +{i Y_2 k_j \ov \om_1 - \om_2} \Ga^{\ur} + i Y_3 \Ga^{\ur \ut \uj}    \ri) \label{kern3}
 \ee
where 
\be \label{yyy}
Y_1 = - q g_{jj}^{-\ha} , \quad Y_2 = - \sC g_{jj}^{-{d+1 \ov 2}}  \sqrt{g_{tt}} , \quad
Y_3 = {1\ov 8} g_{jj}^{-{d \ov 2}} \sC \ .
\ee 
This is the form of $Q^j$ that will be used in the remainder of this calculation. In~\eqref{kern3}, the $\CC$-dependent terms ($\CC$ was given in~\eqref{defcc})  can be interpreted as giving a ``charge renormalization'' resulting from mixing between the gauge field and graviton.

\subsection{Angular integration}

We will now use the spherical symmetry of the underlying system to perform the
angular integration in~\eqref{retx3}.
For this purpose we choose a reference direction, say, with $k_x = k \equiv |\vk|$ and all other spatial components of $\vk$ vanishing. We will denote this direction symbolically as $\th =0$ below. Then from the transformation properties of spinors it is easy to see that
 \bea \label{urr}
 \rho_B (\vk) &=& U (\th) \rho_B (k, \th =0) U^\da (\th), \cr
 \vertexZ_i  (\vk) &=& R_{ij} (\th) U(\th) \vertexZ_j (k, \th=0) U^\da (\th)
 \eea
 where $R_{ij} (\th)$ is the orthogonal matrix which rotate a vector $\vec k$ to $\th=0$ and $U$ is the unitary matrix which does the same rotation on a spinor (i.e. in $\al,\beta$ space). The angular integral in~\eqref{retx3} is reduced to
 \be
 \label{eq:angularintegral}
 {1 \ov (2 \pi)^{d-1}} \int d^{d-2} \th  \, R_{ik} (\th) R_{jl} (\th)
 = C \delta_{ij} \delta_{kl}
 \ee
 where $C$ is a normalization constant and $d^{d-2} \th$ denotes the measure for angular integration. 
 Note that $C = {1 \ov 4 \pi}$ for $d=3$ and $C= { 1 \over 12 \pi^2 } $ for $d=4$. 
 The conductivity can now be written as
  \be
  \sig^{ij}(\Omega)  = \delta^{ij} \sig(\Omega)
  \ee
  with
 \bwt
  \be
 \sig(\Omega)   =   -{C \ov  i \Omega } \int_0^\infty dk   k^{d-2}\int {d \om_1 \ov 2 \pi} {d \om_2 \ov 2 \pi}
 {f(\omega_1) - f(\omega_2) \over \om_1 - \Om  - \om_2-i \ep} 
 \sum_i \rho_B^{\alpha\beta} (\om_1,k) \, \vertexZ_{\beta\gamma}^i (\om_1, \om_2, \Omega,k) \, \rho_B^{\gamma\delta}(\omega_2, k) \,
 \vertexZ_{\delta\alpha}^i (\omega_2, \omega_1, \Omega,k)
 \label{condu3}
 \ee
where $\rho_B (\om, k)$ and $\vertexZ_{\delta\alpha}^i (\omega_2, \omega_1, \Omega,k)$
in~\eqref{condu3} and in all expressions below should be understood as the corresponding quantities evaluated at $\th =0$ as in~\eqref{urr}.

Equation~\eqref{condu3} can now be further simplified in a basis in which $\rho_B$ is diagonal, i.e. $\rho_B^{\al \beta} = \rho_B^\al \delta^{\al \beta}$, leading to
 \be
 \sig(\Omega)
  =   -{C \ov i \Omega } 
  \int_0^\infty d k k^{d-2} \int {d \om_1 \ov 2 \pi}
   {d \om_2 \ov 2 \pi}  \,
  { f(\omega_1) - f(\omega_2) \over \om_1 - \Om  - \om_2-i \ep}
  \, \rho_B^{\alpha} (\om_1 ,k) \, M_{\alpha \gamma} (\om_1, \omega_2, \Omega, k) \, \rho_B^{\gamma} (\om_2,k)
 \label{condu4}
 \ee
 \ewt
where (there is no summation over $\alpha,\gamma$ below)
\be
 M_{\alpha \gamma} (\om_1, \omega_2, \Omega,k)= \sum_i \vertexZ_{\alpha \gamma}^i (\om_1, \omega_2, \Omega,k) \vertexZ_{\gamma \alpha}^i (\om_2, \omega_1, \Omega,k)
 \label{eq:isum}
 \end{equation}
 with $\Lam^i$ given by~\eqref{effZ}. From~\eqref{proZ} we also have
 \be\label{mmer}
  M_{\alpha \gamma}^* (\om_1, \omega_2, \Omega,k) = 
    M_{\alpha \gamma} (\om_1, \omega_2, -\Omega,k) \ .
\ee
The DC conductivity can now be obtained by taking the $\Omega \to 0$ limit in \eqref{condu4}, which can be written as 
\bea
 \label{conduDC}
 \sig_{\rm DC}
  &=&   -{C \ov 2 } \sum_{\alpha,\gamma}  \int_0^\infty d k k^{d-2} \int {d \om \ov 2 \pi}  \, {\p f (\om) \ov \p \om} \,
  \cr&&
  \rho_B^{\alpha} (\om ,k) \,  \sM_{\alpha \gamma} (\om, k) \, \rho_B^{\gamma} (\om ,k) + \SS
 \eea
where
 \be \label{Omlim}
 \sM_{\al \ga} (\omega, k) \equiv \lim_{\Om \to 0} M_{\al \ga} (\omega+\Omega, \omega, \Omega, k) 
 \ee
 is real (from~\eqref{mmer}). Note that the term written explicitly in~\eqref{conduDC} is obtained by taking the imaginary part of ${1 \ov  \om_1 - \Om  - \om_2-i \ep}$
in~\eqref{condu4}.  The rest, i.e. the part proportional to the real part of  ${1 \ov  \om_1 - \Om  - \om_2-i \ep}$, 
is collectively denotes as $\SS$. We will see in Sec.~\ref{sec:optia} that such contribution vanishes in the low temperature limit, so we will neglect it from now on.

Let us now look at the $\Om \to 0$ limit of~\eqref{Omlim},
for which the $\frac{1}{\om_1 - \om_2}$ term in~\eqref{kern3} has to be treated with some care. Naively, it appears 
divergent; however, note that
\bea
&& \lim_{\Om \to 0} {1 \ov \Om}  \overline{\psinorm_\beta} (r; \om + \Om,\vk) \Ga^{\ur}  \psinorm_\gamma (r; \om, \vk) \cr\cr
&&= - \overline{ \psinorm_\beta} (r; \om,\vk) \, \Ga^{\ur} \, \p_{\om} \psinorm_\gamma (r; \om, \vk)
\label{newtr}
\eea
is finite because (see equation~\eqref{varW2} of Appendix~\ref{app:bulkP})
\be
\overline{\psinorm_\beta} (r; \om,\vk) \Ga^{\ur}  \psinorm_\gamma (r; \om, \vk)=0 \ .
\ee
Introducing 
 \be 
 \lam_{\beta\gamma}^{i} (\om, k)  \equiv 
\lim_{\Om \to 0} \vertexZ_{\beta \gamma}^i (\om \pm \Om, \om, \Om,k) 
\ee
we thus have 
\be
 \sM_{\alpha \gamma} (\omega, k)= \sum_i \lam_{\alpha \gamma}^{i} (\om, k) \lam_{\gamma \alpha}^{i} (\om, k)
 \end{equation}
and from~\eqref{newtr} and~\eqref{kern3} 
 \bwt
 \be
 \lam_{\beta\gamma}^{j} (\om, \vk)  
= \int dr \sqrt{-g } \,  K_A (r;\Om=0) \, \overline{\psinorm_\beta} (r;  \om,\vk) \,
 \le(Y_1 \Ga^{\uj} + i Y_3 \Ga^{\ur \ut \uj} - i  k_j Y_2  \Ga^{\ur} \p_\om   \ri) \,
 \psinorm_\gamma (r; \om, \vk)  \ .
 \label{dc10}
 \ee
\ewt

The above expressions~\eqref{condu4} and~\eqref{conduDC} are very general, but we can simplify  them slightly by using some explicit properties of the expression for $\rho_B$. We seek singular low-temperature behavior in the conductivity, which will essentially arise from low-frequency singularities in $\rho_B$. At the holographic Fermi surfaces described in \cite{Faulkner:2009wj}, at discrete momenta $k = k_F$, only {\it one} of the eigenvalues of $\rho_B$, say $\rho_B^1$, develops singular behavior. We can extract the leading singularities in the $T \to 0$ limit by simply taking the term in~\eqref{condu4} proportional to $(\rho_B^1)^2$. Thus \eqref{condu4} simplifies to 
\bwt
\be
 \sig(\Omega)
  =   -{C \ov i \Omega } 
  \int_0^\infty d k k^{d-2} \int {d \om_1 \ov 2 \pi}
   {d \om_2 \ov 2 \pi}  \,
  { f(\omega_1) - f(\omega_2) \over \om_1 - \Om  - \om_2-i \ep}
  \, \rho_B^{1} (\om_1 ,k) \, M_{11} (\om_1, \omega_2, \Omega, k) \, \rho_B^{1} (\om_2,k)
  \label{necon1}
  \ee
and we will only need to calculate $M_{11}$.  Similarly, for the one-loop DC conductivity,
\be
\label{eq:finalsig}
 \sig_{\rm DC}=  -{C \ov 2 }\int_0^\infty d k k^{d-2} \int {d \om \ov 2 \pi}  \, {\p f (\om) \ov \p \om}   \rho_B^{1} (\om ,k) \, \sM_{11} (\om, k) \, \rho_B^{1} (\om ,k) \ .
\ee
\ewt

\subsection{$M_{11}$ in $d=3$}

For definiteness, let us now focus on $d=3$ and choose the following basis of gamma matrices
 \bea
&&  \Ga^\ur = \left( \begin{array}{cc}
-\sigma^3  & 0  \\
0 & -\sigma^3
\end{array} \right), \quad
 \Ga^\ut = \left( \begin{array}{cc}
 i \sigma^1  & 0  \\
0 & i \sigma^1
\end{array} \right), \quad
\cr &&
\Ga^{\underline x} = \left( \begin{array}{cc}
-\sigma^2  & 0  \\
0 & \sigma^2
\end{array} \right) , \quad
 \Ga^{\underline y} = \left( \begin{array}{cc}
 0  & \sigma^2 \\
\sigma^2 & 0
\end{array} \right) \ .
\label{realbasis1}
 \eea
and write
\be \label{emmr}
\psinorm_1 = (- g g^{rr})^{-{1 \ov 4}} \left( \begin{array}{c}
  \Phinorm_1  \\
   0
\end{array} \right), \qquad \psinorm_2 = (- g g^{rr})^{-{1 \ov 4}} \left( \begin{array}{c}
  0 \\
   \Phinorm_2
\end{array} \right)
\ee
where $\Phinorm_{1,2}$ are two-component bulk spinors. As discussed in Appendix~\ref{app:spd}, in the basis~\eqref{realbasis1}, the fermion spectral function is diagonal and the subscript $1,2$ in~\eqref{emmr} can be interpreted as the boundary spinor indices. Also note that  in this basis the Dirac equation is real in momentum space and $\Phinorm_{1,2}$ can be chosen to be real. 

 It then can be checked that $\vertexZ^x$ (evaluated at $\theta=0$) only has diagonal components while $\vertexZ^y$ only has off-diagonal
components. From~\eqref{eq:isum}, we then find that 
\be \label{exm11}
M_{11} =  \vertexZ_{11}^x (\om_1, \omega_2, \Omega,k) \vertexZ_{11}^x (\om_2, \omega_1, \Omega,k) \ .
\ee
where 
\bea  \label{lam11}
&& \vertexZ_{11}^x (\om_1, \omega_2, \Omega,k) = \int dr \sqrt{g_{rr}} \, K_A (\Om) \times  \cr
&& \Phinorm^T_1 (\om_1,k)  \le(Y_1 \sig^3
  -{ i Y_2  k \ov \om_1 - \om_2} \sig^2  + Y_3 \sig^1 \ri) \Phinorm_1 (\om_2,k) \nonumber \\
\eea

Similarly, for the DC conductivity, 
\be \label{dchh}
\sM_{11} =  \lam_{11}^{x} (\om,k) \lam_{11}^{x} (\om,k)
\ee
with 
  \bea
&&  \lam_{11}^{x}  (\om,k)  =  \int dr \sqrt{g_{rr}} \,  K_A (r;\Om=0)  \times \cr 
&&  \Phinorm_1^T (r;  \om,k) \,
 \le(Y_1 \sig^3 +  Y_3  \sig_1+ i k Y_2 \sig^2 \p_\om \ri) \,
 \Phinorm_1 (r; \om, k) \ .  \nonumber \\ 
 \label{zzz1}
 \eea
Equations~\eqref{exm11}--\eqref{zzz1} are a set of a complete and self-contained expressions that can be evaluated numerically if the wave-functions $\Phi_{1,2}$ are known. $Y_{1,2,3}$ were introduced here in~\eqref{yyy}. 

As this was a somewhat lengthy exposition, let us briefly recap: after a great deal of calculation, we find the optical and DC conductivities are given by~\eqref{necon1} and~\eqref{eq:finalsig}, with (for $d=3$) $M_{11}$ given by~\eqref{exm11}--\eqref{lam11} and $\sM_{11}$ given by~\eqref{dchh}--\eqref{zzz1}.

\section{Effective vertices}\label{sec:vertex}

In this section we study in detail the analytic properties of the effective vertices~\eqref{exm11}--\eqref{zzz1} appearing respectively in the expressions for optical and DC conductivities~\eqref{necon1} and~\eqref{eq:finalsig} in the regime of low frequencies and temperatures. We will restrict to $d=3$. 

For simplicity of notations, from now on we will suppress various superscripts and subscripts in $M_{11}, \Lam_{11}^x, \sM_{11}, \lam_{11}^{x}$ and $ \Phinorm_1$, and denote them simply as $M, \Lam, \sM, \lam $ and $\Phinorm$.  Recall that 
under complex conjugation 
 \bea \label{cpxc}
 &&\vertexZ^* (\om_1, \om_2, \Om,k)  = \vertexZ (\om_2 ,\om_1, -\Om,k)  \\
&&  M^* (\om_1, \om_2, \Om,k) 
  = M (\om_1, \om_2, -\Om,k) 
   \eea
and both $\lam (\om,k)$ and $\sM (\om,k)$ are real. Introducing scaling variables 
\be  \label{dnne}
w_1 \equiv {\om_1 \ov T}, \quad w_2 \equiv {\om_2 \ov T}, \quad s \equiv {\Om \ov T}
\ee
we will be interested in the regime
\be  \label{dnne1}
w_{1}, w_2,  s = {\rm fixed}, \qquad T \to 0 \ .
\ee

\subsection{Some preparations}

As in the discussion of Sec.~\ref{app:btoB} it is convenient to separate the radial integral 
in~\eqref{lam11} into two parts, coming from IR and UV region respectively, i.e. 
 \be
 \Lam = \Lam_{IR} + \Lam_{UV}
 \ee
with 
\be \label{lamd1}
\Lam_{UV} = \int_{r_c}^\infty dr \sqrt{g_{rr}}  \cdots , \qquad
\Lam_{IR} = \int_{r_0}^{r_c} dr  \sqrt{g_{rr}}  \cdots
\ee
where $r_0$ is the horizon at a finite temperature and $r_c$ is the crossover radius 
specified in~\eqref{cross1} and~\eqref{cross}. 
In the inner (IR) region it is convenient to use coordinate $\xi$ introduced in~\eqref{nescV}, and then 
\be \label{lamd2}
\Lam_{IR} =  \int_{\xi_c}^{\xi_0} d \xi \sqrt{g_{\xi \xi}} \cdots  
\ee
In the limit~\eqref{dnne}, as discussed around~\eqref{cross}, we can take $r_c \to r_*$ and $\xi_c \to 0$
in the integrations of~\eqref{lamd1} and~\eqref{lamd2}.  Note, however, this limit can only be straightforwardly taken provided that the integrals of~\eqref{lamd1} are convergent as $r_c \to r_*$. 
Below we will see in some parameter range this is not so and the limit should be treated with care.

Now let us look at the integrand of the vertex~\eqref{lam11} in the limit~\eqref{dnne}. For this purpose let us first review the behavior of the vector propagator $K_A$ and spinor wave function $\Phinorm$ in the IR and UV regions, which are discussed respectively in some detail in Sec.~\ref{app:btoB} and  Appendix~\ref{app:spd} (please refer to these sections for definitions of various notations below):
\ben

\item From equation~\eqref{eq:aysolution}, we find in the outer region 
\bea  
K_A (\Om) &= & {\eta_+^{(0)} \ov \ca^{(0)}_+}  +  i O(T^3)   \cr
 & = & {r-r_* \ov r} + O(T) +  i O(T^3)  
\label{kk1}
 \eea
 with the leading term independent of $\Om$ and $T$ and real.  In the above we have also indicated the leading temperature dependence of 
 the imaginary part (from~\eqref{irretr}).  In the inner region from the second line of equation~\eqref{eq:aysolution} 
 we have 
  \be \label{KAte}
 K_A (\Om) = T K_A (s, \xi) + O(T^4)  \cdots 
 \ee
where 
\be 
K_A (s, \xi) = {1 \ov \ca_+^{(0)} } (\eta^+_I + \sG_y  (s) \eta^-_I) 
\ee
has no explicit $T$-dependence.

\item In the outer region, to lowest order in $T$, the normalizable spinor wave function $\Phinorm$ can be expanded in $\om$ as, 
\be 
\Phinorm = \Phinorm^{(0)} + \om  \Phinorm^{(1)} + \cdots 
\ee 
where $\Phinorm^{(0)}$ and  $\Phinorm^{(1)}$ are defined respectively in~\eqref{sout} and~\eqref{sout1} and are $T$-independent. 

In the inner region, to leading order, $\Phinorm$ can be written as  (from~\eqref{sint1})
\be \label{inerB}
\Phinorm (\xi,w)  = {a_+^{(0)} \ov W} T^{-\nu_k} \Psi_I^-  + \cdots 
\ee
where $a_+^{(0)}$  and $W_0$ are some $k$-dependent constants (but independent of $\om$ and $T$), and $\Psi_I^\pm (\xi, w)$ do not have any explicit dependence on $T$. The above expression, however, does not apply near a Fermi surface $k=k_F$ where $a_+^{(0)} (k_F)$ is zero. Near a Fermi surface we have instead (see discussion in Appendix~\ref{app:spd} around~\eqref{nedo})
\bea 
\Phinorm (\xi;w,T)  &= & {1 \ov W} \le[a_+ (k, \om, T)  T^{-\nu_{k_F}}  \Psi_I^- (\xi;w)  \ri. \cr
 && \le. -   a_-^{(0)} (k_F)  T^{\nu_{k_F}} \Psi_I^+ (\xi,w) \ri]
\label{inerB1}
\eea
where 
\bea 
a_+ (k, \om, T) &=& c_1 (k-k_F) - c_2 \om + c_3 T + \cdots  \cr
& = & c_1 (k- k_F (\om, T)) 
\label{aexp1}
\eea
with real coefficients $c_1, c_2, c_3$.   Again in~\eqref{inerB1} all the $T$-dependence has been made manifest.
Here we have also introduced  a ``generalized'' Fermi momentum $k_F (\om, T)$ defined by 
 \be \label{newkf1} 
k_F (\om, T) = k_F + {1 \ov v_F} \om - {c_3 \ov c_1} T + \cdots \  
\ee
with $v_F = {c_1 \ov c_2}$. 

\item We now collect the asymptotic behavior of various functions appearing in the effective vertices~\eqref{lam11} and~\eqref{zzz1}:  
 
\ben 
\item  For $r \to \infty$,  
\bea 
\Phinorm \sim r^{-m R}, \quad K_A (\Om) \sim O(1) \cr
 \sqrt{g_{rr}} \sim {1 \ov r}, \quad Y_1 \sim {1 \ov r}, \quad Y_{2,3} \sim {1 \ov r^d}
\label{largr1}
 \eea
 and thus the UV part of the integrals are always convergent as $r \to \infty$. Note that in our convention $mR > -\ha$ with the negative mass corresponding to the alternative quantization.

\item For $r \to r_*$,  in the outer region, 
\be \label{ro12}
 \sqrt{g_{rr}} \sim {1 \ov r-r_*}, \quad Y_1, Y_3 \sim O(1), \quad Y_2  \sim r-r_* \ .
\ee
From~\eqref{kk1}, $K_A (\Om) \sim r-r_*$. From~\eqref{sout} and~\eqref{sout1},
\be
 \Phinorm^{(0)} =  {1 \ov W} (a_+^{(0)} \Psi^{(0)}_- - a_-^{(0)} \Psi^{(0)}_+), 
\ee 
and 
\be 
 \Phinorm^{(1)} = {1 \ov W} 
 \le(a_+^{(1)} \Psi^{(0)}_-  + a_+^{(0)} \Psi^{(1)}_- - a_-^{(1)} \Psi^{(0)}_+ - a_-^{(0)} \Psi^{(1)}_+\ri) 
 \ee
 where 
 \be \label{ro22}
 \Psi^{(n)}_\pm \sim (r-r_*)^{\pm \nu_k - n} , \quad r \to r_*\ .
 \ee

\item Near the event horizon $\xi \to \xi_0$, 
\be 
K_A (s, \xi) = (\xi_0 - \xi)^{i {s \ov 4 \pi}} \le(1 + \cdots \ri) , \qquad
\ee
and 
\be 
\Psi_I^\pm  (w, \xi) \sim c_\pm  (\xi_0 - \xi)^{i {w \ov 4 \pi}}
+ c_\pm^*  (\xi_0 - \xi)^{-i {w \ov 4 \pi}}
\ee
\een
where $c_\pm$ some $\xi$-independent constant spinors (which depend on $w$ and $k$).
Also note 
\be 
g_{\xi \xi} \sim {1 \ov \xi_0 - \xi}, \quad Y_{1,3} \sim O(1), \quad Y_2 \sim (\xi_0 - \xi)^\ha \ .
\ee
One can then check that the integrals for the IR part~\eqref{lamd2} are always convergent near the horizon $\xi_0$. 

\een

\subsection{Low temperature behavior} 

With the preparations of last subsection, we will now proceed to work out the 
low temperature behavior of the effective vertices~\eqref{lam11} and~\eqref{zzz1},  which in turn will play an essential role in our discussion of the low temperature behavior of the DC and optical conductivities in Sec.~\ref{sec:cond}. 
The stories for~\eqref{lam11} and~\eqref{zzz1} are rather similar. For illustration we will mainly focus on~\eqref{zzz1} and only point out the differences for~\eqref{lam11}.  
The qualitative behavior of the vertices will turn out to depend on the value of $\nu_k$. We will thus treat different cases separately. 

\subsubsection{$\nu_k < \ha$}

Let us first look at the inner region contribution. Equations~\eqref{KAte} and~\eqref{inerB} 
give the leading order temperature dependence as  
\be \label{incon}
\lam \propto (a_+^{(0)})^2 
T^{1- 2 \nu_k}  + \cdots  \ 
\ee
where we have also used that $\sqrt{g_{tt}} \propto T$. The outer region contribution $\lam_{UV}$ starts with order $O(T^0)$ and we can thus ignore~\eqref{incon} at leading order. 
The full vertex can be written as 
\be \label{effr1}
\lam (\om, k) =\lam_0 (k) + O(T^{1-2 \nu_k})  
\ee
where $\lam_0 (k)$ is given by the zero-th order term of $\lam_{UV}$ and can be written as 
\bwt
  \be \label{effDc}
 \lam_0 (k)  =  \int_{r_*}^\infty dr \sqrt{g_{rr}} \, {r-r_* \ov r}  \le[ \Phinorm^{(0) T} \,
 \le( - {q R \ov r} \sig^3 + {\CC R^3 \ov 8 r^3}  \sig_1 \ri)   \Phinorm^{(0)} -i k \CC  \sqrt{h} {R^3  \ov r^3}  \Phinorm^{(0)T} \sig^2 \Phinorm^{(1)}  \,
  \ri]  \
 \ee
\ewt
where we have taken $r_c \to r_*$ in the lower limit of the integral (as commented below~\eqref{lamd2}), and have plugged in the explicit form of $Y_{1,2,3}$ From~\eqref{largr1}--\eqref{ro22} it can also be readily checked that the integral is convergent on both ends. 
Note that $\lam_0 (k)$ is  {\it independent of both $T$ and $\om$} and real.

Similarly for~\eqref{lam11}, one has 
\be 
\Lam (\om_1, \om_2, \Om; k) =  \Lam_0 (w_1, w_2, s; k) + O(T^{1-2 \nu_k})  
\ee 
with the leading term $\Lam_0$ given by
  \be \label{effop}
\Lam_0 (w_1, w_2, s;k) = \lam_0 (k) \ .
 \ee

\subsubsection{$ \nu_k \geq \ha$}

When $\nu_k \geq \ha$, the inner region contribution~\eqref{incon} is no longer negligible 
for generic momentum. Closely related to this, the leading outer region contribution,  which is given by~\eqref{effDc}, now becomes divergent at the lower end~(near $r_*$). More explicitly, 
from equations~\eqref{ro12}--\eqref{ro22} we find that 
as $r \to r_*$, the integrand of~\eqref{effDc} behaves schematically as 
\be \label{outcon}
(a_+^{(0)})^2 (r-r_*)^{-2 \nu_k} + O((r-r_*)^0) + \cdots \ .
\ee
The divergence is of course due to our artificial separation of the whole 
integral into the IR and UV regions and there should be a corresponding divergence in $\lam_{IR}$ in the limit $\xi_c \to 0$ 
to cancel the one from~\eqref{effDc}. What the divergence signals is that the leading contribution to the full effective vertex now comes from the IR region, as the UV region integral is also dominated by 
the IR end. Thus  for a generic momentum $k$, the effective vertex $\lam$ has the leading behavior
\be \label{tlamL}
\lam (\om, k)\sim (a_+^{(0)})^2 
T^{1- 2 \nu_k} + \cdots  \ . 
\ee
A slightly tricky case is $\nu_{k} = \ha$, for which $\lam_0$ has a logarithmic divergence 
and could lead to a $\log T$ contribution once the divergence is canceled. We have not checked its existence carefully,  as it will not affect the leading behavior of the DC and optical conductivities (as will be clear in the discussion of next section).  Thus in what follows, it should be 
understood  that for $\nu_k = \ha$, the $O(T^0)$ behavior in~\eqref{tlamL} could be $\log T$. 

It can also be readily checked that for generic $k$, the effective vertex $\Lam$ has the same temperature scaling as $\lam$. 

At a Fermi surface $k=k_F$, $a_+^{(0)} (k_F) =0$~\cite{Faulkner:2009wj}, for which the leading order term in~\eqref{tlamL} vanishes. Thus near a Fermi surface, which is main interest of this paper, we need also to examine subleading terms. 
Plugging~\eqref{inerB1} into the expression~\eqref{zzz1} for the vertex, we find that near $k_F$ the temperature dependence of  $\lam$ (including both IR and UV contributions) can be written as\footnote{There is also a term proportional to $a_+ (k,\om,T) \p_\om a_+ (k,\om,T) T^{1- 2 \nu_{k_F}}$ from the last term in~\eqref{zzz1}, but its coefficient, being proportional to $\Psi_I^{-T} \sig^2 \Psi_I^{-}$ is zero.}
\be 
\lam  (\om, k)= B (\om,k) (a_+ (k,\om,T))^2 T^{1- 2 \nu_{k}}  +  \lam_{0,{\rm finite}} + O(T) 
 \label{efflo}
\ee
where $B (k,\om)$ is a smooth function of $k$ and non-vanishing near $k_F$. At low temperatures it scales 
with temperature as $O(T^0)$. 
$ \lam_{0,{\rm finite}}$ denotes the finite part of~\eqref{effDc}, which is again $\om$ and $T$-independent, and a smooth function of $k$ (also at $k_F$). For $k  - k_F \lesssim  O(T)$, using~\eqref{aexp1} we can further write~\eqref{efflo} as 
\be \label{efflo1}
\lam  (\om, k)= B (\om,k_F) c_1^2 (k- k_F (\om,T))^2 T^{1- 2 \nu_{k_F}} +  \lam_0 (k_F) + \cdots \ 
\ee
where $k_F (\om, T)$ was the ``generalized Fermi momentum'' introduced in~\eqref{newkf1}.
In~\eqref{efflo1} we have also used 
\be  \label{sedod}
 \lam_{0,{\rm finite}} (k_F)  = \lam_0 (k_F)    \ 
\ee
as from~\eqref{outcon} $\lam_0$ is finite at $k_F$. Note that expression~\eqref{efflo1} applies to all $\nu_{k_F}$ including $\nu_{k_F} < \ha$. 

From~\eqref{efflo1}, note that at the Fermi surface $k=k_F$, 
\be 
\lam (\om, k_F) \sim \bca  O(T^0)  & \nu_{k_F} < {3 \ov 2} \cr
                       O\le(T^{3 - 2 \nu_{k_F}}\ri)  & \nu_{k_F} \geq {3 \ov 2} 
                       \eca 
                       \ee
and the vertex develops singular temperature dependence for $\nu_{k_F} \geq {3 \ov 2}$. 
However, at the generalized Fermi momentum $k_F (\om,T)$, the singular contribution is suppressed. This structure will be important below in understanding the low temperature behavior of the DC and optical conductivities. 

Similarly  the vertex~\eqref{lam11} can be written for $k  - k_F \lesssim  O(T)$ as 
\bwt
\be
\Lam (\om_1, \om_2, \Om,k) = \tilde B (\om_1, \om_2, \Om,k_F) c_1^2 (k- k_F (\om_1,T)) (k- k_F (\om_2,T))  T^{1- 2 \nu_{k_F}} 
 + \lam_0 (k_F)  
 + \cdots 
 \label{opedd}
\ee
\ewt
where $ \tilde B (\om_1, \om_2, \Om,k) $ is a smooth function of $k$, which scales 
with temperature as $O(T^0)$, and we have used~\eqref{effop}. 

To summarize the main results of this section: 

\ben 

\item For $\nu_k < \ha$, the effective vertices are $O(T^0)$ for all momenta. For both the DC and optical 
conductivities they are given by $\lam_0 (k)$ of~\eqref{effDc}, 
which is a smooth function of $k$.   

\item For $ \nu_k \geq \ha$, the vertices develop  singular temperature dependence for generic 
momenta as $T^{1- 2 \nu_k}$. But near the Fermi surface (more precisely at the generalized Fermi momentum $k_F (\om,T)$) the singular contribution is suppressed. 

\item For all values of $\nu_{k_F}$, the vertices for the DC and optical conductivities are given by~\eqref{efflo1} and~\eqref{opedd} respectively.

\een

\section{Evaluation of Conductivities}\label{sec:cond}

With the behavior of the effective vertices in hand we can now finally turn to the main goal of the paper: the leading low temperature 
behavior of the DC and optical conductivities. We will first present the leading temperature scaling and then calculate the numerical prefactors in the last subsection. In the discussion below we will only consider 
a real $\nu_k$. Depending on the values of $q$ and $m$, there could be regions in momentum space 
where $\nu_k$ is imaginary, referred to as oscillatory regions in~\cite{Liu:2009dm,Faulkner:2009wj}. We  
consider the contribution from an oscillatory region in Appendix~\ref{app:osck}. 
We will continue to 
follow the notations introduced in~\eqref{dnne} with below $w = {\om \ov T}$ and $f(w) = {1 \ov e^w +1}$. 

\subsection{DC conductivity} \label{sec:easytemp}

Let us first consider the leading low temperature dependence of the DC conductivity~\eqref{eq:finalsig}
which we copy here for convenience
\be
\label{eq:finalsig1}
 \sig_{\rm DC}=  -\int {d w \ov 2 \pi}  \, {\p f (w) \ov \p w}  I (w,T)
\ee
with $w = {\om \ov T}$ and (dropping all the super and subscripts) 
\be \label{fetr}
I (w,T) \equiv {C \ov 2 }  \int_0^\infty d k k^{d-2}   \rho_B^2 (w ,k,T) \, \lam^2 (w,k,T)  \ .
 \ee
Note that as a function $w$, the Fermi function $f(w)$ is independent of $T$, thus all the $T$-dependence of $\sig_{DC}$ is contained in the momentum 
integral $I (w,T)$. 

As discussed in~\cite{Faulkner:2009wj} (and reviewed in Appendix~\ref{app:spd}), for generic momentum, the spinor spectral function $\rho_B$ has leading low temperature dependence 
\be \label{oufer}
\rho_B \sim T^{2 \nu_k} \ ,
\ee 
Using~\eqref{oufer},~\eqref{effr1} and~\eqref{tlamL}, we find that for a generic momentum (up to possible logarithmic corrections)
\be \label{eorep}
\rho_B^2 \lam^2  \sim \bca T^{4 \nu_k} & \nu_k < \ha \cr
                                T^2   & \nu_k \geq \ha 
                                \eca \ 
                                \ee
where  the leading contribution of the first line (for $\nu_k < \ha$) comes from the UV part of the vertex, while for the second line the leading contribution comes from the IR part of the vertex.

Near a Fermi surface as reviewed at the end of Appendix~\ref{app:spd}
\be \label{rhobF1}
\rho_B =  {2 h_1 \Im \Sig \ov (k-k_F(\om,T) - \Re \Sig)^2 + (\Im \Sig)^2} \ .
\ee
where $ h_1$ is a positive constant, $k_F (\om,T)$ is given by~\eqref{newkf1} 
and 
\be \label{ooen}
 \Sigma = T^{2\nu_{k_F}} g(w) \ .  
\ee
$g (w)$ is a $T$-independent scaling function (depending on ${k_F/\mu}$) which can be obtained from
the retarded function in AdS$_2$ evaluated at $k_F$ (see~\eqref{finiteTspinorG}--\eqref{marF} for explicit expressions).

Now let us consider the momentum integral~\eqref{fetr} near a Fermi surface. For this purpose it is convenient to 
introduce a new integration variable 
\be 
y = k - k_F (\om,T)
\ee
in terms of which~\eqref{fetr} can be written to leading order as 
\bwt
\be \label{niint}
I (w,T) \bigr|_{FS} = 2 C h_1^2 k_F^{d-2} \int_{-\infty}^\infty dy \, \le({\Im \Sig \ov (y- \Re \Sig)^2 + (\Im \Sig)^2} \ri)^2  \; \le( B(k_F) c_1^2 y^2 T^{1 - 2 \nu_{k_F}} + \lam_0 (k_F)\ri)^2  + \cdots
\ee
\ewt
where we have used~\eqref{efflo1}. 
Now the key is that  since $\Sig \sim T^{2 \nu_{k_F}}$, 
by scaling $y \to T^{2 \nu_{k_F}} y$, the term proportional to $A$  in the last parenthesis becomes proportional to $T^{1 + 2 \nu_{k_F}}$ and can  be ignored. 
 Now the integral can be straightforwardly evaluated and we find that 
\be \label{cc0}
I (w,T) \bigr|_{FS} = {C' \ov  2 \Im g(w)}   T^{-2 \nu_{k_F}} 
\ee  
where 
\be\label{eopp}
C' = 2 \pi C h_1^2 k_F^{d-2}  \lam_0^2 (k_F)   \ .
\ee
Clearly~\eqref{cc0} dominates over the contribution from regions of momentum space 
away from a Fermi surface which from~\eqref{eorep} can at most be $O(T^0)$.\footnote{See 
Appendix~\ref{app:osck} for a discussion of the contribution from oscillatory regions which 
is again at most of order $O(T^0)$.} 

Plugging~\eqref{cc0} into~\eqref{eq:finalsig1}, we then find that for all $\nu_{k_F}$ the DC conductivity has the following leading low temperature behavior
\be \label{fineo}
\sig_{DC} = \al  T^{-2 \nu_{k_F}} 
\ee 
where $\al$ is a numerical prefactor given by
\be 
\al  
=  - {C' \ov 2} \int {d w \ov 2 \pi}  \, {\p f (w) \ov \p w }{1 \ov \Im g(w)} \, .
\ee
 We will discuss the numerical evaluation of $\al$ in Sec.~\ref{sec:num}. 
Note that since both $-{\p f \ov \p w}$ and $\Im g (w)$ are positive and even, the integral 
in the above expression is manifestly positive. 
 
We emphasize that in the above derivation it is crucial that the same $k - k_F (\om, T)$ appears in 
both the spectral function~\eqref{rhobF1} and the effective vertex~\eqref{efflo1}. As a result 
the leading contribution to the DC conductivity is dominated by the UV part of the effective 
vertex due to suppression at $k_F (\om, T)$, despite that for $\nu_{k} > \ha$ the vertex is generically dominated by the IR part.

Finally note that in writing down~\eqref{fineo} we have assumed there is a single Fermi surface. 
In the presence of multiple Fermi surfaces, the contribution from each of them can be simply added  together and the one with the largest $\nu_{k_F}$ dominates.

\subsection{Optical conductivity} \label{sec:optia}

Let us now look at the optical conductivity, which from~\eqref{necon1} can be written as 
\be
 \sig(\Omega)
  =   -{C \ov i \Om } 
 \int {d \om_1 \ov 2 \pi}
   {d \om_2 \ov 2 \pi}  \,
  { f(\om_1) - f(\om_2) \over \om_1 - \Om  - \om_2-i \ep} \, I (\om_1,\om_2, \Om,T)
  \label{necon2}
  \ee
where 
\be \label{imopI}
 I (\om_1,\om_2, \Om,T) \equiv  \int_0^\infty d k k^{d-2}  \, \sI (\om_1,\om_2,\Om,k) 
 \ee
with
 \bea
&& \sI (\om_1,\om_2,\Om,k)   \cr
 & &=  \rho_B (\om_1 ,k) \, \Lam (\om_1, \omega_2, \Omega, k) \,  \Lam (\om_2, \omega_1, \Omega, k)  \rho_B (\om_2,k) \ . \nonumber \\ 
 \eea
Recall that the vertex $\Lam$ 
satisfies~\eqref{cpxc} which implies that 
\be 
\sig (\Om) = \sig^* (-\Om)
\ee 
as one would expect since the system has time-reversal symmetry. 

\subsubsection{Temperature scaling}

All the temperature dependence of~\eqref{necon2} is in $I (\om_1,\om_2, \Om,T)$ which we examine first.  
As in~\eqref{oufer}--\eqref{eorep} one finds that away from a Fermi surface 
 $\sI$ has the following 
leading $T$-dependence 
\be \label{eorep1}
\sI (\om_1, \om_2, \Om, k)  \sim \bca T^{4 \nu_k} & \nu_k < \ha \cr
                                T^2   & \nu_k \geq \ha 
                                \eca \ 
                                \ee
where for $\nu_k < \ha$ the leading contribution comes from the UV part of the vertex, while for the second line the leading contribution comes from the IR part of the vertex.    
Thus one could at most get 
\be 
 I (\om_1,\om_2, \Om,T) \sim O(T^0)                         
\ee
from regions of momentum space away from a Fermi surface. 
                                
Near a Fermi surface  $\Lam$ has the low temperature expansion~\eqref{opedd}  and 
$\rho_B$ is given by~\eqref{rhobF1}--\eqref{ooen}. Introducing $y = k - k_F (\om_1, T)$, then the integral has the following structure 
\bwt
\bea 
I (\om_1,\om_2, \Om,T) |_{\rm FS} &\sim & \int dy 
\le({\Im \Sig_1 \ov (y- \Re \Sig_1)^2 + (\Im \Sig_1)^2} \ri) \le({\Im \Sig_2 \ov (y + \de - \Re \Sig_2)^2 + (\Im \Sig_2)^2} \ri)  \cr
&& \times \; \le( b_1 y (y + \de)  T^{1 - 2 \nu_{k_F}} + \lam_0 (k_F)\ri) 
\le( b_2 y (y + \de)  T^{1 - 2 \nu_{k_F}} + \lam_0 (k_F)\ri)
\label{hene}
\eea
\ewt
where 
\be
\de \equiv {1 \ov v_F} (\om_1 - \om_2) , 
\ee
 $\Sig_{1,2} \equiv \Sig (\om_{1,2})$, and 
$b_{1,2}$ are some $y$-independent functions of $\om_1, \om_2, \Om$ which scale with temperature 
as $O(T^0)$.  For $y \sim O(T^0)$ (i.e. away from the Fermi surface) the integrand scales as~\eqref{eorep1}. 
Near the Fermi surface, i.e.  in the range $y \lesssim O(T)$,  as in the analysis
of~\eqref{niint}, due to that $\Sig_{1,2} \sim T^{2 \nu_{k_F}}$, the dominant contribution in the $y$-integral  comes from the region $ y \sim O(T^{2 \nu_{k_F}})$. One then finds  from a simple scaling that the term proportional to $\lam_0^2 (k_F)$ (i.e. the UV part of the vertex)  is dominating.
The corresponding temperature scaling of~\eqref{hene} depends on the range of $\de$. 
For $\delta \sim O(T^{2 \nu_{k_F}})$, one has 
\be \label{jjee}
I (\om_1,\om_2, \Om,T) |_{\rm FS}   \sim T^{- 2 \nu_{k_F}}, \quad \delta \sim O(T^{2 \nu_{k_F}})
\ee
while for $\de \sim O(T)$, one finds 
\be \label{eorep2}
I (\om_1,\om_2, \Om,T) |_{\rm FS}   \sim \bca \lam_0^2 T^{-2 \nu_{k_F}} & \nu_{k_F} < \ha \cr
                                \lam_0^2 T^{2 \nu_{k_F}-2} 
                                & \nu_{k_F} \geq \ha 
                               \eca \ .
                               \ee

To summarize,  the contribution from near the Fermi surface is given by 
\bwt
\be \label{fincond}
 \sig(\Omega)
  =   -{C \lam_0^2 (k_F) \ov i \Om }  \int dk k^{d-2}  \int {d \om_1 \ov 2 \pi}
   {d \om_2 \ov 2 \pi}  \,
  { f(\om_1) - f(\om_2) \over \om_1 - \Om  - \om_2-i \ep}  \,  \rho_B (\om_1, k) \rho_B (\om_2, k)  + \cdots 
  \ee
\ewt
which is of the form of that for a Fermi liquid in the absence of vortex corrections.




\subsubsection{Contribution from Fermi surface} 

Now let us look at the contribution  from the Fermi surface in detail and work out the explicit frequency dependence. As discussed above we only need include the UV part of the effective vertex, which gives   
\be \label{imexps}
I^{(\rm FS)} (\om_1,\om_2, \Om,T) = \lam_0^2 (k_F)
k_F^{d-2} \int dk \rho_B (\om_1, k) \rho_B (\om_2, k)
\ee
The latter integral can be done straightforwardly (see Appendix~\ref{sec:integrals} for details) and gives 
\bea 
 I^{(\rm FS)} (\om_1,\om_2, \Om,T) 
&=&  2 \pi h_1  \lam_0^2 (k_F) k_F^{d-2} \rho_B (\om_2, K_2)  \cr
&=&  2 \pi h_1  \lam_0^2 (k_F) k_F^{d-2} \rho_B (\om_1, K_1) \nonumber \\
\label{uiep}
\eea
where $K_2 \equiv k_F (\om_1 , T) + \Sig^* (\om_1)$ and $K_1 \equiv k_F (\om_2 , T) + \Sig^* (\om_2)$. 
We now plug~\eqref{uiep} into~\eqref{necon2} and evaluate one of the frequency integral as follows. 
Split the integrand into two terms; in the one with the $f(\omega_2)$, we use the second line of~\eqref{uiep}
and do the $\om_1$ integral, which can be written as
\bea 
 && \int {d \om_1 \ov 2 \pi}
  {\rho_B (\om_1, K_1) \over \om_1 - \Om  - \om_2-i \ep} = G^R (\om_2 + \Om, K_1) \cr
&& = {h_1 \ov  -{\Om \ov v_F}+  \Sig^* (\om_2) -  \Sig(\om_2 + \Om)}
\eea
where in the first line we used the spectral decomposition of the boundary fermionic retarded function $G^R$ 
and the second line used~\eqref{boundre}.  Similarly for the term with $f(\omega_1)$ we can 
use the first line of~\eqref{uiep} and do the $\om_2$ integral, which gives 
\be 
  \int {d \om_2 \ov 2 \pi}
  {\rho_B (\om_2, K_2) \over \om_1 - \Om  - \om_2-i \ep} 
 = {h_1 \ov  {\Om \ov v_F}+  \Sig (\om_1) -  \Sig^*(\om_1 - \Om)} \ .
\ee
Combining them together we thus find that  
 \be \label{finopt}
 \sig(\Omega)
  =   {C' \ov i \Om } 
 \int {d \om \ov 2 \pi}
  { f(\om) - f(\om + \Om ) \over -{\Om \ov v_F} + \Sig^* (\om) - \Sig (\om + \Om)}
  \ee
where $C'$ was introduced before in~\eqref{eopp}. It is now manifest from the above equation that 
in the $\Om \to 0$, we recover~\eqref{fineo}. This confirms the claim below~\eqref{Omlim} that $\SS$ 
in~\eqref{conduDC} vanishes at leading order at low temperatures. 

We now work out the qualitative $\Om$-dependence of~\eqref{finopt} which has a rich structure depending on the 
value of $\nu_{k_F}$. As stated earlier we work in the low temperature limit $T \to 0$ with $s = {\Om \ov T}$ fixed. 

\smallskip
\noindent  {\bf 1. $\nu_{k_F} < \ha$:}
in this case given~\eqref{ooen}, to leading order we can ignore the term proportional to $\Om$ in the downstairs of the integrand of~\eqref{finopt}. Then $\sig (\Om)$ can be written in a scaling form 
\be \label{ss1}
\sigma (\Om) = T^{-2\nu_{k_F}}  F_1( \Omega/T)
\ee
with $F_1 (s) $ a universal scaling function given by  
\be \label{f1def}
 F_1 (s) =   C'  \int {d w \ov 2 \pi}
 \,
  { f(w + s) - f(w) \over i s} \, {1 \ov   g (w + s) - g^* (w) } \ 
 \ee
with $g$ given by~\eqref{gfund}.  
In the $s \to 0$ limit we recover the DC conductivity~\eqref{fineo}. In the limit 
$s \to \infty$, which corresponds to the regime $T \ll \Om  \ll \mu$, using~\eqref{laowl} we find that\footnote{Note that in our setup 
$\mu$ is a UV cutoff scale, thus we always assume $\Om \ll \mu$.}
 \be \label{opfa}
 \sig (\Om) = C''  (-i \Om)^{- 2\nu_{k_F}}
 \ee
 where $C''$ is a real constant given by 
 \be
 C'' = - {C' \ov 4 \pi i h_2} \int_{-1}^1 {dy \ov c(k_F) (1+y)^{2 \nu_{k_F}} -   c^* (k_F) (1-y)^{2 \nu_{k_F}}}
  \ee
with $c(k_F)$ given by~\eqref{defck}. In obtaining~\eqref{opfa} we have made a change of variable 
$w = {s \ov 2} (y-1)$ in~\eqref{f1def} and taken the large $s$ limit. 
 Note that the falloff in~\eqref{opfa} is much slower than the Lorentzian form
familiar from Drude theory. The behavior~(\ref{ss1}) and~\eqref{opfa} are indicative of  a system without a scale and with no quasiparticles. 

\smallskip
\noindent  {\bf 2. $\nu_{k_F} > \ha$:} in this case there are two regimes: 

\bi

\item[{\bf 2a.}] with $u = {\Om \ov T^{2 \nu_{k_F}}} = {\rm fixed}$ and $ s = u T^{2 \nu_{k_F} -1} \to 0$,
we find~\eqref{finopt} becomes
 \be
 \sig (\Om) = T^{-2 \nu_{k_F}} F_2 (u)
 \ee
with
\be
F_2 (u) ={C' \ov 2 \pi i} \int d w {\p f(w) \ov \p w} {1 \ov {u \ov v_F} + 2i \Im g (w)}
\ee
Since ${\p f \ov \p w}$ is peaked around $w=0$, we can approximate the above expression by setting 
$g(w)$ to its value at $w=0$, leading to a Drude form 
 \be\label{drude}
 \sig (\Om) \approx {i C' T^{-2 \nu_{k_F}} \ov 2 \pi } {1 \ov {u \ov v_F} + 2 i \Im g (0)}
 = {\om_p^2 \ov {1 \ov \tau} - i \Om}
 \ee
 with
 \be
\om_p^2 \equiv {v_F C' \ov 2 \pi}, \qquad {1 \ov \tau} \equiv 2 \Im g (0) v_F T^{2 \nu_{k_F}} \ .
  \ee
This behavior is consistent with charge transport from quasiparticles with a transport scattering rate given 
by $\tau \propto T^{-2 \nu_{k_F}}$. Furthermore we could interpret $C'$ as proportional to the quasiparticle
density. Indeed from~\eqref{eopp} it is proportional to the area of the Fermi surface. Note that $\lam_0 (k_F)$ in 
$C'$ can be interpreted as the effective charge of the quasiparticles.

\item[{\bf 2b.}] For $s = {\Om \ov T} ={\rm fixed}$, the two $\Sig$ terms in the downstairs of the integrand
of~\eqref{finopt} are  much smaller than the $\Om$ term, and we can then expand in power series of 
$\Sig$, with the lowest two terms given by
\be \label{irnm}
\sig (\Om) = {i \om_p^2 \ov \Om} \le(1+  T^{2 \nu_{k_F}-1} k(s)
+ \cdots \ri)
\ee
with 
\be 
k (s) = {v_F \ov s^2} \int dw \, (f (w) - f(w+s)) (g^*(w)- g (w+s))  \ .
\ee
In the large $s$ limit using~\eqref{laowl} we find 
\be 
k(s) \to - a (- 2 i s)^{2 \nu_{k_F}-1}   ,\quad
a = {4 v_F h_2 \Im c(k_F) \ov 2 \nu_{k_F} +1}
\ee
in which case $\sig (\Om)$ (i.e. for $T \ll \Om \ll \mu$) can be written as 
  \be \label{ueop1}
  \sig (\Om) ={i \om_p^2 \ov \Om} -  2 a  \om_p^2  (- 2i \Om)^{2 \nu_{k_F}-2} + \cdots  \ .
  \ee
The leading $1/\Om$ piece in~\eqref{irnm} gives rise to a term proportional to $\delta (\Om)$ with a weight consistent with~\eqref{drude}.  The subleading scaling behavior may be interpreted as 
contribution from the leading irrelevant operator. 

\ei
Note that in both regimes discussed above the temperature scalings are consistent with those identified earlier 
in~\eqref{jjee}--\eqref{eorep2}. For real part of $\sig (\Om)$ we have $\delta \propto \om_1 - \om_2 \propto \Om$ as constrained by the delta function resulting from the imaginary part of ${1 \ov \om_1 - \Om  - \om_2-i \ep}$
in~\eqref{fincond}, while for the imaginary part of $\sig (\Om)$ the dominant term (i.e. the term proportioal to ${i \ov \Om}$) comes from the region $\om_1 - \om_2 \sim O(T^{2 \nu_{k_F}})$.

\smallskip

\noindent  {\bf 3. $\nu_{k_F} = \ha$:}  the Marginal Fermi liquid, for which the $\Om$ term in the 
downstairs of~\eqref{finopt} is of the same order as $\Sig$, and we have 
 \be
 \sig(\Om) = T^{-1} F_3 \le({\Om \ov T}, \log {T \ov \mu} \ri)
 \ee
where $F_2$ can be written as 
 \be
  F_3 =  {C'  }   \int {d w \ov 2 \pi}
  { f(w + s) - f(w) \over i s} \, {1 \ov  {s \ov v_F} + g (w + s) - g^* (w)}
 \ee
with $g (w)$ now given by~\eqref{marF}.  Due to time
reversal symmetry the real part $\sig_1 (\Om)$ of $\sig (\Om) $ is an even function in $\Om$ 
and thus for $\Om/T < 1$, one can again approximate $\sig (\Om)$ by a Drude form with
the transport scattering time $\tau \propto {1 \ov T}$. 
 For $\Om \gg T$, using~\eqref{marF1} we find that 
 \be
 \sig (\Om) =  {1 \ov \Om} {C'\ov 2 \pi i  c_1}  \le({1 \ov   \log {\Om \ov T}}
 + {1 \ov  (\log {\Om \ov T})^2} \le({1 + i \pi \ov 2 } + {1 \ov v_F c_1}\ri)\ri) + \cdots 
 \ee
which is analogous to~\eqref{opfa}, but with logarithmic modifications. 
Recall that there are no logarithmic corrections for the DC conductivity~\eqref{fineo}.

\subsection{Numerical coefficients} \label{sec:num}

In this section, we discuss the numerical computation of the conductivities in 2+1 boundary dimension ($d=3$).

\subsubsection{Optical conductivity}

The optical conductivity is given by~\eqref{finopt} that we copy here for convenience
\be   
 \label{numericaleq1}
 \sig(\Omega)
  =   {C' \ov i \Om }
 \int {d \om \ov 2 \pi}
  { f(\om) - f(\om + \Om ) \over -{\Om \ov v_F} + \Sig^* (\om) - \Sig (\om + \Om)}
  \ee
where $C' =2 \pi C h_1^2  k_F^{d-2} \lam_0^2 (k_F)  $. The formula implicitly depends on the bulk fermion mass $m$ and charge $q$.
We are using this version since it only contains one $\om$ integral and it is easier to evaluate than~\eqref{condu4}. In order to compute $\sig(\Omega)$ for a fixed $m$ and $q$, we need the following quantities:
\begin{itemize}

  \item Fermi momentum: $k_F(m,q)$.

At $T=0$, $\Re \, G^{-1}(k, \om=0)$ changes its sign at the Fermi momentum. We determine the location of this sign change using the Newton method (up to 40 iterations). The algorithm needs an initial $k$ value where the search starts. This initial value was set by empirical linear fits on $k_F(m,q)$. When there were multiple Fermi surfaces, we picked the primary Fermi surface (the one with the largest $k_F$).

Computing $G^{-1}$ involves solving the Dirac equation in the bulk.  We used Mathematica's \texttt{NDSolve} to solve the differential equation using \texttt{AccuracyGoal/PrecisionGoal = 12\ldots 22}, and \texttt{WorkingPrecision = 70}. Typical IR and UV cutoffs are \texttt{$10^{-12}\ldots 10^{-20}$} and \texttt{$10^{-25}\ldots 10^{-40}$}, respectively. The resulting $k_F$ values are typically accurate to the $10^{th}$ digit.

  \item Numerator of the Green's function: $h_1(m,q)$.

The numerator of the fermionic Green's function is determined by fitting a parabola on six data points of $G^{-1}(k, \omega=0)$ near the Fermi surface (i.e. $k = k_F-10^{-5} \ldots k_F + 10^{-5}$), and then taking the derivative of the parabola at $k=k_F$. The computation of redundant data points makes the resulting $h_1$ value somewhat more accurate, but its main function is to monitor the stability of the numerics: whenever the six points are not forming an approximately straight line, we know that the $k_F$ finding algorithm has failed. In this case, we need to go back and ``manually'' obtain the value of the Fermi momentum.

  \item Self-energies: $ \Sigma(\omega; m, q, T)$.

Let $\widetilde\Sigma(\omega)$ denote the self-energy at the Fermi surface with the linear ${\om \ov v_F}$ term included. Then,
\be
  G_R(\omega, k) = {h_1 \over   (k-k_F) - \widetilde\Sigma(\omega) }
\ee
Since we already know $h_1$ and $k_F$, we determine $\widetilde\Sigma$ by computing the fermionic Green's function at $k=k_F$. Computing both $\widetilde \Sigma^*(\omega; m, q, T)$ and $ \widetilde\Sigma(\omega+\Omega; m, q, T)$ then gives the denominator of~\eqref{numericaleq1}.


  \item Effective vertex: $\Lambda(\omega_1, \omega_2, \Omega, k; m,q)$.

The numerical code computed the frequency-dependent $\Lambda^i_{\alpha\beta}(\omega_1, \omega_2, \Omega, k; m,q)$ (see~\eqref{effZ} for an explicit formula) instead of the simpler $\lam_0(k_F)$.



We determine $\Lambda(\omega_1, \omega_2, \Omega, k)$ by first numerically computing $K_A(r,\Om)$, which is the bulk-to-boundary gauge field propagator with ingoing boundary conditions at the horizon. (Note that at $\Om=0$ this may be done analytically.)
We then compute the spinor propagator with normalizable UV boundary conditions at both $\omega_1$ and $\omega_2$ and also compute the $\Lambda$ integral using a single \texttt{NDSolve} call. The integration proceeds towards the horizon where it oscillates somewhat before converging.


\end{itemize}

\begin{figure}[h!]
\begin{center}
\includegraphics[scale=0.44]{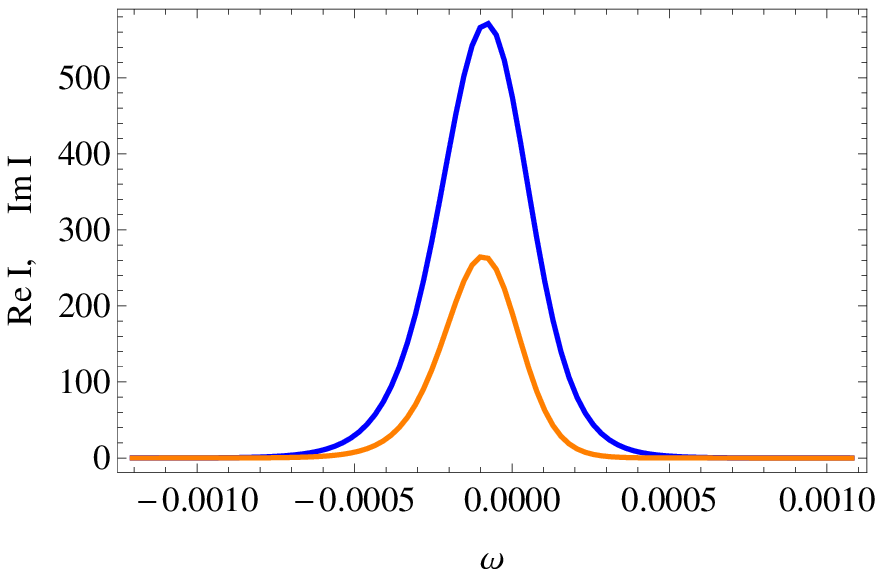}
\includegraphics[scale=0.44]{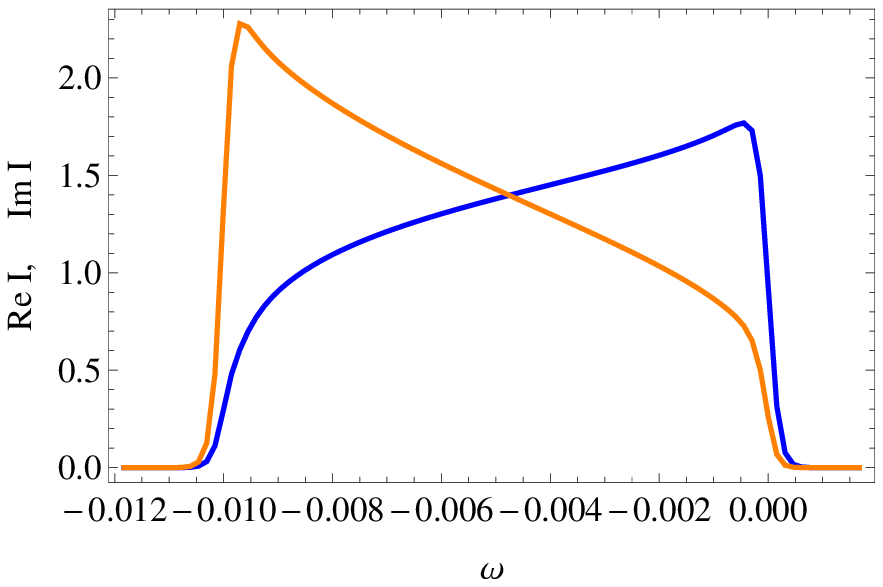}
\caption{Typical functions whose integral gives the conductivity: $\sigma(\Om) = \int_{-\infty}^{\infty} I(\om) d\om $. The two figures correspond to $\Om \sim T$ and $\Om \gg T$. The real and imaginary parts are indicated by blue and orange colors, respectively.}
 \label{fig:numintegrand}
\end{center}
\end{figure}

By using the above quantities, we compute the conductivity at a fixed $\Omega$ and $T$ by performing the integral over $\omega$ in~\eqref{numericaleq1}. The Fermi functions suppress the integral exponentially outside a certain window set by the parameters, see FIG.~\ref{fig:numintegrand}. The size of this window can be determined and is used to automatically set the integration limits.
The integrand is computed at $15\ldots 30$ points. We used Mathematica's parallel computing capabilities in order to compute three data points at the same time.

\begin{figure}[h!]
\begin{center}
\includegraphics[scale=0.7]{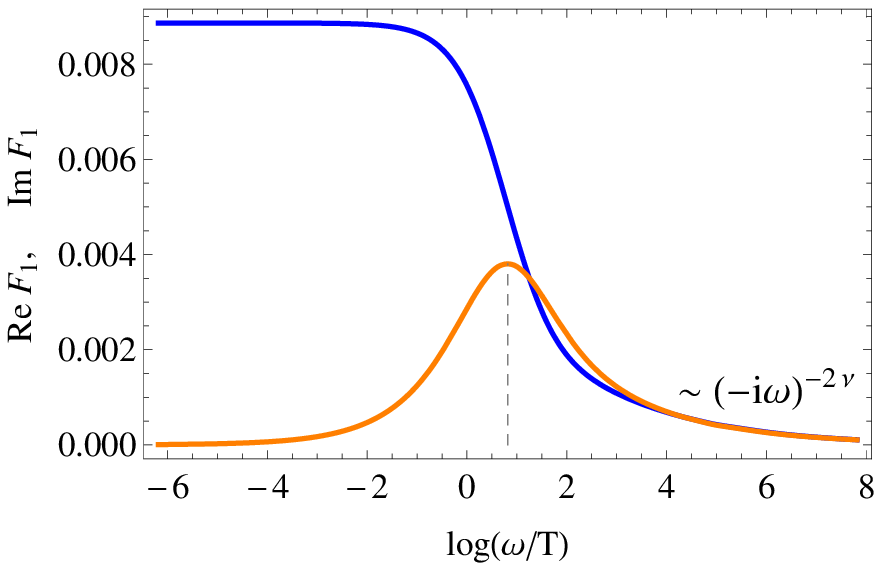}
\includegraphics[scale=0.7]{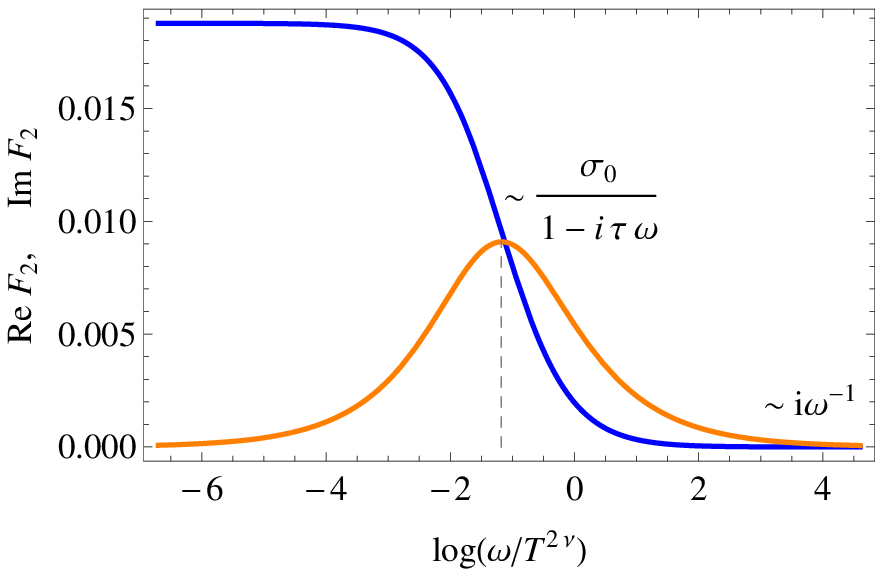}
\caption{These figures show the scaling functions for the optical conductivity.
{\bf Top:} Real and imaginary parts of the scaling function $F_1$ for $m=0, q=1$, where the IR fermion exponent is $\nu_{k_F} \approx 0.24 $.
{\bf Bottom:} Real and imaginary parts of the scaling function $F_2$
for $m=0, q=2$, where the IR fermion exponent is
$\nu_{k_F} \approx 0.73 $ and hence we are in the regime with a stable quasiparticle.
As indicated in the figure, both real and imaginary parts resemble the Drude behavior. 
\label{fig:scaling_functions}
}\end{center}
\end{figure}

Fig.~\ref{fig:scaling_functions} show the scaling functions
$F_{1,2}$ defined in the previous section for $\nu_{k_F}< \half$ and $\nu_{k_F} > \half$. The behavior in limiting cases agrees with the analysis above.

\subsubsection{DC conductivity}


The DC conductivity can be computed using~\eqref{fineo}.
\be
  \sigma_{DC}( T) =   - { k_F C \ov 2}\int d\omega { \p f\le({\omega \ov T}\ri) \over \p\omega} {h_1^2 \, \lambda^2(\omega; T) \over \Im \Sigma(\omega, k_F, T)  }
\ee

The computation of $k_F$, $h_1$ and $\Sigma$ was detailed in the previous subsection.
The $\omega$ derivative inside $\lambda(\omega; T)$ is computed by taking the difference of the wavefunctions with $\Delta\omega = 10^{-6}\ldots 10^{-8}$.
The numerical AC conductivity in the zero frequency limit matched the output of the DC conductivity code.

\begin{figure}[h!]
\begin{center}
\includegraphics[scale=0.7]{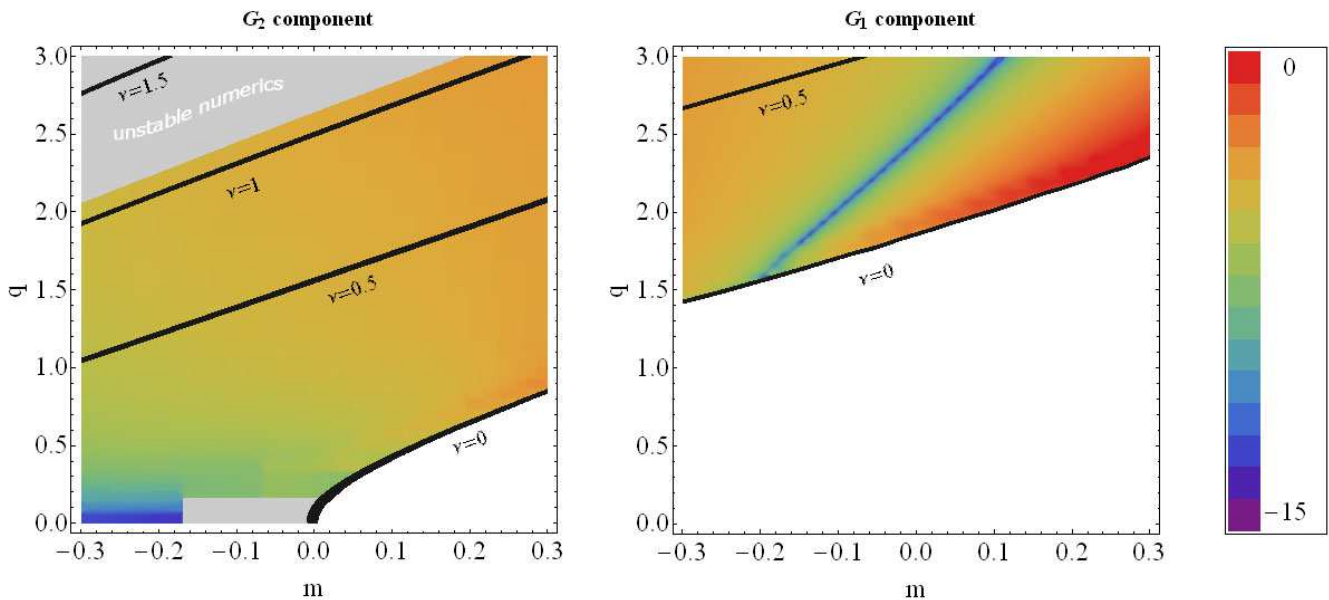}
\caption{The plots show the coefficient $\log \alpha(m,q)$ for the primary Fermi surface for both spinor components. In the white regions there is no Fermi surface. Black lines indicate half-integer $\nu_{k_F}(m,q)$. In the two gray regions in $G_2$ the numerical computations were unreliable (with error greater than 3\%). \label{fig:alpha1}
}\end{center}
\end{figure}

For a given pair $(m,q)$, the DC conductivity is computed at different temperatures between $T = 10^{-4} \ldots 10^{-8}$.  Then, the temperature-independent $\alpha$ coefficient is computed by a fit using $\alpha(m,q) = \sigma_{DC}(m,q,T) T^{2\nu_{k_F}}$.
In the $m-q$ space, the resolution was $45 \times 45$ with computation time approx. 22 hours. The results are seen in FIG.~\ref{fig:alpha1}. The plot shows  $\log \alpha(m,q)$ using a color code. The numerically unstable areas (with error larger than 3\%) are colored gray in the figure.
For $\nu_{k_F}>1.1$ (in the $G_2$ component) the numerical inaccuracies became too large.
For $q<0.3$, the automated $k_F$ finding algorithm typically failed and we had to determine $k_F$ manually.


Note the deep blue line in the $G_1$ spinor component.  At these points, the effective vertex $\lam_0 (k_F)$ changes sign and therefore the leading contribution to the DC conductivity vanishes (so does the leading contribution to the optical conductivity since it is also proportional to $\lam_0$).
 Since the DC effective vertex is real, this happens along a codimension one line in the $m-q$ plane. This `bad metal' line crosses the $\nu_{k_F}=1/2$ line at around $m \sim 0.18$ and $q \sim 3.4$ (not in the figure).

\bwt

\ewt

\section{Discussion and conclusions}

\begin{figure}[h]
\begin{center}
\includegraphics[scale=0.50]{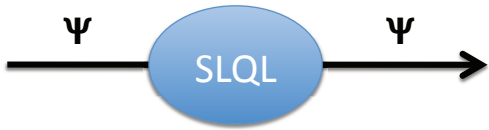}
\end{center}
\caption{
\label{fig:critcoupling}
The system can be described by a low energy effective action where fermionic excitations $\Psi$ around a free fermion Fermi surface hybridize with those in a strongly coupled sector (labelled as SLQL in the figure) described on the gravity side by the AdS$_2$ region~\cite{Faulkner:2009wj,Faulkner:2010tq,Faulkner:2010jy} (see~\cite{Iqbal:2011ae} for a more extensive review).}
\end{figure}

Despite the complexity of the intermediate steps, the result that we find for the DC conductivity is very simple. One can package all radial integrals into effective vertices in a way that makes it manifest that the actual conductivity is completely determined by the lifetime of the one-particle excitations, as is clear from the formula~\eqref{fincond}. 

We should stress that from a field-theoretical point of view this conclusion is {\it not} a priori obvious, as the single-particle lifetime measures the time needed for the particle to decay, whereas the conductivity is sensitive to the {\it way} in which it decays.
For example, if the Fermi quasiparticles are coupled to a gapless boson (as is the case in many field-theoretical constructions of non-Fermi-liquids; see e.g. \cite{varmareview,reizer,baym,Holstein:1973zz,Polchinski:1993ii,Nayak:1993uh,Halperin:1992mh,altshuler:1994,Schafer:2004zf,
Boyanovsky,sungsik-2009,nagaosa,Patrick,nave,Metlitski:2010pd,Metlitski:2010vm,Mross:2010rd,Hartnoll:2011ic} and references therein), small-momentum scattering is strongly preferred because of the larger phase space available to the gapless boson at smaller momenta. However, this small-momentum scattering does {\it not} degrade the current and so contributes differently to the conductivity than it does to the single-particle lifetime, meaning that the resistivity grows with temperature with a higher power than the single-particle scattering rate \cite{nagaosa}.  Such systems are therefore better metals
than one would have guessed from the single-particle lifetime.

In our calculation, the current dissipation is more efficient. 
To understand why, note that in our gravity treatment the role played by the gapless boson in the above example is instead filled by the AdS$_2$ region. From a field theoretical point of view, our system can be described by a low energy effective action~\cite{Faulkner:2009wj,Faulkner:2010tq,Faulkner:2010jy} in which fermionic excitations $\Psi$ around a free fermion Fermi surface 
hybridize with those of a strongly coupled sector, which can be considered as the field theory dual 
of the AdS$_2$ region and was referred to as a semi-local quantum liquid (SLQL) in~\cite{Iqbal:2011in}.
See Fig.~\ref{fig:critcoupling}. The SLQL provides a set of {\it fermionic} gapless modes to which the excitations around the Fermi surface can decay. In our bulk treatment this process has a nice geometric interpretation in terms of the fermion falling into the black hole, as in Fig.~\ref{adiss}. 
The crucial point is that because of the semi-local nature of the SLQL--as exhibited by the self-energy~\eqref{ooen}--there are gapless fermionic modes {\it for any momentum}\footnote{This is similar to that postulated for the bosonic fluctuation spectrum in the MFL description of the cuprates~\cite{varma}. But an important distinction is that here the gapless modes are fermionic.}.  Thus the phase space for scattering is not sensitive to the momentum transfer, and the conductivity is determined by the one-particle lifetime. 
Note that this scenario for strange metal transport is rather similar to that discussed before in~\cite{Ye,george} as reviewed in~\cite{Sachdev:2010um}.

We find that  the conductivity at the Marginal Fermi Liquid point $\nu_{k_F} = \ha$ -- when the single-particle spectral function takes the MFL form -- is consistent with a {\it linear resistivity}, just as is observed in the strange metals. The correlation between the single-particle spectral function and the collective behavior and transport properties is a strong and robust prediction of our framework. While it is fascinating that this set of results is self-consistent, we do stress that the marginal $\nu_{k_F} = \ha$ point is not special from our gravity treatment, and more work needs to be done to understand if there is a way to single it out in holography.\footnote
{See~\cite{Jensen:2011su,Hartnoll:2012rj}
for recent work in this direction.}


\begin{appendix}

\vspace{0.2in}   \centerline{\bf{Acknowledgements}} \vspace{0.2in}

We thank M.~Barkeshli, E.~Fradkin, T.~Grover, S.~Hartnoll, G.~Kotliar, P.~Lee,
J.~Maldacena, W.~Metzner, S.~Sachdev, T.~Senthil, and B.~Swingle for valuable discussions and encouragement. 
During its long gestation, this work was supported in part by funds provided by the U.S. Department of Energy
(D.O.E.) under cooperative research agreements DE-FG0205ER41360, DE-FG02-92ER40697,
DE-FG0205ER41360,
and DE-SC0009919,	
and the OJI program,
in part by the Alfred P. Sloan Foundation,
and in part by the National Science Foundation under Grant No.~NSF PHY05-51164
and PHY11-25915.
All the authors acknowledge the hospitality of the KITP, Santa Barbara at various times,
in particular
the miniprogram on {\it Quantum Criticality and the AdS/CFT Correspondence} in July, 2009,
and the 
program {\it Holographic Duality and Condensed Matter Physics} in Fall 2011.
TF was also supported by NSF grant PHY 0969448, PHY05-51164, the UCSB physics department and the Stanford Institute for Theoretical Physics.
DV was also supported by the Simons Institute for Geometry and Physics
and by a COFUND Marie Curie Fellowship,
and acknowledges the Galileo Galilei Institute for Theoretical Physics in Florence
for hospitality during part of this work.

\section{Resistivity in clean systems}
\label{sec:deltaargument}


In a translation-invariant and
boost-invariant system
at finite charge density (without disorder
or any other mechanism by which the charge-carriers
can give away their momentum),
the DC resistivity is zero.
An applied electric field will accelerate the charges.
This statement is well-known but we feel that some
clarification will be useful.
It can be understood as follows.
Start with a uniform charge density at rest and in equilibrium, in a frame where
\be  j^t \equiv \rho  \neq 0, ~~ T^{tt} = \epsilon \neq 0,
~~T^{ii} = P, ~~~~~j^i = 0, ~~\pi^i \equiv T^{ti}= 0 ~~;\ee
$\epsilon$ is the energy density and $P$ is the pressure.
Boost\footnote
{We perform a small boost $u^i \ll c$,
so we can use the Galilean transformation, even if the system is
relativistic.}
by a velocity $u^i$ to a frame where
\be j^i = u^i \rho, ~~~~\pi^i = u^i (\epsilon + P )  ~.\ee
This gives
\be j^i = { \rho \over \epsilon + P } \pi^i  ~~,\ee
which is effectively a constitutive relation.
In a non-relativistic system, the enthalpy $\epsilon+P$
reduces to the mass of the particles.
Combining this with
conservation of momentum (Newton's law)
\be \partial_ t \pi^i = \rho E^i \ee
and Fourier transforming gives
\be j^i(\Omega) = {i\over \Omega} {  \rho^2 \over (\epsilon + P) } E^i (\Omega) \ee
and hence
\be
\label{accelerationcond}
 \Re \sigma(\Omega) = {\pi \rho^2 \over \epsilon + P } \delta(\Omega)
 \ee
plus, in general, dissipative contributions.
Note that in systems where the relation $ \vec J = {\rho \over \epsilon +P} \vec \pi$ is
an operator equation, momentum conservation
implies that there are no dissipative contributions,
and the conductivity is exactly given by~\eqref{accelerationcond}. This is indeed consistent with the leading term we obtained in~\eqref{ballis}. 

The Fermi surface contribution which is main result of the paper does {\it not} contain a delta function in $\Omega$, as the
Fermi surface current can dissipate via interactions with the
$O(N^2)$ bath. Although the total momentum 
(of the Fermi surface plus bath) is conserved, 
the time it takes the bath to 
return momentum given to it by the Fermi surface degrees of freedom
is parametrically large in $N$, 
as in probe-brane conductivity calculations~\cite{Karch:2007pd}.
The DC conductivity we obtain
is averaged over a long time that is of ${\cal O}(N^0)$.

Our discussion here is somewhat heuristic, but a more careful hydrodynamic analysis that also takes into account the leading frequency dependence was performed \eg~in \cite{deltaCond}, where it was explicitly shown that the presence of impurities broadens this delta function into a Drude peak.

\section{Mixing between graviton and vector field} \label{app:mix}

In this section we construct the tree-level equations of motion for the coupled vector-graviton fluctuations about the charged black brane background. The action can be written in the usual form,
 \be \label{grac}
 S = {1 \ov 2 \kappa^2} \int d^{d+1} x \,
 \sqrt{-g} \le[\sR - 2 \Lam - {R^2 \ov g_F^2} F_{\mu \nu} F^{\mu \nu} \ri]
 \ee
with background metric given by
 \be \label{metrM2}
 ds^2 = -g_{tt} dt^2 + g_{rr} dr^2 + g_{ii} dx_i^2
 \ee
and a nonzero background profile $A_0(r)$. It is convenient to work with the radial background electric field $E_r = \p_r A_0$, which satisfies the equation of motion
 \be
  \p_r (E_r \sqrt{-g} g^{tt} g^{rr}) = 0 \ .
  \ee
We will denote the $r$-independent quantity
 \be
 \Q \equiv  -E_r \sqrt{-g} g^{tt} g^{rr} = {\kappa^2 \rho} {g_F^2 \ov 2 R^2}
 \ee
where we have used~\eqref{chard}.

Now consider small fluctuations
 \be
 A_M \to  A_0 \delta_{M0} + a_M, \qquad g_{MN} \to  g_{MN} + h_{MN} \ .
 \ee
We seek to determine the equations of motion for these fluctuations; we first use our gauge freedom to set
\be
h_{rM} = a_r = 0 \ .
\ee

At quadratic level in fluctuations the Maxwell action can now be written as
\bea
 S_\text{EM} &= & -{C_1 } \int d^{d+1}x \, \le[
   {1 \ov 4} \sqrt{-g} f_{MN} f^{MN}  
 \ri. \cr && \le. 
- 
 \ha E_r \le(\sqrt {-g} g^{tt} g^{rr} + \sqrt{-g} g^{tr} g^{tr} \ri)_{(2)} E_r  
   \ri. \cr  && \qquad \le.
- E_r (g^{tt} g^{rr} \sqrt{-g})_{(1)} f_{0r} 
   \ri. \cr  && \qquad \le.
 - \Q \le( h_{t}^i
  f_{ir} +  h_{r}^i f_{ti} \ri) \ri]
 \eea
 with
 \be
 C_1 = {2 R^2 \ov g_F^2 \kappa^2 } , \qquad f_{MN} = \p_M a_N - \p_N a_M
 \ee
The canonical momentum for $a_\mu$ is then given by
 \be
 \pi^i = {1 \ov C_1} {\delta S \ov \delta \p_r a_i} = -  \sqrt{-g} g^{rr}  g^{ii} f_{ri} -
  \Q \, h_{t}^i
 \ee
\be
\pi^t =  \sqrt{-g} g^{rr} g^{rr} f_{0r} + \le(\sqrt{-g} g^{rr} g^{rr} \ri)_{(1)} E_r
\ee
The equation for $a_r$ is essentially the Gauss law constraint in the bulk and leads to the conservation of this canonical momentum,
 \be
 \p_\mu \pi^\mu =0
 \ee
Finally, the dynamical equations for $a_i$ are
 \be
 - \p_r \pi^\mu + \sqrt{-g} \p_\nu f^{\nu \mu} = 0
 \ee

We turn now to the gravitational fluctuations. At this point it is helpful to specialize to the zero momentum limit, i.e. all fluctuations depend only on $t$ and $r$. Now all spatial directions are the same, and so we pick one direction (calling it $y$) and focus only on $h_\alpha^y$, with $ \al = (t,r)$, and where the indices are raised by the background metric. We will then find a set of coupled equations for $a_y$ and $h^y_t$(and $a_r$ and $h^y_r$, which will be set to zero in the end).
The relevant equations then become
   \be
 \pi^y  = -  \sqrt{-g} g^{rr}  g^{yy} a_y' - \Q h^y_t
 \ee
 \be \label{eq:1}
  \p_r \pi^y - \sqrt{-g} g^{yy}  g^{tt} \om^2  a_y= 0
 \ee
 \be
\label{eq:ahrelation}
      2 \kappa^2 C_1  \Q  a_y   = - \sqrt{-g} g_{yy} g^{rr} g^{tt} \p_r h^y_t
 \ee
Taking a derivative of the first equation with respect to $r$ one can derive an equation for $a_y$ alone
    \be \label{massivegaugefield}
  \p_r (\sqrt{-g} g^{rr}  g^{yy} a_y') + \le({2 \kappa^2 C_1 \Q^2 } { g_{rr} g_{tt} \ov \sqrt{-g} g_{yy} }  - \sqrt{-g} g^{yy}  g^{tt} \om^2  \ri) a_y    = 0
 \ee
Note now that using
 \be
 g_{tt} =  f r^2 , \qquad g_{rr} = {1 \ov r^2 f}, \qquad g_{ii} = r^2
 \ee
 we  find that ~\eqref{massivegaugefield} becomes 
    \be \label{3}
  \p_r \(r^{d-1} f a_y' \) + \le( \CC \Q r^{- d-1} -  \frac{\om^2 r^{d-5}}{f} \ri) a_y    = 0  
 \ee
In the last expression 
we introduced the constant
\be 
\CC \equiv 2 \kappa^2 {C_1 } \Q  = 2 \kappa^2 \rho \
.\ee
In $d=3$, $\CC|_{d=3}= { 4 \sqrt{3} \over g_F} \left( {r_* \over R} \right)^{2} $.

Equation \eqref{eq:ahrelation} implies a corresponding relation between
 the bulk-to-boundary propagators $K_a, K_h$ of the metric and gauge field, which is important
 for our calculation:
 \be
 \CC K_a = - \sqrt{-g} g_{yy} g^{rr} g^{tt} \partial_r K_h~.
 \ee

\section{Spinor bulk-to-bulk propagator} \label{app:bulkP}

In this appendix we derive the spinor bulk-to-bulk propagator. For simplicity of exposition we focus on the case when the dimension $d$ of the boundary theory is odd; the correspondence between bulk and boundary spinors is different when $d$ is even, and though a parallel treatment can be done we shall not perform it here. We denote by $\sN$ the dimension of the bulk spinor representation; in the case that $d$ is odd we have $\sN = 2^{\frac{d+1}{2}}$. Our treatment will essentially apply to any asymptotically AdS spacetime with planar slicing and a horizon in the interior; the criterion of asymptotically AdS is important only in the precise choice of UV boundary conditions and can be easily modified if necessary.

\subsection{Spinor equations} \label{sec:backs}

We begin with the bulk spinor action:
\begin{equation}
\label{eq:diracaction}
S = - i \int d^{d+1} x \sqrt{-g} \bar{\psi} \left( \Gamma^M \sD_M - m \right) \psi
\end{equation}
where $\bar{\psi} \equiv \psi^\dagger \Gamma^{\ut}$. From here we can derive the usual Dirac equation,
\be
(\Ga^M \sD_M  - m) \psi = 0, \label{diraceqn}
\ee
where the derivative $\sD_M$ is understood to include both the spin connection and couplings to background gauge fields
\be
 \sD_M  = \p_M + {1 \ov 4} \om_{ab M} \Ga^{ab} - i q A_M \ . 
 \ee
The abstract spacetime indices are $M , N \cdots$ and
the abstract tangent space indices are $a,b, \cdots$. The index with an underline
denotes that in tangent space. Thus $\Ga^a$ to denote gamma matrices
in the tangent frame and $\Ga^M$ those in curved coordinates. Note that
 \be
 \Ga^M = \Ga^a e_a{^M}
 \ee
The nonzero spin connections for~\eqref{bhmetric1b}--\eqref{bhga2b} are given by 
 \be
 \om_{\ut \ur} = -\ha {g_{tt}' \ov g_{tt}} \sqrt{g^{rr}} e^\ut , \qquad
 \om_{\ui \ur} =   \ha{g_{ii}'\ov g_{ii}} \sqrt{g^{rr}} e^\ui, 
 \ee
with
 \be
e^\ut = g_{tt}^\ha dt , \qquad e^\ui = g_{ii}^\ha dx^i \ .
\ee
From the above one finds that 
\be \label{ueo}
{1 \ov 4} \om_{ab M} \Ga^M \Ga^{ab} = {\Ga^r \ov 4} \p_r \log (-g g^{rr}) \equiv U (r) \Ga^r  \ .
\ee
In momentum space the Dirac equation~\eqref{diraceqn} can then be written explicitly as 
\be \label{diraceqnE}
\le[- i (\om + q A_t) \Ga^t + i k_i \Ga^i + \Ga^r (\p_r + U) - m \ri] \psi (\om, \vec k; r) = 0
\ee
whose conjugate can be written as 
\be \label{diraceqnE1}
\bar \psi \le[- i (\om + q A_t) \Ga^t + i k_i \Ga^i - (\overleftarrow{\p_r} + U) \Ga^r  - m \ri] = 0 \ .
\ee
Applying~\eqref{diraceqnE} and~\eqref{diraceqnE1} to $\psinorm_\ga$ and $\bar\psinorm_\beta$ 
in~\eqref{inepw} respectively and using~\eqref{ueo}, one can readily derive~\eqref{inepw}.

\subsection{Green functions} 

We define the retarded and advanced bulk-to-bulk propagator as 
\bea 
D_R (t,\vec x; r,r') &= & i \theta (t) \vev{\{\psi (t,\vec x, r), \bar \psi (0,r') \}} \\
D_A (t,\vec x; r,r') &= & - i \theta (-t) \vev{\{\psi (t,\vec x, r), \bar \psi (0,r') \}} 
\eea
whose Fourier transform along boundary directions satisfy the equation 
\be \label{defeqn}
(\Ga^M \sD_M  - m)D_{R,A} (r,r';\om,\vk) = -\frac{i}{\sqrt{-g}}\delta(r-r') \ . 
\ee
In the above equations we have suppressed the bulk spinor indices which we will do throughout the paper. 
The spectral function $\rho(r_1,r_2;\omega,\vk)$ is defined by 
\be \label{sper1}
\rho(r,r';\om,\vk) = -i(D_R(r,r';\om, \vk) - D_A(r,r';\om,\vk)) \ .
\ee
The Euclidean two-point function is related to $D_R$ by the standard analytic continuation 
\be 
D_E(r,r';i\om_m,\vk)  = D_R (r,r'; \om = i \om_m ,\vk) 
\ee
and satisfies the spectral decomposition 
 \be
D_E(r,r';i\om_m,\vk) = \int \frac{d{\omega}}{2\pi} \frac{\rho(r,r';\omega,\vk) }{i\om_m - \omega} \ .
\ee

We define the corresponding {\it boundary} retarded and advanced Green functions as follows
  \bea \label{g1}
 G^R_{\al \beta} (t, \vec x) &= & i \th (t) \vev{ \{\sO_\al (t, \vec x), \sO^\da_\beta (0) \}} \\
 G^A_{\al \beta} (t, \vec x) &=& -i \th (-t) \vev{ \{\sO_\al (t, \vec x), \sO^\da_\beta (0) \} }
 \eea
 where $\sO$ is the boundary operator dual to the bulk field $\psi$ and $\al, \beta$ are boundary spinor indices. 
 The boundary spectral function $\rho_B$ is defined by 
 \be \label{sper2}
 G^R (\om, \vk)  - G^A (\om, \vk) = i \rho_B (\om, \vk), 
 \ee
and is Hermitian 
\be 
 \rho^\da_B = \rho_B\ .
\ee
From~\eqref{g1} the linear response relation is  
\be \label{linNe}
\vev{\sO (k) } = G_{R} (k) \ga^t \chi (k)
\ee
where $\chi$ denotes a source, $\ga^t$ is the boundary gamma matrix, and we have suppressed the spinor indices. 

Our convention for bulk Gamma matrices is that 
\be 
(\Ga^{a})^\da = \Ga^{\ut} \Ga^a \Ga^{\ut} , \quad (\Ga^{\ut})^2 = -1
\ee
and for boundary ones 
\be 
(\ga^t)^2 = -1, \quad (\ga^t)^\da = - \ga^t \ .
\ee
As mentioned earlier we will focus on odd $d$, for which case, there is also a $\Ga^5$ in the bulk 
which anticommutes with all the $\Ga^a$'s and satisfies 
\be \label{gamma5}
(\Ga^5)^\da = \Ga^5, \qquad (\Ga^5)^2 = 1 \ .
\ee

\subsection{Bulk solutions}

We begin by recalling how to obtain the boundary retarded Green function and 
some properties of the solutions to the Dirac equation~\eqref{diraceqn} (see also~\cite{Iqbal:2009fd}). 

Near the horizon $r_0$, it is convenient to choose the in-falling and out-going solutions as the basis of wave functions 
\begin{equation} \label{defSi}
\psi^{in,out}_{a}(r; \omega,\vk) \to  \xi^{in,out}_{a}  e^{\pm i \om \sig (r)} , \quad
r \to r_0
\ee
where $\sig (r) \equiv - \int dr \sqrt{g_{rr} g^{tt}}$, 
and $\xi_a$ are constant basis spinors, which  
satisfy the constraint
\begin{equation} \label{hocos}
\left( 1 \mp \Gamma^{\ut} \Gamma^{\ur} \right) \xi^{in,out}_{a}  = 0~.
\end{equation}
The index $a$ labels different independent solutions. From the above equation clearly we have $a \in \le\{1..\frac{\sN}{2}\ri\}$.
Equation~\eqref{hocos} also implies that for any $a,b$
\be 
\bar \psi^{in}_a \Ga^{\ur} \psi^{out}_b = 0 \ .
\ee
We will normalize 
\be 
\xi^{in \,\da}_a \xi_b^{in}  = \delta_{ab}, \qquad \xi^{out \,\da}_a \xi_b^{out}  = \delta_{ab} \ .
\ee

Near the boundary $r \to \infty$ it is convenient to consider purely normalizable $\psinorm$ and purely non-normalizable $\Psinon$ solutions defined respectively by
\bea\label{normsoln} 
\psinorm_{\alpha}(r \to \infty) & \to & \zeta_{\alpha}^{-} r^{- m R-d/2} \\ 
\Psinon_{\alpha}(r \to \infty) & \to  & \zeta_{\alpha}^{+} r^{+ m R-d/2}~~~ \label{normsoln2}
\eea
where $\zeta_{\alpha}^{\pm}$ are constant spinors which satisfy 
\begin{equation}
 \left(1 \mp \Gamma^{\ur} \right) \zeta_{\alpha}^{\pm}  = 0 ~ \ .
 \end{equation}
Again index $\al$ labels different solutions and runs from $1$ to $\frac{\sN}{2}$ (as the two different eigenspaces of $\Gamma^{\ur}$ span the full spinor space). Since the normalizable and non-normalizable solutions correspond to boundary operator and source respectively, $\al$ can be 
interpreted as the {\it boundary} theory spinor index. We choose the normalization 
\begin{equation}
\label{eq:normsofzeta}
\zeta^{\dagger \pm}_{\alpha} \zeta^{\pm}_{\beta} = \de_{\alpha\beta}\ 
\end{equation}
and have the following completeness relation,
\begin{equation}
\sum_{\alpha} \zeta^{\pm}_{\alpha} \zeta^{\dagger \pm}_{\alpha} = \frac{1}{2}\left(1 \pm \Gamma^{\ur} \right)
~.
\end{equation}
It is also convenient to choose 
\be \label{eorn}
\ze^+_\al = \Ga^5 \ze^-_\al 
\ee
where $\Ga^5$ was introduced earlier around~\eqref{gamma5}. 
The boundary gamma matrices can then be defined as 
\be \label{gatd}
\ga^\mu_{\al \beta} = -i  (\ze^-_\al)^\da \Ga^{\underline\mu} \ze^+_\beta \ . 
\ee

Now expand the in-falling solutions in terms of of $\psinorm_{\alpha}$ and the $\Psinon_{\al}$
\be
\psi^{in}_{a}  
=   \Psinon_{\alpha}   A_{\al a} 
+  \psinorm_{\alpha} B_{\alpha a } 
\label{infexp}
\ee
where $A$ and $B$ are both $\frac{\sN}{2} \times \frac{\sN}{2}$ matrices that connect the infalling and boundary solutions. 
Identifying $A$ with the source $\chi$~\eqref{linNe}, with $\ga^t$ defined as in~\eqref{gatd}, 
one can then check\footnote{As discussed e.g.  in~\cite{Iqbal:2009fd}, $\vev{\sO}$ should be identified with the boundary value of the canonical momentum conjugate to $\psi$.} that $B$ can be identified precisely with $\vev{\sO}$. It then follows that the boundary theory spinor retarded Green's function $G_R$ can 
be written as
\be
(G_R \ga^t)_{\al\beta} = (B A^{-1})_{\al\beta} \  \label{boundthy}
\ee
with $\ga^t$ the boundary theory gamma matrix. 
This is the covariant generalization of 
expressions given previously for the boundary fermion Green's function~\cite{Iqbal:2009fd},
and will be useful in what follows. One can find the {\it advanced} boundary theory correlator by using outgoing solutions and their corresponding outgoing expansion coefficient matrices $B,A$ in \eqref{boundthy}. 

We now compute some Wronskians that we will need later. Note first that by using the Dirac equation \eqref{diraceqn} we can show that for {\it any} two radial solutions $\psi_1(r)$, $\psi_2(r)$ evaluated at the same frequency and momentum, the Wronskian $W[\psi_1,\psi_2]$ defined as
\be \label{wronsk}
W[\psi_1,\psi_2] \equiv \sqrt{-gg^{rr}} \overline{\psi_1}(\om,k) \Ga^{\ur} \psi_2(\om,k) 
\ee
is a {\it radial} invariant, i.e. $\p_r W = 0$. Using~\eqref{normsoln} and~\eqref{normsoln2} at $ r=\infty$ one then finds that 
\be  \label{wronk1}
W [\psinorm_{\alpha},  \Psinon_{\beta}] = \bar \zeta^-_\al \Ga^{\ur} \zeta^+_\beta = 
 \bar \zeta^-_\al  \zeta^+_\beta  = i \ga^t_{\al \beta} = - W [\Psinon_{\alpha},  \psinorm_{\beta}]
\ee
where  we have used~\eqref{gatd}. It can also be readily checked that 
\be \label{varW2}
W [\psinorm_{\alpha},  \psinorm_{\beta}] = W [\Psinon_{\alpha},  \Psinon_{\beta}]  = 0
\ee
and 
\be \label{varW1}
W[\psi^{in}, \psi^{out}] =0, \quad W[\psi^{in}_a, \psi^{in}_b] =\de_{ab}= -
W[\psi^{out}_a, \psi^{out}_b] . 
\ee
Also note that 
\be 
W [\psinorm_{\alpha}, \psi^{in}_a] = i (\ga^t A)_{\al a}, \quad
W [\Psinon_{\alpha}, \psi^{in}_a] = - i (\ga^t B)_{\al a}
\ee

We can also expand the outgoing solutions as
\be
\psi^{out}_{a}  
=   \Psinon_{\alpha}   \tilde A_{\al a} 
+  \psinorm_{\alpha} \tilde B_{\alpha a }  \ .
\label{outexp}
\ee
with
\be 
W [\psinorm_{\alpha}, \psi^{out}_a] = i (\ga^t \tilde A)_{\al a}, \quad
W [\Psinon_{\alpha}, \psi^{out}_a] = - i (\ga^t \tilde B)_{\al a}
\ee
Using the above Wronskians we can also write 
\be 
\psinorm_{\alpha} = i \psi^{in}_a (A^\da \ga^t)_{a \al} - i  \psi^{out}_a (\tilde A^\da \ga^t)_{a \al}
\ .
\ee

\subsection{Constructing the propagator}

We are now ready to construct the bulk-to-bulk retarded propagator $D_R$ which satisfies the equation~\eqref{defeqn}
together with the boundary conditions that as either argument $r$ or $r' \to r_0$ the propagator should behave like an in-falling wave in~\eqref{defSi}, and similarly as $r$ or $r' \to \infty$ the propagator should be normalizable as in~\eqref{normsoln}. Note that the Dirac operator in~\eqref{defeqn}
above acts only on the left index of the propagator (which is a matrix in spinor space) and on the argument $r$; if we can demonstrate that the propagator indeed satisfies this equation then it will also satisfy the corresponding equation with the differential operator acting from the right and as a function of $r'$, by the equality of left and right inverses. Thus we will only explicitly show that the operator satisfies the equation in $r$. For the advanced propagator $D_A$, the only difference is that the propagator should behave like a outgoing wave at the horizon. 

With the benefit of hindsight, we now simply write down the answer for the bulk-to-bulk retarded and advanced propagator
\bwt
\be
D_{R,A}(r,r';\om,k) = \psinorm_{\al}(r)G_{\al\beta}^{R,A} (\om,k)\overline{\psinorm_{\beta}}(r') - \begin{cases} \Psinon_{\al}(r) \ga^t_{\al\beta}\overline{\psinorm_\beta}(r') \qquad & r < r' \\ \psinorm_{\al}(r) \ga^t_{\al\beta}\overline{\Psinon_{\beta}}(r') ~ & r > r'\end{cases} \label{propform}
\ee
\ewt

We now set out to prove that the above propagators have all of the properties required of them, very few of which are manifest in this form. We will discuss $D_R$ explicitly, with exactly parallel 
story for $D_A$.  For  $r > r'$ we have
\be
D_R(r,r';\om,k) = \psinorm_{\al}(r)\le(G_{\al\beta}^R(\om,k)\overline{\psinorm_{\beta}}(r') - \ga^t_{\al\beta}\overline{\Psinon_{\beta}}(r')\ri) 
\ee
which satisfies~\eqref{defeqn} in $r$, as well as the boundary condition that the solution be normalizable as $r \to \infty$, as the dependence on $r$ is simply that of the normalizable solution $\psinorm_{\al}$.  For $r < r'$ we have
\be
D_R(r,r';\om,k) = \le(\psinorm_{\al}(r)G_{\al\beta}^R(\om,k) - \Psinon_{\al}(r) \ga^t_{\al\beta}\ri)\overline{\psinorm_\beta}(r')
\ee
Now using~\eqref{boundthy} and~\eqref{infexp} we can write the above equation as
\be
D_R(r,r';\om,k) =- \psi^{in}_a(r)(A^{-1} \ga^t)_{a\beta}\overline{\psinorm_\beta}(r')
\ee
which satisfies both the defining equation~\eqref{defeqn} and the infalling boundary condition for $r < r'$, as the dependence on $r$ is now simply that of the in-falling solution. 

We now verify that the discontinuity across $r = r'$ is consistent with the delta function 
in~\eqref{defeqn}, which when integrated across $r = r'$ becomes 
\be
\sqrt{-gg^{rr}}\Ga^{\ur}\le(D_R(r+\ep,r) - D_R(r,r+\ep)\ri) =- i \ .
\ee
Inserting \eqref{propform} into this equation we thus  need to show
\be
- i \sqrt{-gg^{rr}}\Ga^{\ur}\le(\psinorm_{\al}(r)\ga^t_{\al\beta}\overline{\Psinon_\beta}(r) - \Psinon_{\al}(r) \ga^t_{\al\beta}\overline{\psinorm_\beta}(r)\ri) = {\bf 1} , \label{disc2}
\ee
where the right hand side is an identity matrix in the bulk spinor space.  To show it  we first contract both sides from the left with $\overline{\Psinon_{\sig}}(r)$. The right-hand side becomes just $\overline{\Psinon_{\sig}}$. The left-hand side then becomes a sum of two Wronskians~\eqref{wronsk}; the Wronskian of $\Psinon$ with itself vanishes as in~\eqref{varW2}, and we find then for left-hand side
\be
- i W[\Psinon_{\sig},\psinorm_{\al}] \ga^{t}_{\al\beta}\overline{\Psinon_{\beta}} = - (\ga^t)^2_ {\sig \beta}\overline{\Psinon_\beta} = \overline{\Psinon_\sig},
\ee
where in the first equality we have used \eqref{wronk1}. This is then consistent with \eqref{disc2}. 
Similarly contracting~\eqref{disc2} to the left with $\overline{\psinorm_{\sig}}$ we find
\be
i W[\psinorm_{\sig},\Psinon_{\al}](\ga^t)^{\al\beta}\overline{\psinorm_\beta} = \overline{\psinorm_\sig},
\ee
which is again satisfied. Note that since $\psinorm_{\sig}$ and $\Psinon_{\al}$ altogether form a complete basis, we have now verified the full matrix equation~\eqref{disc2}, and thus the propagator proposed in~\eqref{propform} is indeed correct. 

Now given~\eqref{propform}, taking the difference between $D_R$ and $D_A$, from~\eqref{sper1} and~\eqref{sper2} we thus find that 
\be
\rho(r,r';\om,k) = \psinorm_{\al}(r)\rho^B_{\al\beta}(\om,k)\overline{\psinorm_{\beta}}(r') \label{ImGspinorbulk}
\ee
where $\rho$ and $\rho_B$ are respectively the bulk and boundary spectral density.  This is the expression used in \eqref{specBB}. 

\section{Boundary spinor spectral functions} \label{app:spd}

In this appendix we specialize the discussion of the previous appendix to $d=3$ in an explicit basis and review the boundary retarded Green function derived in~\cite{Faulkner:2009wj}.

We choose the following basis of bulk Gamma matrices
 \bea
 && \Ga^\ur = \left( \begin{array}{cc}
-\sigma^3  & 0  \\
0 & -\sigma^3 
\end{array} \right), \;\;
 \Ga^\ut = \left( \begin{array}{cc}
 i \sigma^1 & 0  \\
0 & i \sigma^1
\end{array} \right), \cr
&&
\Ga^{\underline x} = \left( \begin{array}{cc}
-\sigma^2  & 0  \\
0 & \sigma^2
\end{array} \right) , \quad
\Gamma^{\uy} = \begin{pmatrix} 0 & \sigma^2 \\ \sigma^2 & 0 \end{pmatrix} \ 
\label{realbasis}
 \eea
with
\be
\Gamma^{5} = \begin{pmatrix} 0 & i \sigma^2 \\ -i \sigma^2 & 0 \end{pmatrix} \ .
\ee
Writing
\be \label{tranS}
\psi =  (- g g^{rr})^{-{1 \ov 4}} e^{-i \om t + i k_i x^i}   \left(\begin{matrix} \Psi_1 \cr \Psi_2 \end{matrix}\right)
\ee
and choosing the momentum to be along the $x$-direction with $k_x =k$, 
the corresponding Dirac equation~\eqref{diraceqn}
can be written as
\be\label{spinorequation}
\left(\sqrt{g^{rr}} \partial_r + m \sigma^3 \right) \Psi_\al = \left(i \sqrt{g^{tt}} \sigma^2 u  + (-1)^\al \sqrt{g^{ii}}  k  \sigma^1
\right) \Psi_\al \  
\ee
with $u = \om + q A_t$ and $\al =1,2$. Note that~\eqref{realbasis} is chosen so that $\Psi_{1,2}$ decouple from each other and equation~\eqref{spinorequation} is  {\it real} for real $\om, k$.

The in-falling solutions $\psi^{in}_{1,2}$ can be written in terms of those of~\eqref{spinorequation}
\bea 
\psi^{in}_{1} &= &  (- g g^{rr})^{-{1 \ov 4}} e^{-i \om t + i k_i x^i}  
\left(\begin{matrix} \Psi_1^{in} \cr 0 \end{matrix}\right), \cr
\psi^{in}_{2} &= &  (- g g^{rr})^{-{1 \ov 4}} e^{-i \om t + i k_i x^i}  
\left(\begin{matrix} 0 \cr \Psi_2^{in} \end{matrix}\right)
\eea
 and $\Psi^{in}_\al$ can in turn be expanded near the boundary as 
\be\label{outerasymptoticspinor}
\Psi^{in}_a \buildrel{r \to \infty}\over {\approx}
A_a r^{mR} \left( \begin{matrix} 0 \cr  1 \end{matrix}\right) 
+ B_a r^{-mR} \left( \begin{matrix}  1  \cr 0 \end{matrix}\right)
 \qquad
a = 1,2  \ .
\ee
We choose the constant spinors in~\eqref{normsoln}--\eqref{normsoln2} to satisfy~\eqref{eorn}  
\be 
\zeta_1^- = \left( \begin{matrix} 1 \cr 0 \cr 0 \cr 0 \end{matrix}\right) , \quad 
\zeta_2^- = \left( \begin{matrix} 0 \cr 0 \cr 1 \cr 0 \end{matrix}\right) , \quad 
\zeta_1^+ = \left( \begin{matrix} 0 \cr 0 \cr 0 \cr 1 \end{matrix}\right) , \quad 
\zeta_2^+ = \left( \begin{matrix} 0 \cr -1 \cr 0 \cr 0 \end{matrix}\right) 
\ee
and the corresponding boundary Gamma matrices~\eqref{gatd} are given by
\be 
\ga^t = - i \sig^2 , \qquad \ga^x = -\sig^1, \qquad \ga^y = - \sig^3 \ .
\ee 
The matrices $A$ and $B$ introduced in~\eqref{infexp} are then given by
\be 
A = \left( \begin{matrix} 0 & A_2 \cr -A_1 & 0 \end{matrix}\right), \quad 
B = \left( \begin{matrix} B_1 & 0 \cr 0& B_2 \end{matrix}\right)
\ee
and from~\eqref{boundthy} the boundary retarded function is diagonal with components given by 
\be \label{bdG}
G_{\alpha \alpha}^R (\om,k)=  {B_\alpha\over A_\alpha} \ , \quad \al =1,2 \ .
\ee

The set of normalizable and non-normalizable solutions introduced in~\eqref{normsoln}--\eqref{normsoln2} can be written more explicitly as 
\be \label{emmr1}
\psinorm_1 = (- g g^{rr})^{-{1 \ov 4}} \left( \begin{array}{c}
  \Phinorm_1  \\
   0
\end{array} \right), \quad \psinorm_2 = (- g g^{rr})^{-{1 \ov 4}} \left( \begin{array}{c}
  0 \\
   \Phinorm_2
\end{array} \right)
\ee
and 
\be \label{emmr2}
\Psinon_1 = (- g g^{rr})^{-{1 \ov 4}} \left( \begin{array}{c}
  0 \\
   \phi_2
\end{array} \right),  \quad
\Psinon_2 = - (- g g^{rr})^{-{1 \ov 4}} \left( \begin{array}{c}
  \phi_1  \\
   0
\end{array} \right)
\ee
where $\Phinorm_{1,2}$ and $\phi_{1,2}$ are two-component bulk spinors defined by 
\bea \label{norm1}
\Phinorm_{\alpha}(r \to \infty)  &\to&  \left( \begin{matrix}  1  \cr 0 \end{matrix}\right) r^{- m R} \\
\phi_{\alpha}(r \to \infty)  &\to&  \left( \begin{matrix} 0 \cr  1 \end{matrix}\right) r^{+ m R} \ .
\label{norm2}
\eea

Let us now briefly summarize the low temperature and frequency behavior 
of $\psinorm_{\alpha}$ and $G^R$~\cite{Faulkner:2009wj} which are needed for understanding the scaling behavior of the effective vertex and conductivities. The regime we are interested in is 
\be \label{scek}
T \to 0, \quad {\rm with} \quad w ={\om \ov T} = {\rm fixed}  \ . 
\ee
The discussion proceeds by dividing the radial direction into inner and outer regions, which is rather similar to that 
of the vector field in Sec.~\ref{app:btoB}. For definiteness below we will consider $\al=1$ in~\eqref{spinorequation} and drop the subscript $1$.

\subsection{Boundary retarded function} 

To leading order in $T$ in the limit of~\eqref{scek}, the Dirac equation~\eqref{spinorequation} in the inner region reduces to that in the near-horizon metric~\eqref{ads2T1} with $w$ as the frequency conjugate to $\tau$. In particular, the spinor operator develops an IR scaling dimension given by 
 \be \label{spinornu}
\nu_k \equiv \sqrt{m_k^2 R_2^2 - e_d^2 q^2 - i \epsilon} , \quad m_k^2 \equiv m^2 +
{k^2 R^2 \ov r_*^2} \ .
 \ee
Near the boundary of the inner region (i.e. $\xi \to 0$), the solutions to~\eqref{spinorequation} behave as $\xi^{\pm \nu_k}$ and we can choose 
the basis of solutions specified by their behavior near $\xi \to 0$ (which also fixes their normalization)
\be \label{innerBF}
\Psi_I^\pm \to  v_\mp  \le({T R_2^2 \ov r-r_*} \ri)^{ \mp \nu_k} 
= v_\mp \xi^{\mp \nu_k}, \qquad \xi \to 0 \ .
\ee
where $v_\pm$ are some constant spinors (independent of $\xi$ and $\om$). 
The retarded solution for the inner region can be written as~\cite{Faulkner:2009wj}
\be
\label{eq:retardedIR1}
\Psi_I^{\rm (ret)} (\xi;w)  =  \Psi_I^+  + \sG_k (w) \Psi_I^- ~.
\ee
where $\sG_k (w)$ is the retarded function for the spinor in the AdS$_2$ region~\cite{Faulkner:2009wj}
and will be reviewed  at the end of this section.\footnote{Note that due to normalization difference $\sG_k (w)$ defined here differs from (D28) of~\cite{Faulkner:2009wj}) by a factor $T^{2 \nu_k}$. }

In the outer region we can expand the solutions to~\eqref{spinorequation} 
in terms of analytic series in $\om$ and $T$. In particular, the zero-th order equation is 
obtained by setting $\om =0$ and $T=0$ (i.e. the background metric becomes that of the extremal black hole). Examining the behavior the resulting equation near $r=r_*$, one finds that  $\Psi \sim (r-r_*)^{\pm \nu_k}$, which matches with those of the inner region in the crossover region~\eqref{cross}. It is convenient to use the basis 
which are specified by the boundary condition 
 \be \label{spinZ}
 \Psi_\pm^{(0)} \to  v_\mp \le(r-r_* \ov R_2^2 \ri)^{\pm \nu_k}  \qquad r \to r_*  \ .
 \ee
 Once the zero-th order solutions are specified, higher order solutions $\Psi_{\pm}^{(n)}(r)$ can then be determined {\it uniquely} from $\Psi^{(0)}_{\pm}$ using perturbation theory, and the two linearly-independent solutions $\Psi_\pm$ can be written as\footnote{Note that as $r \to r_*$,
 $\Psi_{\pm}^{(n)}(r) \sim (r-r_*)^{\pm \nu_k - n}$.}
\be
\label{eq:ay1}
\Psi_{\pm}(r) = \sum_{n}^\infty  \omega^n \Psi_{\pm}^{(n)}(r)
\ee
where for economy of notation, we have left implicit the expansion in $T$. 
 Comparing~\eqref{innerBF} and~\eqref{spinZ}, in the overlapping region we have the matching 
 \be \label{eq:mat2}
 \Psi_\pm \leftrightarrow  T^{\pm \nu_k} \Psi_I^\pm \ .
 \ee

$\Psi_\pm$ can be expressed in terms of the set of normalizable and non-normalizable solutions introduced in~\eqref{norm1}--\eqref{norm2} (recall that all quantities here refer to $\al =1$)
\be \label{rorp}
\Psi_\pm = b_\pm \Phinorm + a_\pm \phi
\ee
where from~\eqref{eq:ay1}, $a_\pm, b_\pm$ can expanded in perturbative series in $\om$ and $T$,
with the zero-th order expressions denoted by $a_\pm^{(0)}, b_\pm^{(0)}$ which are functions of $k$ only. 

Using~\eqref{eq:retardedIR1},~\eqref{eq:mat2},~\eqref{bdG} and~\eqref{rorp}, now the full retarded boundary Green function can then be written as~\cite{Faulkner:2009wj}
\be \label{spgr}
G^R (\om, k) =  {b_+ + \sG_k T^{2 \nu_k} b_- \ov a_+ + \sG_k T^{2 \nu_k} a_-} \ 
\ee
which implies that the corresponding spectral function scales with temperature as 
\be 
\rho_B \equiv 2 \Im G^R \sim T^{2 \nu_k} \ .
\ee

\subsection{Normalizable solution} 

Let us now turn to the low energy behavior of the bulk normalizable solution $\Phinorm$.
Using~\eqref{rorp},  $\Phinorm$ can be written in the outer region as 
\be \label{manor}
\Phinorm (r;\om)= {1 \ov W} (a_+ (\om) \Psi_- (r;\om) - a_- (\om) \Psi_+ (r;\om))
\ee
where 
\be
W \equiv  a_+ b_- - a_- b_+ \ .
\ee

The Wronskian for equation~\eqref{spinorequation} is 
\be 
W [\eta_1, \eta_2] = \eta^T_1 \sig^2 \eta_2 
\ee
where $\eta_{1,2}$ are two solutions.  Applying it to $\Psi_\pm$ we find that 
\be \label{wros}
W [\Psi_+, \Psi_-] = {\rm const} 
\ee
Normalizing $\Psi_\pm$ so that the constant on the right hand side of the above equation is $\om$-independent, then after inserting the $\om$ expansion~\eqref{eq:ay1} of $\Psi_\pm$, equation~\eqref{wros} must be saturated by the zero-th order term and 
all the coefficients of higher order terms on the left hand side must be zero, e.g. at first order in $\om$,
\be 
\Psi^{(0)T}_+ \sig^2 \Psi^{(1)}_- + \Psi^{(1)T}_+ \sig^2 \Psi^{(0)}_- = 0 \ .
\ee
Furthermore, equating the value of $W[\Psi_+, \Psi_-]$ at $r=r_*$ and at $r = \infty$ we conclude that 
\be 
W = - i v_+^T \sig^2 v_-
\ee
which is $\om$-independent. 

Expanding~\eqref{manor} in $\om$ we find that  in the outer region $\Phinorm$ can be written as 
\be 
\Phinorm = \Phinorm^{(0)} + \om \Phinorm^{(1)} + \cdots 
\ee
where 
\be \label{sout}
 \Phinorm^{(0)} =  {1 \ov W} (a_+^{(0)} \Psi^{(0)}_- - a_-^{(0)} \Psi^{(0)}_+), 
\ee 
and 
\be 
 \Phinorm^{(1)} = {1 \ov W} 
 \le(a_+^{(1)} \Psi^{(0)}_-  + a_+^{(0)} \Psi^{(1)}_- - a_-^{(1)} \Psi^{(0)}_+ - a_-^{(0)} \Psi^{(1)}_+\ri) \ .
 \label{sout1}
 \ee 
The expression for $\Phinorm$ in the inner region can then be obtained from matching as 
\be  \label{sint}
\Phinorm (\xi; w,T)  = {1 \ov W } (a_+ T^{-\nu_k} \Psi_I^- (\xi;w) - a_- T^{\nu_k} \Psi_I^+ (\xi,w)) 
\ee
with the lowest order term given by 
\be  \label{sint1}
\Phinorm (\xi;w,T)  = {a_+^{(0)}  \ov W}  T^{-\nu_k} \Psi_I^- + \cdots \ .
\ee 

\subsection{Near a Fermi surface}

At a Fermi surface $k=k_F$ we have~\cite{Faulkner:2009wj}
\be \label{FFe}
a_+^{(0)} (k_F) = 0
\ee 
and~\eqref{sint1} does not apply. Near $k_F$ we have the expansion 
\be \label{aexp}
a_+ (k, \om, T) = c_1 (k-k_F) - c_2 \om + c_3 T + \cdots 
\ee
where $c_1 = \p_k a_+^{(0)} (k_F)$, $c_2 = - a_+^{(1)} (k_F)$.  Thus near $k_F$, in the  inner region the leading behavior for $\Phinorm (\xi;w,T)$ 
becomes 
\bea 
\Phinorm (\xi;w,T)  &= & {1 \ov W} \le[ a_+ (k, \om, T) T^{-\nu_{k_F}}  \Psi_I^- (\xi;w)  \ri. \cr
 && \le. -   a_-^{(0)} (k_F)  T^{\nu_{k_F}} \Psi_I^+ (\xi,w) \ri]
 \label{nedo}
\eea
where the coefficient of the first term $a_+ (k, \om, T)$ should be now understood as given by~\eqref{aexp}.

Finally let us look at the behavior of the retarded Green function~\eqref{spgr}
near a Fermi surface~\eqref{FFe}, which can be written as 
\be \label{boundre}
G^R = {h_1 \ov k-  k_F (\om,T) - \Sig (\om, k)}  \ .
\ee
$k_F (\om,T)$ in~\eqref{boundre} is defined as the zero of~\eqref{aexp}, i.e. $a_+ (k_F (\om, T)) =0$  and can be considered as a generalized Fermi momentum
 \be \label{newkf}
k_F (\om, T) \equiv k_F +{1 \ov v_F} \om - {c_3 \ov c_1} T + \cdots \ 
\ee
where $v_F \equiv  {c_1 \ov c_2}$ is positive for $\nu_{k_F} > \ha$. $\Sig (\om, k)$ is given by
\be \label{verp}
\Sig =  h_2 T^{2 \nu_{k_F}} \sG_{k_F} \le({\om \ov T} \ri)  
\ee 
and $h_1, h_2$ are positive constants whose values are known numerically. 
The spectral function can be written as 
\be \label{rhobF}
\rho_B = 2 \, \Im \, G^R =  {2 h_1 \Im \Sig \ov (k-k_F (\om, T)- \Re \Sig)^2 + (\Im \Sig)^2} \ .
\ee

For notational convenience  we write 
\be\label{finiteTspinorG}
 \Sig (\om, T; k_F) =  T^{2 \nu_{k_F}} g \le({\om \ov T}; {k_F \ov \mu} \ri) 
 \ee
where the explicit expression for $g$ can be obtained from that of $\sG_k$ given in Appendix D of~\cite{Faulkner:2009wj} 
\be \label{gfund}
 g \( {\omega \over T}, {k \over \mu} \) = h_2  (4 \pi)^{2 \nu_{k}} c(k) {\Gamma (\ha+\nu_k -\frac{i \omega }{2 \pi T }+i {q e_d} ) \ov \Gamma \left(\frac{1}{2}-\nu_k -\frac{i \omega }{2 \pi T}+i
{q e_d} \right)}
\ee
with $c(k)$ given by 
\be \label{defck}
c(k) = \frac{\Gamma (-2 \nu_k ) \, \Gamma \left(1+\nu_k -i {q e_d} \right)}{\Gamma (2 \nu_k )\,   \Gamma \left(1-\nu_k -i {q e_d }\right)} \frac{ \eta_k - \nu_k}
{ \eta_k +  \nu_k} 
\ee
and $\eta_k \equiv {mR_2}  + i {k R R_2 \ov r_*} - i {q e_d}$.\footnote{Note that the sign of the second term in $\eta_k$ depends on which component of the spinor we are looking at. Here the sign is for the first component.
Also note that the definitions of $c(k)$ and $h_2$ differ by a phase factor from those used in~\cite{Faulkner:2009wj}. In particular, the definition of $h_2$ in~\eqref{verp} ensures it is positive as discussed in Appendix D4 of~\cite{Faulkner:2009wj}.} 
$g$ approaches a constant as $w = \om/T \to 0$ and as $w \to \infty$
\be \label{laowl}
g  (w)\to  h_2 \, e^{- i \pi \nu_k} \, c(k) \, (2 w)^{2 \nu_k} \ .  
\ee
For the Marginal Fermi Liquid case, $\nu_{k_F} = \half$, the above expressions should be modified. Instead 
one finds that 
\be  \label{marF} 
g =  2 \pi i d_1 u  -  \pi  c_1 \le(2 u \log{T \ov \mu} + 2 u
 \psi( - i u) + i \pi u + i \ri)  + \cdots
\ee
where $u \equiv {\omega \ov 2 \pi T} - {q e_d}$, $\psi$ is the digamma function, $c_1, d_1$ are positive 
constants\footnote{They are related by 
$${d_1\over c_1 } =\half \le( \pi +  2 \Im \psi \le( \half + i qe_d \ri) \ri) = { \pi \over 1 + e^{ - 2 \pi qe_d } } \ .
 $$}, 
and $\cdots$ denotes terms which are real and analytic in $\om$ and $T$. In the limit $w = \om/T \to \infty$ 
equation~\eqref{marF} becomes 
\be \label{marF1}
g (w)=  i d_1 w  - c_1 w \log w  + \cdots \ .
\ee

\section{Couplings to graviton and vector field} \label{app:cou}

In this section we determine the couplings of a spinor to graviton and gauge field fluctuations; these are necessary to construct the bulk vertex. We consider a free spinor field with the action
 \be \label{Dirac2}
 S = - \int d^{d+1} x \sqrt{-g} \, i (\bar \psi \Ga^M \sD_M \psi  - m \bpsi \psi)
 = \int d^{d+1} \sqrt{-g} \, \sL
 \ee
where $\bpsi = \psi^\da \Ga^\ut$.
We now consider a perturbed metric of the form
 \be
 ds^2 = - \tilde g_{tt} dt^2 + h (dy + b dt)^2 + g_{rr} dr^2 + h dx_i^2
 \ee
with
 \be
 b \equiv h_t^y, \qquad \tilde g_{tt} = g_{tt} + h b^2
 \ee
The new spin connections are given by
 \bea
 && \om_{\ut \uy} = f_2  e^{\ur}  \\
 && \om_{\ut \ur} = - \tilde f_0 e^\ut +  f_2 e^{\uy}
 \\
 && \om_{\uy \ur } = f_1 e^{\uy} + f_2 e^{\ut}
   \\
 && \om_{\ui \ur } = f_1 e^{\ui}
 \eea
with
\be
f_1 \equiv \ha{h' \ov h} \sqrt{g^{rr}}, \quad
f_2 \equiv \ha \sqrt{h \ov g_{rr} \tilde g_{tt}} b', \quad
\tilde f_0 \equiv \ha {\tilde g_{tt}' \ov \tilde g_{tt}} \sqrt{g^{rr}}
\ee
and
 \be
 e^\ut =
  \tilde g_{tt}^\ha dt , \quad e^{\uy} = h^\ha (dy + bdt), \quad
  e^{\ur} = g_{rr}^\ha \quad e^\ui = h^\ha dx^i
\ee
Also note that
 \be
 \Ga^t = \tilde g_{tt}^{-\ha} \Ga^{\ut}, \qquad
 \Ga^y = -  \tilde g_{tt}^{-\ha} b \Ga^{\ut} + h^{-\ha} \Ga^{\uy}
 \ee

We thus find the corrections to the Dirac action are given by (with $a \equiv a_y$):
at cubic order
 \be
 \label{cubiccoupling}
 \delta \sL_3 =- i \bar \psi \le(-  g_{tt}^{-\ha} h_t^y \Ga^{\ut} \p_y + {1 \ov 4}
 f_2 \Ga^{\ur \ut \uy} - i h^{-\ha} q a_y \Ga^{\uy} \ri) \psi \ .
 \ee
In \eqref{cubiccoupling} we have restored the indices on $b \equiv h_t^y, a_y$
because they make the covariant nature of the expression manifest.

At quartic order there are both $b^2 \psi^2$ and $b a \psi^2$ terms. For completeness we list them, although they are not required for our calculation. 
The couplings of the bulk spinor
which are quadratic in the bosonic bulk modes
(altogether, quartic in fluctuations) are
\bwt
\be {\cal L}_4 = \sqrt{-g} i \bar \psi \left[
{ hb^2 \over 2 g_{tt}} \( \Gamma^M {\cal D}_M - m\)_{(0)}
- {b\over \sqrt{g_{tt}}}  \Gamma^{\underline t}
\(  {h b \over 2g_{tt}} D_t - i q a \)
+ {1\over 4} \Gamma^{\underline r} \( \tilde f_0 \)_{(2)} \right]\psi
\ee
\ewt
where
\be\( \tilde f_0 \)_{(2)} = \half \sqrt{g^{rr}} \partial_r \left( {hb^2 \over g_{tt}} \right) \ .\ee

\section{Other contributions} \label{app:other}

In the main text we concentrated on the contributions from a Fermi surface
in Fig.~\ref{fig:polarization_ads}. 
Here we consider various other contributions to the conductivities which we neglected in the main text. These include the contributions from seagull diagrams depicted in Figure~\ref{fig:seagull} which arise from quartic couplings involving the graviton (schematically, terms like $h^2 \bpsi\psi$ and $h A \bpsi \psi$ in the Lagrangian), and 
contributions from the oscillatory region, i.e. the region in momentum space where the IR dimension for  
the fermionic operator is imaginary. 
We justify our neglect of these contributions by showing that they are nonsingular in temperature and thus 
are subleading compared to those considered in the main text. Our discussion will be schematic. 


\subsection{Seagull diagrams}

We write the schematic form of a seagull diagram $S$ with external Euclidean frequency $\Om_l$:
 \bea
&& S^{ij}(\Om_l) =  T \sum_{i\om_m}  \int {d^{d-1} k  \ov (2 \pi)^{d-1}} \int dr_1 \sqrt{g(r_1)}  \times \nonumber \\
  && \tr \le(P^j (r_1; -i\Om_l, \vk) D_E(r_1,r_1;i\om_m,\vec{k}) P^i (r_1; i\Om_l, \vk)\ri)  \nonumber \\
\label{seagull}
 \eea
Here $P^i$ contains the information of the graviton or gauge field propagators and vertex and is deliberately left vague.  It is shown in equation \eqref{matsumF} in Appendix \ref{app:useF} that the Matsubara sum can be rewritten in terms of an integral over the bulk spectral density
\be
T\sum_{i\om_m} D_E(r_1,r_2;i\om_m,\vk) = \int \frac{d\om}{2\pi}\tanh\le(\frac{\beta\om}{2}\ri) \rho(r_1,r_2;\om,\vk)
\ee
Now as before we express the bulk spectral density in terms of the boundary spectral density $\rho_B$ and bulk normalizable wave functions $\psinorm_{a}(r)$: $\rho(r,r', \omega, k) = \psinorm_{\alpha}(r,k) \, \rho_B^{\alpha \beta} (\omega, k)\, \overline{\psinorm_{\beta}} (r',k)$.  Away from the Fermi surface the discussion of~\eqref{seagull} parallels to that of the main text. In particular, the potential singular $T$ dependence coming from the IR part of the vertex is compensated by $T$-dependence of the spectral function, and as a result is non-singular. 
Near a Fermi surface, the eigenvalues of the boundary spectral density matrix take the form
\eqref{rhobF1}--\eqref{ooen}. As we take $T \to 0$, since all the other factors in~\eqref{seagull} are analytic in momentum $k$, the $k$-integral can be schematically written as 
\be 
\int dk  {\Im \Sig \ov (k-k_F(\om,T) - \Re \Sig)^2 + (\Im \Sig)^2}  \times \cdots 
\ee
with all the other factors evaluated at $k=k_F (\om, T)$. The above integral can then be straightforwardly 
integrated and yields a contribution of order $O(T^0)$. Also similar to the discussion in the main text, the potential singular contribution from the effective vertex is suppressed at $k = k_F (\om, T)$, resulting a non-singular contribution.

\subsection{Oscillatory region contribution}\label{app:osck}

We return to the expression \eqref{necon1} for the conductivity as an integral over $k$.
In the previous sections we have studied the temperature-dependence of the region of $k$ near a Fermi surface at $k_F$.
Here we ask whether the ``oscillatory region''
(values of $k$ such that particle production occurs in the $AdS_2$ region of the geometry)
make significant contributions to the conductivity.
We will find that their contribution is finite at $T = 0$,
and hence subleading compared to the $T^{-2\nu}$ behavior of a Fermi surface.
We will not worry about numerical factors here.

For illustration, let us look at the DC conductivity~\eqref{eq:finalsig1}--\eqref{fetr}, which we copy here for convenience
\be
\label{osccond}
\sig_{\rm DC}
  =   -{C \ov 2 } \int_0^\infty d k k^{d-2} \int {d \om \ov 2 \pi}  \, {\p f (\om) \ov \p \om} \,
  \rho_B^2 (\om ,k) \, \lam^2 (\om ,k, T)  \ .
  \ee
In low temperature limit, the fermion spectral density
in the oscillatory region
may be written
\bea
\label{oscdentistry}
\rho_{{\rm osc}}(\omega, k ) = \Im \frac{e^{ i \theta} |c| \omega^{i \lambda} + 1 }{e^{ i \theta'} |c| \omega^{i \lambda} + 1 }~~.
\eea
where $ c \omega^{i \lambda}$ is the IR Green's function
(with the IR dimension imaginary) at $T=0$. $e^{ i \theta, \theta'}$ are  phases.
This expression is valid in the oscillatory regime $ k < k_{{\rm osc}} = \sqrt{ q^2 e_d^2 - m_k^2 R_2^2}$
(see eqn (68) of \cite{Faulkner:2009wj}).
The important point now is that as a function of $\omega$, the object
\eqref{oscdentistry}
 is {\it bounded}.
In fact, it can be bounded uniformly in $k$ (\ie\ we can find a constant $\AA$
such that $\AA > \rho_{{\rm osc}}(\omega, k)$ for all $k < k_{{\rm osc}}$).
Numerical evidence for this statement is figure 7 of \cite{Liu:2009dm}.
In the oscillatory region~\eqref{effr1} still applies except that $\nu_k$ is now imaginary. 
Thus we see that in the oscillatory region the effective vertex $\lam$ is also nonsingular 
in the limit $T \to 0$.

 We thus conclude that the contribution from the oscillatory region to $\sig_{DC}$ 
is nonsingular in the low temperature limit.

\section{Some useful formulas} \label{app:useF}

Here we compile some standard and useful identities that are used in the main text. 

\subsection{How to do Matsubara sums} \label{sec:mat}

A standard trick is perform a Matsubara sum over discrete imaginary Euclidean frequencies is to rewrite the sum over  frequencies as a contour integral (we consider fermionic frequencies here)
\be
T \sum_{i\om_m} \to \frac{1}{2\pi i}\int_C d\om \frac{1}{2}\tanh\le(\frac{\beta \om}{2}\ri) \label{tanhf}
\ee
where we take the contour $C$ to encircle all the poles. A convenient deformation of the contour is to make it into two lines, one running left to right just above the real axis and the other running right to left just below. In the fermionic case this encircles all the poles. Exactly parallel manipulations can be used to obtain
the identity ($\Om_l = {2 \pi l \ov \beta}$ with $l$ an integer)
 \be \label{fnrn}
 T\sum_{\om_m} {1 \ov i(\om_m + \Om_l)- \om_1}\frac{1}{i\om_m  - \om_2}
 = \pm {f (\om_1 ) - f(\om_2) \ov \om_1 - i \Om_l - \om_2}
 \ee
with
 \be
 f(\om) = {1 \ov e^{\beta \om} \pm 1}
 \ee
where the upper (lower) sign is for fermion (boson).

One can apply this kind of technique for the frequency sums involving  spinor bulk-to-bulk propagator.
As an example consider the spectral decomposition of a Euclidean correlation function
\be
D^E (i \om_n) = \int \frac{d{\Om}}{2\pi} \frac{\rho (\Om)}{i\om_n + \Om}  \ .
\label{spectE}
\ee
We then find that 
\bea
 T \sum_{i\om_m} D_E(i\om_m) && =  \frac{1}{2\pi i}\int\frac{d\Om}{2\pi}\int_{-\infty}^{\infty}
d\om\frac{1}{2}\tanh\le(\frac{\beta\om}{2}\ri)\rho(\Om)
\cr
&\times &\left[\frac{1}{\om + i\epsilon - \Om}-\frac{1}{\om - i\epsilon - \Om}\ri]
\eea
The bracketed factor reduces to a delta function, and we find 
\be
T \sum_{i\om_m} D_E(;i\om_m) = \ha \int \frac{d\Om}{2\pi}\tanh\le(\frac{\beta\Om}{2}\ri) \rho(\Om) \ . \label{matsumF}
\ee
Similarly consider
\bea
\nonumber
S(\Om_l) &\equiv & T\sum_{\om_m} D^E_1( \om_m + \Om_l) D^E_2(\om_m )
 \cr
 &= & T\sum_{\om_m}\int \frac{d\om_1}{2\pi}\frac{d\om_2}{2\pi} \frac{\rho_1(\om_1)}{i(\om_m + \Om_l) - \om_1}\frac{\rho_2(\om_2)}{i\om_m  - \om_2},
\eea
where $\om_m = {2 \pi m \ov \beta}$ with $m$ a half integer (an integer) for fermions (bosons), while $\Om_l = {2 \pi l \ov \beta}$ with $l$ an integer.
Then using~\eqref{fnrn} we find that
\be
S(\Om_l) = \pm \int \frac{d\om_1}{2\pi}\frac{d\om_2}{2\pi} \frac{f(\om_1) - f(\om_2)}{\om_1 -i \Om_l - \om_2 }\, \rho_1(\om_1)\, \rho_2(\om_2) \ . \label{sum_ident}
\ee

\subsection{Useful integrals} \label{sec:integrals}

We now give details for some integrals which we encountered in the main text. First consider the integral in~\eqref{imexps} 
\be 
I_B \equiv  \int dk \, \rho_B (\om_1, k) \rho_B (\om_2, k)
\ee
where 
\be 
\rho_B (\om,k) = 2 \Im \le({h_1 \ov k-  k_F (\om,T) - \Sig} \ri)
\ee
The above integral has the form 
\be 
I (a,b) = - \int dk \, \le({1 \ov k-a} - {1 \ov k-a^*} \ri)\le({1 \ov k-b} - {1 \ov k-b^*} \ri)
\ee
which can be carried out straightforwardly by opening the parenthesis and evaluating each term using contour integration. Note that since both $a$ and $b$ lie in the upper half plane, only two among the four terms contribute and we find 
\be 
I(a,b)= 4 \pi \Im \le({1 \ov b^*- a} \ri) 
\ee
We thus find that 
\bea 
I_B &=&  4 \pi \Im \le({h_1^2 \ov {1 \ov v_F} (\om_1 - \om_2) + \Sig^* (\om_1) - \Sig (\om_2)} \ri)  \cr
& = & 2 \pi h_1 \rho_B (\om_1, K_1) = 2 \pi h_1 \rho_B (\om_2, K_2) 
\eea
with $K_1 = k_F (\om_2,T) + \Sig^* (\om_2)$ and $K_2 =  k_F (\om_1,T)  + \Sig^* (\om_1)$.

\end{appendix}

\end{document}